\documentclass{emulateapj}

\usepackage{latexsym}  
\usepackage{epsfig}
\usepackage{graphics}
\usepackage{natbib}
\usepackage{color}
\usepackage{amsmath}
\usepackage{hyperref} 

\newcommand{\Xe}{{X_{\rm e}}}
\newcommand{\Xefid}{{X_{\rm e}^{\rm fid}}}

\newcommand{\ns}{n_{\rm s}}

\newcommand{\Planck}{{\sc Planck}}
\newcommand{\intd}{{\rm d}}
\def\avrg#1{{\langle #1 \rangle}}

\slugcomment{Submitted to ApJ}
\shorttitle{Eigen-analyses of recombination histories}

\shortauthors{Farhang et al.}
\begin{document}

\title{Semi-blind Eigen-analyses of Recombination Histories Using CMB Data}

\author{M. Farhang\altaffilmark{1,2}, J.~R. Bond\altaffilmark{1}  and J. Chluba\altaffilmark{1} }
\altaffiltext{1}{Canadian Institute for Theoretical Astrophysics, 60 St George, Toronto ON, M5S 3H8}
\altaffiltext{2}{ Department of Astronomy and Astrophysics, University of Toronto, 50 St George, Toronto ON, M5S 3H4}

\begin{abstract}
Cosmological parameter measurements from CMB experiments such as
Planck, ACTpol, SPTpol and other high resolution follow-ons
fundamentally rely on the accuracy of the assumed recombination model,
or one with well prescribed uncertainties. Deviations from the
standard recombination history might suggest new particle physics or
modified atomic physics. Here we treat possible perturbative
fluctuations in the free electron fraction, $\Xe(z)$, by a semi-blind
expansion in densely-packed modes in redshift. From these we construct
parameter eigenmodes, which we rank order so that the lowest modes
provide the most power to probe the $\Xe(z)$ with CMB
measurements. Since the eigenmodes are effectively weighed by the
fiducial $\Xe$ history, they are localized around the differential
visibility peak, allowing for an excellent probe of hydrogen
recombination, but a weaker probe of the higher redshift helium
recombination and the lower redshift highly neutral freeze-out
tail. We use an information-based criterion to truncate the mode
hierarchy, and show that with even a few modes the method goes a long
way towards morphing a fiducial older {\sc Recfast} $X_{\rm e,i} (z)$
into the new and improved {\sc CosmoRec} and {\sc HyRec} $X_{\rm e,f}
(z)$ in the hydrogen recombination regime, though not well in the
helium regime. Without such a correction, the derived cosmic
parameters are biased. We discuss an iterative approach for updating
the eigenmodes to further hone in on $X_{\rm e,f} (z)$ if large
deviations are indeed found. We also introduce control parameters that
downweight the attention on the visibility peak structure, e.g.,
focusing the eigenmode probes more strongly on the $\Xe (z)$
freeze-out tail, as would be appropriate when looking for the $\Xe$
signature of annihilating or decaying elementary particles.
\end{abstract}

\section{Introduction}
The {\Planck} Surveyor\footnote{\url{http://www.rssd.esa.int/Planck}}
is now well into its mission, observing the temperature and
polarization anisotropies of the cosmic microwave background (CMB)
with unprecedented accuracy \citep{hfi11,lfi11}.  Both {\sc ACT}
\citep[e.g., see][]{haj10, dun10, das11} and {\sc SPT} \citep{lue10,
  van10} are pushing the frontier of $TT$ CMB power spectra at small
scales, and in the near future {\sc
  SPTpol}\footnote{\url{http://pole.uchicago.edu/}} \citep{SPTpol} and
{\sc ACTPol}\footnote{\url{http://www.physics.princeton.edu/act/}}
\citep{nie10} will provide additional small scale $E$-mode
polarization data, complementing the polarization power spectra
obtained with {\Planck} and further increasing the significance of
$TT$ power spectra.

Using these datasets, cosmologists will be able to determine the key
cosmological parameters with high precision \citep{PlanckBlueBook,
  Tauber2010}, making it possible to distinguish between {various
  models} of {\it inflation} \citep[e.g. see][for recent constraints
  from WMAP]{Komatsu2010} by measuring the precise value of the
spectral index of scalar perturbations, $n_{\rm s}$, and constraining
its possible running, $n_{\rm run}$, as well as the tensor-to-scalar
ratio, $r$.  In addition many non-standard extensions of the minimal
inflationary model are under discussion, and the observability of
these possibilities with {\Planck}\citep{PlanckBlueBook} and future
CMB experiment is being considered.

These encouraging observational prospects have motivated various
independent groups \citep[e.g. see][]{Dubrovich2005, Chluba2006,
  Kholu2006, Switzer2007I, Wong2007, Jose2008, Karshenboim2008,
  Hirata2008, Chluba2008a, Jentschura2009, Labzowsky2009, Grin2009,
  Yacine2010} to assess how uncertainties in the theoretical treatment
of the cosmological recombination process could affect the science
return of {\sc Planck} and future CMB experiments.  The precise
evolution of the free electron fraction, $\Xe$, with time influences
the shape and position of the peak of the Thomson visibility function,
which defines the last scattering surface \citep{Sunyaev1970,
  Peebles1970}, and hence controls how photons and baryons decouple as
electrons recombine to form neutral helium and hydrogen atoms.
Consequently, the ionization history changes the acoustic oscillations
in the photon-baryon fluid during recombination and therefore directly
affects the CMB temperature and polarization power spectra.  For the
analysis of future CMB data this implies that in particular close to
$z\sim 1100$ the ionization history better be understood at the $\sim
0.1\%$ level.

Probing the ionization history in time is equivalent to probing it in
space with the light cone relating the two.  Thus what we try to do in
this paper, namely to come up with optimized probing functions for the
recombination history, is quite akin to creating probes of the spatial
structure of the boundary between HII and neutral hydrogen
regions. Here of course we look from neutral to ionized, the
cosmological recombination problem being an inside-out HII region,
except in a predominantly electron scattering regime with a very large
photon to baryon ratio which lowers the transition temperature between
ionized and neutral.

The old recombination standard was set by {\sc Recfast}
\citep{sea99, Seager2000}, but its reliability for the
precision cosmology was brought into question, e.g., by
\citet{Seljak2003} .  For the standard six parameter cosmology in
particular our ability to measure the precise value of $n_{\rm s}$ and
the baryon content of our Universe may be compromised if modifications
to the recombination model of {\sc Recfast} are neglected
\citep{rub10, sha11}, introducing biases of a few $\sigma$ for
      {\Planck}.

Currently it appears that {\it all} important corrections to the {\it
  standard} recombination scenario (SRS hereafter) have been
identified \citep[e.g., see][for an overview]{Fendt2009, rub10}.  The
new recombination codes, {\sc CosmoRec} \citep{Chluba2010b} and {\sc
  HyRec} \citep{Yacine2010c} both account for these modifications to the
SRS, superseding the physical model of {\sc Recfast} and allowing fast
and accurate computation of the ionization history on a model-by-model
basis.  {\sc CosmoRec} and {\sc HyRec} presently agree at a level of
$\sim 0.1\%-0.2\%$ during hydrogen recombination, so that from
standard recombination physics little room for big surprises seems to
be left.

However, {\it what if something non-standard happened? What if
  something was overlooked in the standard recombination scenario?}
From the scientific point of view the ionization history is a
theoretical ingredient to the cosmological model, which usually is
assumed to be precisely known and not subject to direct measurement.
Clearly, it is important to estimate the possible level of uncertainty
in the recombination model and to confront our understanding of the
recombination problem with direct observational evidence.  Here we
describe how well future cosmological data alone are able to constrain
possible deviations from the SRS.

In the past, several non-standard extensions of the recombination
scenario have been considered.  These include models of {\it delayed
  recombination}, in which hypothetical sources of extra photons that
can lead to ionizations or excitations of atoms are introduced using
simple parametrizations \citep{Peebles2000}.  In particular, models of
{\it decaying} \citep[e.g., see][]{che04, zha07} and {\it annihilating
  particles} \citep[e.g., see][]{pad05, zha06, hue09, sla09, hue11}
were discussed.  In addition to extra photons, {\it varying
  fundamental constants} \citep[e.g., see][]{Kaplinghat1999,
  Scoccola2009, Galli2009b} could affect the recombination dynamics in
subtle ways.

All these ideas rely on a specific model for the (physical) process
under consideration, with the derived constraints depending on the
chosen parametrization.  This minimizes the number of additional
parameters, but does not allow us to answer questions about more
general perturbations around the SRS and how well they can actually be
constrained.

Here we approach this problem in a different way.  We introduce
perturbations to the SRS over a wide range of redshifts around
hydrogen ($z\sim 1100$) and helium ($z\sim 1800$) recombination, using
different basis functions.  We then compute the corresponding signals
in the CMB power spectra and perform a principle component
decomposition to obtain eigenmode functions, ordered with respect to
the level at which they can be constrained by the data.  We study in
detail how the eigenmodes depend on different experimental settings,
the fiducial model, as well as the chosen parametrization for the
recombination perturbations.

Our method is similar to the one used by \citet{mor08}, where the
eigenmodes for different reionization scenarios ($6\lesssim z\lesssim
30$) were constructed. However, here we explicitly construct the mode
functions at redshifts $z\gtrsim 200$, with particular attention to
the dependence of the eigenmodes on different assumptions.  We
investigate how to use our prior knowledge of possible perturbations
of the ionization history to choose the parametrization which is more
preferred by the data.  We also carry out a careful convergence study
and show the equivalence of different basis functions (e.g.,
triangles, Gaussian bumps, Fourier series and Chebyshev polynomials).
We particularly focus on the helium recombination problem, showing
that in the absence of very tight constraints on the hydrogen
recombination, we will not be able to unravel well remaining
uncertainties in helium recombination with CMB data.

Similarly, small changes in the freeze-out tail of recombination are
only weakly constrained, if possible ambiguities during hydrogen
recombination are included.

Details of the general methodology to construct the eigenmodes for
perturbations to ionization history are given in \S~\ref{sec:method}.
In \S~\ref{sec:rec_EMs} we computed different eigenmodes over a rather
wide redshift range ($z\in[200,3000]$).  At the end of
\S~\ref{sec:rec_EMs}, we develop a criterion which allows us to
truncate the hierarchy of the eigenmodes based on their information
content.  In \S~\ref{sec:measure} the modes are applied to two
specific examples of ionization scenarios, illustrating how the method
should be used with real CMB data. At the end of this section, we also
discuss how the approach should be iterated if hints toward a
considerable difference between the assumed and true model of
recombination are indicated by the data.  We close the paper by a
brief discussion.

\section{Methodology}
\label{sec:method}
In this section we introduce the approach and parametrization used to
construct the principle components, or the eigenmodes, which will be
used to describe possible corrections to the recombination
scenario. Our method is mainly driven by the assumption of {\it small
  relative} perturbations around the fiducial model computed with the
{\sc Recfast} code (\cite{sea99}; see \cite{won08} for recent
updates).  As an example we have in mind the recombination corrections
obtained with refined recombination models \citep{Chluba2010b,
  Yacine2010c}. However we also briefly discuss the possibility to
constrain significant changes in the freeze-out tail of recombination
and modes that mainly focus on helium recombination.

Throughout this paper the cosmic parameters, which will be referred to
as the standard (cosmological) parameters, are $(\Omega_{\rm b}h^2,
\Omega_{\rm dm}h^2,H_0,\tau,n_{\rm s},A_{\rm s})$ as measured by
WMAP7\footnote{http://lambda.gsfc.nasa.gov/product/map/dr4/params
  \\ /lcdm\_sz\_lens\_wmap7.cfm}, unless stated otherwise.  In several
cases we also vary $Y_{\rm p}$ as a seventh parameter. Lensing is
included in all simulations if not explicitly stated otherwise.\\

\begin{figure*} 
\begin{center}
\includegraphics[scale=0.7]{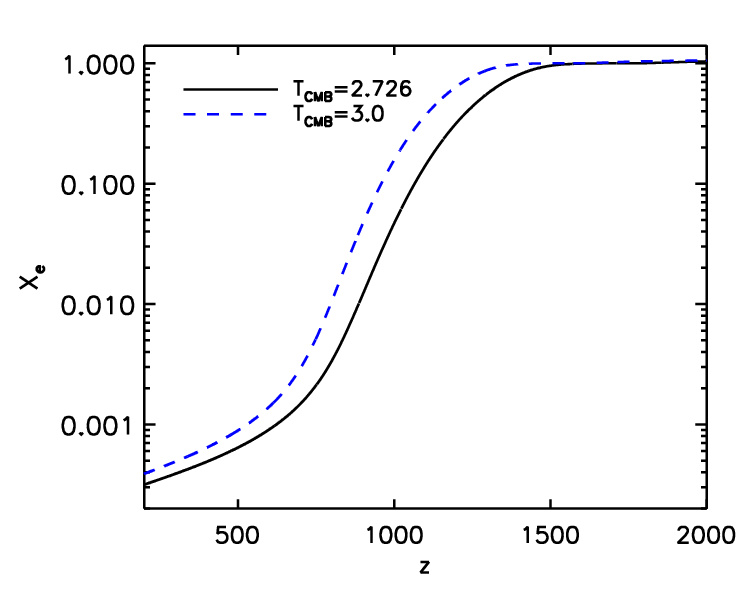}
\includegraphics[scale=0.7]{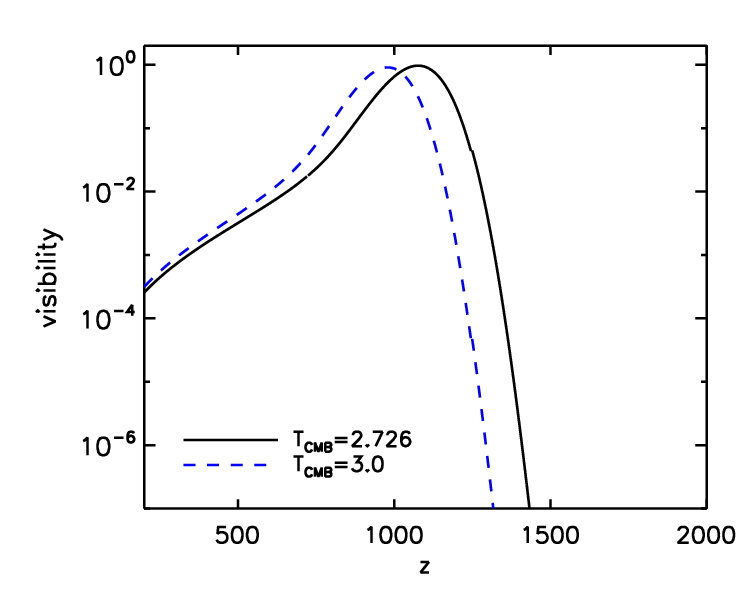}
\end{center}
\caption{The cosmological ionization history (left), $\Xe\equiv N_{\rm
    e}/[N_{\rm p}+N_{\rm HI}]$, and differential visibility function
  (right) for the standard recombination scenario with $T_{\rm
    CMB}=2.726$K \citep{fix09} contrasted to a case with $T_{\rm
    CMB}=3$K.  Here $N_{\rm p}$ and $N_{\rm HI}$ represent the number
  density of ionized and neutral hydrogen, while $N_{\rm e}$ denotes
  the number density of free electrons.}
\label{xevis_fid}
\end{figure*}
\subsection{The standard recombination scenario}
\label{sec:SRS}
As explained in the introduction, the cosmological recombination
history is one of the major theoretical inputs for computations of the
CMB anisotropies. Consequently, high precision unbiased cosmic
parameter measurements from current and future CMB experiments require
a sufficiently accurate model for hydrogen and helium recombination.

The ionization fraction for the SRS is shown in the left panel of
Fig.~\ref{xevis_fid}. It was calculated using {\sc Recfast v1.4.2},
which accounts for some of the modification to helium recombination
\citep{Kholupenko2007, Switzer2007I, Jose2008, Chluba2009c} using
fudge parameters, but neglects detailed radiative transfer corrections
\citep[see][and references therein]{Chluba2010b, Yacine2010c} around
$z\sim 1100$.  The solid curve corresponds to an ionization fraction
with the measured temperature of the CMB radiation, $T_{\rm CMB} \sim
2.726 {\rm K}$ \citep{fix09}.  For comparison and to illustrate the
temperature dependence of the ionization history, the ionization
fraction corresponding to $T_{\rm CMB}=3 {\rm K}$ is also plotted
(dashed curve).  A larger value of $T_{\rm CMB}$ means more photons in
the Wien tail of the CMB blackbody, so that the matter is kept ionized
until lower redshift.

On the right the corresponding differential visibility functions (or
visibility functions for short) are plotted:
\begin{equation}
g(z)\equiv-\frac{{\rm d} e^{-\tau(z)}}{{\rm d}\eta},
\end{equation}
where $\eta$ is the conformal time and $\tau$ is the Thomson
scattering optical depth from redshift $z$ to now.

The visibility function describes the probability that a photon we
observe today last scattered off free electrons at a certain position
along the line of sight.  The CMB anisotropies formed mainly during
the epoch of hydrogen recombination defined by the peak of the
visibility function located at redshift $z\sim 1100$.  They are thus
most sensitive to changes around the maximum of visibility.  For
example, an increase in the width of the visibility bump corresponds
to a more extended or slower recombination process, leading to more
Thomson scatterings of photons off free electrons.  These scatterings
lead to the cancellation of the CMB anisotropies along the line of
sight on scales comparable and smaller than the recombination width,
while enhancing the polarization signal on larger scales.  The
location of the maximum of the visibility function for an assumed
cosmological model, on the other hand, determines the distance to the
last scattering surface.  This in turn determines the positions of the
peaks of the CMB power spectra. Similarly, any change in the
ionization history, through affecting the visibility, would lead to
(possibly measurable) changes in the CMB power spectra.

As the right panel of Fig.~\ref{xevis_fid} indicates, at high
redshifts $z\gtrsim 1400$ the visibility function falls off very
quickly. At those times the number of free electrons is still so large
that scatterings occur very frequently and the mean free path is very
short. Consequently, the part of the ionization history which is
connected to helium recombination mainly affects the damping tail of
the CMB anisotropies, but even there the effect is rather moderate, as
in the redshift range \footnote{The recombination of doubly ionized
  helium ends around redshift $z\sim 5000$.} $1400 \lesssim z\lesssim
3000$ helium can at most alter the number of electrons by $\sim 8\%$.

\subsection{Choice of perturbation parametrization}
\label{sec:motivation}
We now introduce small perturbations to the SRS that allow
construction of the eigenmodes.

There are different ways to parametrize the perturbations to the
ionization history in a (semi-)model-independent way.  For example, to
study how well the low redshift ionization history ($z$ in the
interval $[6,30]$) can be constrained by future CMB data, \cite{hu03}
and \cite{mor08} used the changes in the ionization fraction in
different redshift bins, $\delta \Xe(z)={\rm const}$, to parametrize
the uncertainties. This is a valid choice for the low redshift region,
because our ignorance of the underlying model of reionization does not
suggest any preferred non-uniform weighting of the perturbations at
different redshifts.  In this regime $\delta\Xe(z)$ probes the
ionization fraction itself and not perturbations guided by a fiducial
model.  The results from this choice of parametrization are shown to
be fiducial model-independent which is expected due to the weak signal
from the reionization process.

In contrast to this, at high redshifts ($z\sim 1100$) there is strong
theoretical support for the exhaustively studied model of
recombination in the realm of standard atomic physics and radiative
processes.  Also as it was mentioned in the introduction, the current
generation of CMB data is sensitive to changes in $\Xe$ at the level
of a few percent.  Therefore the main assumption in this paper is that
the {\it fiducial} model for the ionization history, $\Xefid(z)$, is
close to the true underlying history, $\Xe(z)$, which we are looking
for.  We call this method {\it semi-blind} emphasizing our belief in
the SRS as the framework of recombination, with the search for
deviations being limited to small perturbations around this reference
model.  The goal is to detect or place upper limits on these potential
small deviations. Clearly, if data point toward significant deviations
from the SRS, an iterative approach should be adopted, as will be
discussed in \S~\ref{sec:beyond_small}.

With small deviations in mind we can write:
\begin{equation*}
\Xe(z)=\Xefid(z)+\delta \Xe(z), 
\end{equation*}
with $|\delta \Xe|/\Xefid \ll 1$. A natural parameter to describe the
perturbation is then the relative deviation in the ionization
fraction:
\begin{equation}
\label{eq:dXe_param_1}
\delta u (z)\equiv \delta \Xe(z)/\Xefid(z) \quad\text{with} \quad |\delta u (z)| \ll 1.
\end{equation}
This parametrization has the advantage of always satisfying the
necessary condition $\Xe \ge 0$. It is also straightforward to fulfill
the $\Xe \le X_{\rm e,max}$ condition in the simulations, where
$X_{\rm e,max}$ is determined by $Y_{\rm p}$, the primordial helium
mass abundance, through $X_{\rm e,max} \simeq 1+Y_{\rm p}/2(1-Y_{\rm
  p}) $. The parametrization in Eq.~\eqref{eq:dXe_param_1} weights
possible perturbations at different redshifts by the fiducial
ionization fraction.  This implies that for $\delta u (z)=\rm const$
the absolute change in the ionization fraction $|\delta \Xe|$ is
down-weighted in the freeze-out tail of $\Xe$ ($z\lesssim 800$; see
Fig.~\ref{xevis_fid}), compared to perturbations around maximum
visibility ($z\sim 1100$) where $\Xe\sim 1$. Throughout this paper
$\delta u (z)$ as defined in Eq.~\eqref{eq:dXe_param_1} will be our
main choice of parametrization.

A more general parameter which includes the above parametrization as a
special case is given by:
\begin{equation}
\label{eq:sig}
\delta u (z) \equiv \delta \Xe(z)/ [\Xefid(z)+\sigma(z)],
\end{equation}
where $\sigma(z) \ge 0$ can be a constant or otherwise convenient
function of redshift allowing to focus on different redshift ranges of
interest.  In particular, when considering possible modifications to
the ionization history introduced by {\it energy injection},
e.g. because of annihilating dark matter, or decaying relic particles
\citep{che04, pad05, zha06, zha07, hue09, sla09,hue11}, where the
freeze-out tail of recombination is disturbed, a value of $\sigma\gg
\Xefid$ would be a proper choice, giving higher weight to the
perturbations in the lower redshift part (see \S~\ref{sec:param} and
Fig.~\ref{recon}).  In the limit of a high value of $\sigma$ relative
to the fiducial $\Xe$ the parameters approach $\delta u (z)=\delta
\Xe(z)$ which uniformly weights perturbations at different redshifts.
This, as already discussed, is a good choice for regions where there
is no strong {\it a priori} belief in the underlying model or if the
redshifts of interest have comparatively low $\Xe$ where $\delta u
(z)$ with $\sigma=0$ does not lead to strong enough signals to probe.
In principle, a conveniently chosen redshift dependent $\sigma(z)$ is
a tool to effectively incorporate our prior knowledge of the
ionization history in the parametrization of its perturbations. For
example, with $\delta u (z)$ defined by Eq.~\eqref{eq:sig} one can
smoothly interpolate between relative and absolute perturbations to
$\Xe$, at high and low redshifts respectively. Also it is clear that
one can focus on different parts of the recombination history by
limiting the redshift range over which the eigenmodes are constructed
[e.g., just on reionization ($0\lesssim z\lesssim 30$) or helium
  recombination ($1400\lesssim z\lesssim 3000$)].

\subsubsection{Alternative parametrizations}
\label{sec:alt_parametrizations}
Instead of directly perturbing the ionization fraction, as we chose
here, it is plausible to parametrize possible changes in the physical
{\it sources} of perturbation to the ionization history, such as
energy injection in the medium which leads to excitation or ionization
of atoms, or the ${\rm Ly}\alpha$ escape probability during
recombination (see introduction).  For example \cite{mit10} chose the
number of photons in the IGM per baryon in collapsed objects as the
parameter to study the low redshift ionization history.
Alternatively, one could modify the fudge factors or functions in {\sc
  Recfast}, or alter the expansion rate given by the Hubble factor,
$H(z)$.

Each of these possibilities implies different priors on the regions
that can be altered and, e.g., in the case of $H(z)$ other aspects of
the cosmological model are also affected. They also cover, in general,
only a limited class of changes to the recombination history.  When
interested in perturbations to the ionization history, $\Xe$ is the
physical quantity which, via the visibility function and the optical
depth, most directly enters the Boltzmann equations which describe
radiation anisotropies and can be solved using the Boltzmann codes
such as {\sc Camb} \citep{CAMB} or {\sc Cmbfast} \citep{CMBFAST}.  The
ionization fraction has the additional advantage, over the visibility
and the optical depth, of being straightforward to limit to physically
allowed values. The nearly direct mathematical encounter of $\Xe$ with
CMB anisotropies guarantees that any perturbation in the plasma that
would lead to changes in the radiation anisotropies should go through
and thus be reflected in $\Xe$.  Therefore the relative changes in
$\Xe$ constitute our preferred physical parameters.

We close by mentioning that, it is also theoretically possible to
consider different variables for time such as (conformal) time,
optical depth and scale factor. However, for our purpose we choose to
work with redshift to describe temporal dependence.  In principle
different parametrizations, if they cover the same range of physical
perturbations, can be transformed to one another with the proper
change of the {\it a priori} distribution of parameters. Here, in the
absence of physically motivated constraints, a uniform prior is
assumed for perturbations at different redshifts regardless of
parametrization (here, e.g., for various values of $\sigma$ in
Eq.~\eqref{eq:sig}).  If the perturbation is strongly constrained by
data, the choice of the prior would not play a major role.

\begin{figure*} 
\begin{center}
\includegraphics[scale=0.72]{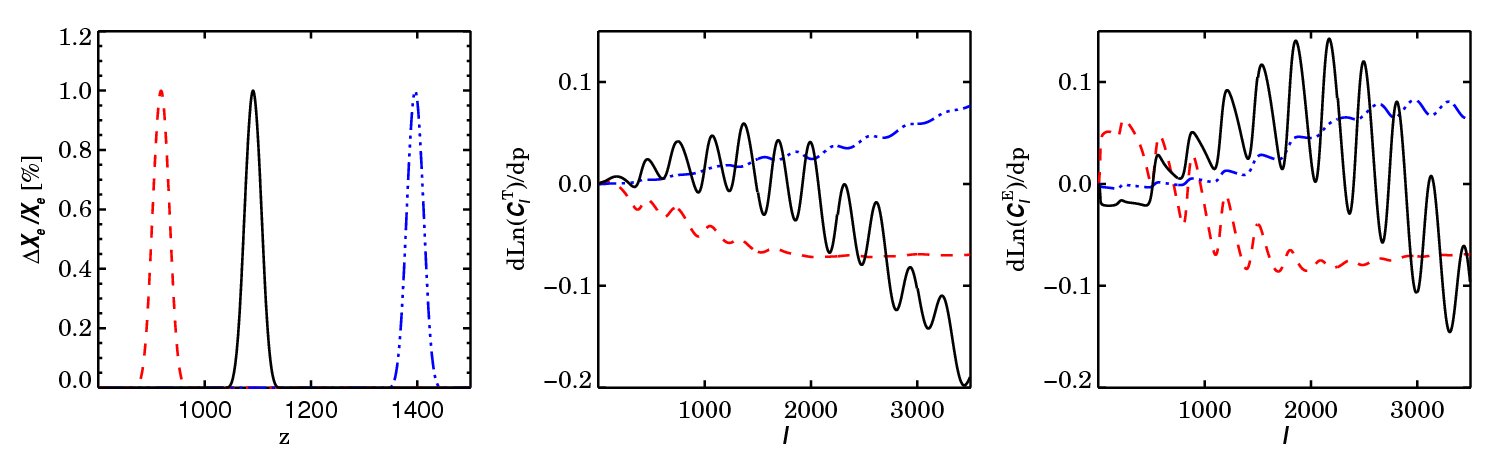}
\end{center}
\caption{Perturbations in the $\Xe$ history, in the form of $M_4$
  splines (left) and the derivatives of the $C_\ell$'s with respect to
  the amplitude of each perturbation ($TT$ power spectrum in the
  center, $EE$ on the right). }
\label{m4_dlnxdlncl}
\end{figure*}

\subsection{Basis functions and their different characteristics}
\label{sec:basis_functions}
Having chosen the parametrization, we now expand the perturbations in
a discrete set of mode (or basis) functions, $\varphi_i(z)$:
\begin{equation}
\delta u (z)=\sum_{i=1}^N  y_i \,  \varphi_i(z) + r (z) ~~~~~~ z_{\rm min}\le z \le z_{\rm max}
\label{paramexpand}
\end{equation}
and $\delta u (z)=0$ elsewhere. Here $r(z)$ is the residual and
$y_i$'s are the parameters defining the strength on the mode
$\varphi_i(z)$.  Often we take $\varphi_i(z)$ to be localized in $z$
about a {\it knot} value $z_i$, but this is not necessary. We can, for
example, choose the $\varphi_i(z)$ to form a complete orthonormal set
in which case $N\rightarrow \infty$ and the residual $r$ would be
zero.  Below, we will discuss different possibilities for the choice
of the mode functions.

\subsubsection{Localized basis functions}
\label{sec:Gaussians}
We first investigated localized Gaussian and triangular bumps as mode
functions. Both can be considered as approximations to the Dirac
$\delta$-function.  We define the $i$th basis function centered at
redshift $z_i$ and having width $\sigma_i$ by:
\begin{equation}
\label{eq:Gauss_bump}
 \varphi_i(z)\propto \exp \left(-{\frac{[z-z_i]^2}{2\sigma_i^2}}\right)
\end{equation}
for the Gaussian case and by
\begin{equation} \label{eq:tri_bump}
  \varphi_i(z) \propto \left\{ \begin{array}{ll} 
   1-\frac{|z-z_i|}{\sigma_i}  & \mbox{ $|z-z_i|< \sigma_i$},\\
   0 & \mbox{otherwise}, \end{array} \right.
\end{equation} 
for the triangles.  Triangular bumps were used earlier in the
principle component analysis of different reionization scenarios
\citep{hu03,mor08}.  In some circumstances, the sharp edges in the
triangles could cause numerical problems.  Smoothed localized
functions such as Gaussians and the $M_4$ splines introduced below
have numerical advantage.

As basis functions, it is more convenient if the set of $\varphi_i$'s
is an orthogonal set.  For this, there should be no overlap between
different bumps. On the other hand, there is no way to cover the whole
redshift range -a necessary condition for completeness- with a finite
number of non-overlapping bumps.  However, depending on the problem of
interest, the width and separation of the (overlapping) bumps can be
properly chosen to ensure all points in the redshift interval have
been covered, while at the same time the orthogonality is not strongly
violated.

 Instead of Gaussian and triangular bumps, one can also adopt an
 approach similar to that used in Smoothed-Particle Hydrodynamics
 (SPH), and think of the basis functions as window functions (or
 kernels) used to interpolate the properties of particles to any point
 in the medium. For us, the {\it particles} would be the spline {\it
   knots} \citep[e.g. see][]{spline} at the specific $z_i$ with the
 associated magnitude $y_i$.  There is a smoothing length $h$
 associated with the kernel over which the properties of the particles
 are smoothed. A commonly used kernel (other than the Gaussian
 considered above) is the {\it cubic $M_4$ spline}
 \citep[e.g.][]{mon05}, defined by:
\begin{align}
 \varphi_i(z) & \propto  M_4(|z-z_i|)\notag \\
    &  =\left\{  \begin{array}{ll} \frac{1}{6}[(2-q)^3-4(1-q)^3] & \mbox{~~$0 \leq q \leq 1$}; \\
   \frac{1}{6}(2-q)^3 & \mbox{~~ $1 \le q \le 2$};\\
   0 & \mbox{~~ $q> 2$ };\end{array} \right.     \label{eq:M4_spline}
\end{align}
where $q=|z-z_i|/h$.  Whereas the Gaussian kernel has non-zero
contributions from every redshift (though the range is usually
truncated beyond about $3\sigma$), the cubic spline is compact,
reaching zero for particles beyond $2h$. (As discussed in
\S~\ref{sec:converg}, we have used the smoothing length $h=1.5\delta
z$ where $\delta z$ is the {\it particle} separation).

We modified the publicly available code {\sc
  Camb}\footnote{http://camb.info/} to simulate CMB power spectra for
a more general recombination scenario that includes perturbations on
top of the SRS.  Introducing narrow features into the ionization
history also required an increase in the redshift sampling of
$\Xe$. We checked the numerical convergence and stability of the
results by using high accuracy settings.

As examples for localized perturbations, Fig.~\ref{m4_dlnxdlncl} shows
three perturbation functions $\delta u (z)=\delta \ln (\Xe)$ based on
$M_4$ splines (left panel) and the corresponding $C_\ell$ response in
$TT$ and $EE$.  The perturbations are located at different redshifts
and have equal widths.  We see that the amplitude of the response
typically increases at smaller scales indicating a change in the
duration of the recombination epoch (i.e., the effective width of the
visibility function). The $C_\ell$ response also has an oscillatory
component similar to a change of the position of the visibility
peak. These oscillations are most noticeable for the perturbations
close to the visibility peak ($z\sim 1100$) and together with the
overall slope of the $C_\ell$ response demonstrate that the
perturbation effectively changes the duration as well as the redshift
of the recombination.

\begin{figure*}
\begin{center}
\includegraphics[scale=0.72]{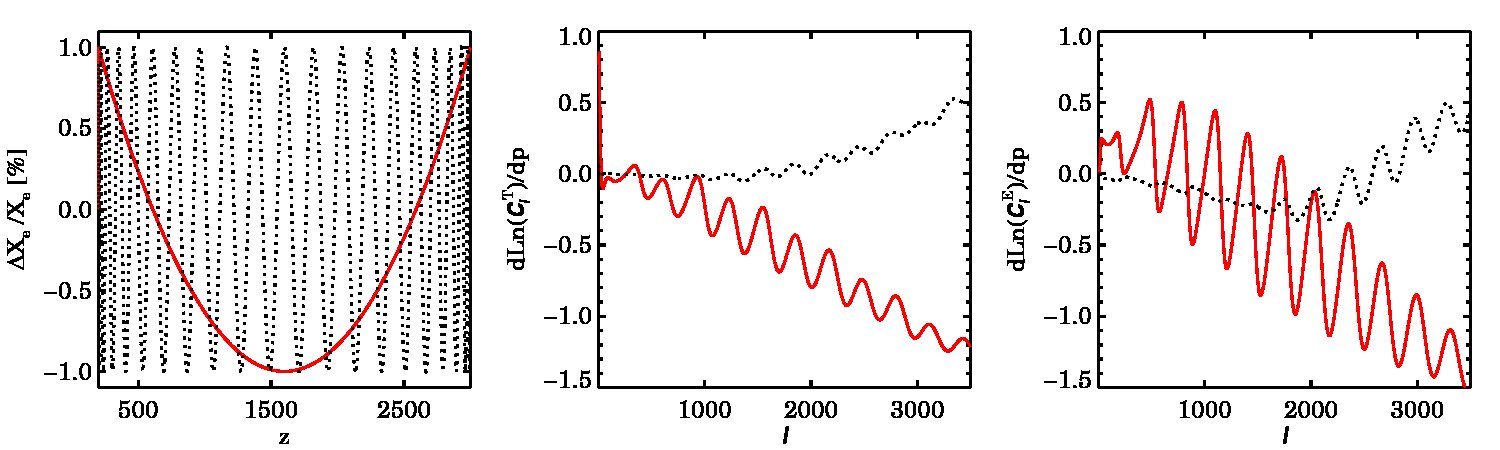}
\end{center}
\caption{Similar to Fig.~\ref{m4_dlnxdlncl} but for
 perturbations in the form of Chebyshev
 polynomials.}
\label{c_dlnxdlncl}
\end{figure*}
\subsubsection{Non-localized basis functions}
\label{sec:Fourier}
In this part we expand the perturbations in terms of two non-localized
basis functions which, unlike the localized case (with finite number
of basis functions), do not suffer from non-orthogonality.  If
sufficiently many functions are taken into account, these basis
functions, similar to the localized ones, can also be considered
complete in practice.
  
The most commonly used set of non-localized basis functions is the
{\it Fourier series}:
\begin{subequations}
\label{eq:fourier}
\begin{eqnarray}
\label{eq:fourier_a}
 &\varphi_i(z) \propto  \cos (i\pi y) ~~~~~~~i&=0,1,2, ...
\\
\label{eq:fourier_b}
&\varphi_i(z) \propto  \sin (i\pi y)   ~~~~~~~ i&=1,2, ...
\\
\label{eq:fourier_c}
&y=\frac{z-z_{\rm mid}}{\Delta z/2}
\end{eqnarray}
\end{subequations}
where $\Delta z$ and $z_{\rm mid}$ are the width and central point in
the redshift range of interest. Thus we have $|y| \le 1$ as is
required for Fourier expansion.

Alternatively, we can use {\it Chebyshev polynomials} of the first
kind, $T_i$, to form the basis.  These modes are constructed using the
recursion formula:
\begin{eqnarray*}
T_{i+1}(x)=2x\,T_i(x)-T_{i-1}(x)
\end{eqnarray*}
with initial conditions $T_0(x)=1$ and $T_1(x)=x$. In this case, the
variable $x$ is replaced by $y$ as given by Eq.~\eqref{eq:fourier_c}.
Chebyshev polynomials of the first kind are orthogonal with respect to
the weight function $w(x)=1/\sqrt{1-x^2}$.

The non-local basis functions are very different in nature from the
localized functions discussed in \S~\ref{sec:Gaussians}.  Therefore
the response of the observables (here the $C_\ell$'s) to the
perturbation $\delta u (z)$ in the form of these functions is also
expected to be rather different.  Figure~\ref{c_dlnxdlncl} shows the
$C_\ell$ responses when perturbing the ionization history using
Chebyshev polynomials with different frequencies.  We see that
perturbations with low frequencies, covering a large redshift range,
lead to $C_\ell$ responses with much larger mean amplitudes when
compared to the perturbations in the form of local bumps
(Fig.~\ref{m4_dlnxdlncl}).  However, as the frequency of the
oscillations of the basis function increases, the response becomes
weaker and its oscillations damp away. That is because neighbouring
oscillations lead to similar responses in the $C_\ell$'s with opposite
signs and can partially cancel out each other. In other words, the CMB
power spectra are less sensitive to the very high frequency
perturbations in the ionization history.

 In principle, in the limit of large mode number, all bases work well.
 However, we have found that for the recombination history, although
 non-localized basis sets have their virtues, the $z$-localized bases
 are better, especially if we are trying to describe narrow features
 in redshift.  We will return to this point in \S~\ref{sec:basis}.

\subsection{Constructing the eigenmodes}\label{sec:em_const}
\label{sec:PCA}
We have so far introduced our choice of parametrization for
characterizing potential perturbations to the ionization fraction and
illustrated how the perturbations in the form of these parameters
affect the CMB power spectra when expanded in different bases.  In
principle all the $q_i$'s (in Eq.~\eqref{paramexpand}) are needed for
a (nearly) complete reconstruction of a general perturbation $\delta u
(z)$. However, in practice data cannot constrain the perturbations in
detail in many cases.  As we saw in Fig.~\ref{c_dlnxdlncl}, very high
frequency perturbations are expected to lead to much smaller signals.

\subsubsection{Eigenmodes with fixed standard parameters (XeMs)}
\label{sec:PCA_std_const}
To avoid dealing with many correlated (and possibly weakly
constrained) parameters (i.e., the $q_i$'s), we construct a set of
their linear combinations which are uncorrelated with each other and
only keep those combinations that are most constrained by data.  This
procedure provides a hierarchy of mode functions and their
corresponding signals in the CMB temperature and polarization power
spectra.  Exclusion of the weakly constrained eigenmodes does not
affect the rest of the measurements since the eigenmodes describing
the recombination perturbations are by construction uncorrelated.

If $N$ parameters are used to characterize the perturbations, i.e., $1
\leq i \leq N$, the $N$ uncorrelated parameters will be determined by
the eigenmodes of the ($N\times N$) Fisher information matrix:
\begin{equation*}
F_{ij}\equiv -\left<\frac{\partial^2 \ln p_{\rm f}}{\partial  q_i \partial
  q_j}\right>,
\end{equation*}
where in the language of Bayesian analysis, $p_{\rm f}\equiv p({\bf
  q}|d,{\cal T})$ (with ${\bf q}=(q_1,...,q_N)$) describes the
posterior probability of the parameters $q_i$'s for the given data $d$
in the theory space ${\cal T}$, i.e., an update from the prior
probability distribution of the parameters, $p_{\rm i}=p({\bf q}|{\cal
  T})$ driven by the likelihood $p_{\rm f}= {\cal L}({\bf q}|d,{\cal
  T})p_{\rm i}/{\cal E}$ where ${\cal L}({\bf q}|d,{\cal T})\equiv
p(d|{\bf q},{\cal T})$ and the evidence ${\cal E}\equiv p(d|{\cal T})$
is a normalization factor.  We include the ${\cal T}$ in the notation
only if there is ambiguity in the theory space under consideration.

 Under the assumption of uniform priors for
the $ q_i$'s, the Fisher matrix reduces to:
\begin{equation*}
F_{ij}= -\left<\frac{\partial^2 \ln {\cal L}}{\partial  q_i \partial
  q_j}\right>.
\end{equation*}
 The derivatives are calculated at the fiducial values of the
 parameters, in this case $q_1=...=q_N=0$. The ensemble average
 $\left<..\right>$ is over realizations of the CMB sky and instrument
 noise. In the standard CMB analysis with Gaussian signal and noise,
 we have ${\cal L}= \exp (-\Delta^\dagger {\bf C}^{-1} \Delta/2)
 /\sqrt{2\pi |{\bf C}|}$.  Here $\Delta$ represents the temperature
 and polarization maps including both CMB signal and instrumental
 noise and ${\bf C}=\left<\Delta \Delta ^\dag \right>$ is the
 theoretical pixel-pixel covariance matrix.  With this likelihood
 function, the Fisher matrix simplifies to:
\begin{equation*}
F_{ij}=\frac{1}{2}{\rm Tr}\left({\bf C}^{-1} \frac{\partial {\bf C}}{\partial
   q_i}{\bf C}^{-1} \frac{\partial {\bf C}}{\partial q_j}\right).
\end{equation*}
In the limit of full sky observation, or in cut-sky cases where
coupling between modes of different scales can be ignored, ${\bf F}$
can be written as:
\begin{equation}
\label{fisher}
F_{ij}= f_{\rm sky} \sum_{\ell=2}^{\ell_{\rm max}} \frac{2\ell+1 }{2} \frac{\partial
  {\bf C}_\ell}{\partial  q_i} {\bf C}_\ell^{-1}
  \frac{\partial {\bf C}_\ell}{\partial q_j}{\bf C}_\ell^{-1}
\end{equation}
with
\[ 
\bf{C_\ell}=
\left( {\begin{array}{cc}
C_\ell^T e^{-\ell^2 \sigma^2} + N_\ell^T& C_\ell^{TE}e^{-\ell^2 \sigma^2}  \\
C_\ell^{TE}e^{-\ell^2 \sigma^2} & C_\ell^{E}e^{-\ell^2 \sigma^2} + N_\ell^E \\
\end{array}} \right),\]
where we have included CMB temperature $T$, $E$-mode polarization and their
cross correlation $TE$. Here $N_\ell^{T,E}$ stands for instrumental noise
contribution to the power spectra and $\sigma$ is the width of the
Gaussian beam. The effect of incomplete sky coverage has been
naively taken into account by the $f_{\rm sky}$ multiplier which reduces
the effective number of observed modes .

The Fisher matrix for $N$ parameters, as any other $N\times N$ real
symmetric matrix, has $N$ independent eigenvectors which can be chosen
to be orthogonal to each other and normalized to one. So ${\bf F}$ can
be decomposed as ${\bf F}={\bf S f} {\bf S}^T $ where the columns of
${\bf S}$ are the eigenvectors of ${\bf F}$ with their corresponding
(non-negative) eigenvalues on the diagonal of the real diagonal matrix
${\bf f}$.  The eigenmodes we are looking for can now be constructed
using these eigenvectors of the Fisher matrix and the basis functions
we started with:
\begin{equation}
E_k(z)=\sum_{i=1}^NS_{ik} \varphi_i(z).
\label{Eiz}
\end{equation}
If the $\varphi_i$'s happen to be orthonormal, then the eigenmodes
$E_k(z)$ will be:
\begin{equation}
\int_{z_{\rm min}}^{z_{\rm max}} E_{k}(z)E_{k'}(z) w(z) {\rm d}z=\delta_{kk'}.
\label{norm}
\end{equation} 
Here $w(z)$ is the weight function with respect to which $\varphi_i$'s
are orthonormal.  Since Eq.~\eqref{norm} is not necessarily fulfilled,
we enforce the $E_k(z)$'s to be normalized to unity (as a matter of
convenience), which is equivalent to a renormalization of the
eigenvectors of ${\bf F}$. Although in general this could change the
rank ordering of the modes, in our case a reordering was not required.
Now, instead of the original $\varphi_i$'s, the set of the eigenmodes
can serve as basis functions for the expansion of perturbations
(compare with Eq.~\eqref{paramexpand}):
\begin{equation}\label{eq:mu_i}
\delta u (z)=\sum_{k=1}^N \mu_k \, E_k(z).
\end{equation}
To be more specific, we will refer to these parameter eigenmodes which
describe perturbations to the $\Xe$ history by XeMs.  In
\S~\ref{sec:recon} we will demonstrate two examples of perturbation
reconstruction with different numbers of eigenmodes taken into account
(Fig.~\ref{recon}). We will see how well these eigenmodes serve as
basis functions and also which features of the original perturbations
are restored (or lost) if only a subset of the eigenmodes are used in
the reconstruction process.

The inverse square of the eigenvalues of the Fisher matrix can be used
to forecast the error bars of the eigenmodes, i.e.,
$f_{ij}=\sigma^{-2}_i \delta_{ij}$, assuming the probability
distribution of the parameters is multivariate Gaussian close to the
maximum. For non-Gaussian likelihoods, the $\sigma_i$'s give the lower
bound for the errors. In the rest of this paper we use the term {\it
  error} for the $\sigma_i$'s, as the Gaussianity of the likelihood
function close to its maximum is usually a good assumption.  If the
modes are sorted in descending order of eigenvalues, the first few
(with smallest $\sigma_i$'s) will be most constrainable. Thus, the
constrainable part of the perturbations to the ionization history can
be described by the eigenmodes which have reasonably small
uncertainties (i.e., high eigenvalues), while the rest is practically
unconstrainable by the dataset under consideration.

This procedure of using an orthogonal transformation to replace the
parameters of the problem with a set of uncorrelated variables is
called {\it principal component analysis} or PCA for short. (The
parameter eigenmodes were used for CMB in \cite{bon96} and
subsequently in \cite{bon97} and many subsequent papers.)  In general,
the PCA needs to be applied to the whole history of ionization
simultaneously (as well as the standard cosmic parameters), since we
do not know {\it a priori} how the ambiguity in one epoch affects the
measurements of the perturbations in other epochs.  However, if it
turned out that a particular period of $\Xe$ history could be
relatively well constrained, e.g. by other probes, one could leave
that epoch out of the perturbations.  Moreover, choosing a suitable
parametrization, potentially changing over time to properly take into
account the different physics at different epochs, is a necessary but
not straightforward task.  In this work, the focus will be on the
epoch of recombination since that is where the main CMB signal is
coming from.  A more complete analysis for the whole ionization
history or where different parts of it are considered simultaneously
is for future work.

\begin{figure*} 
\begin{center}
\includegraphics[scale=0.88]{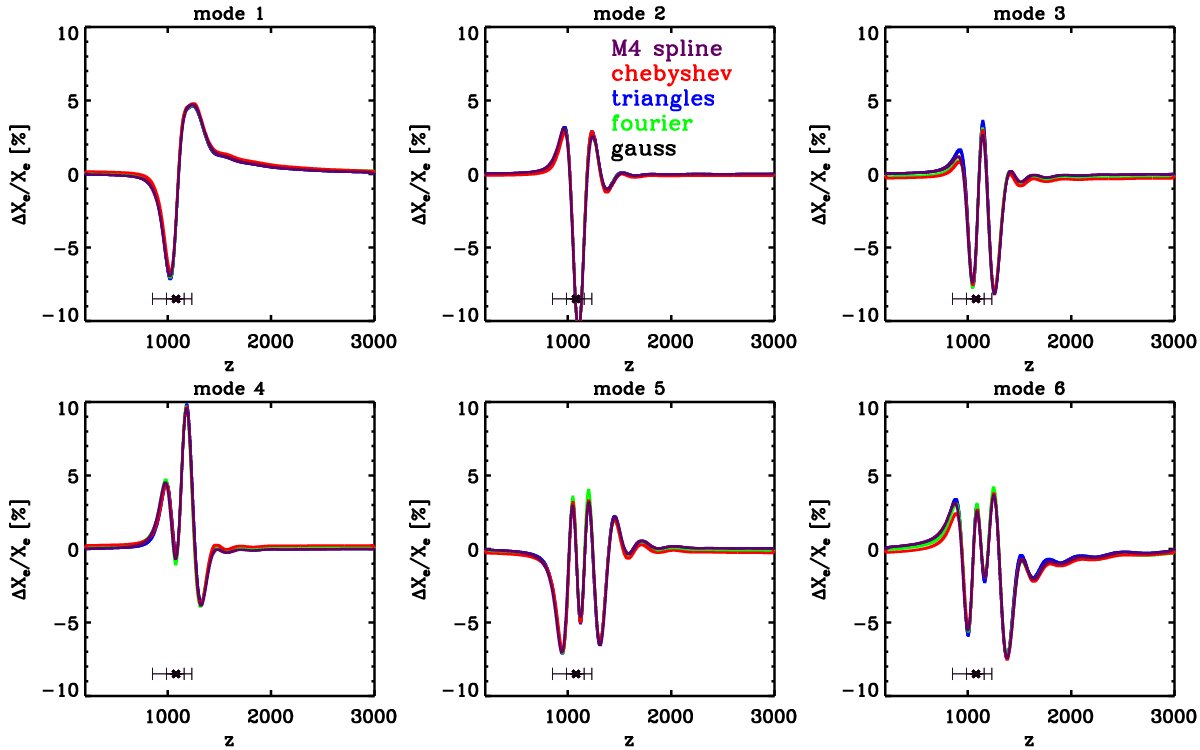}
\end{center}
\caption{The six most constrained XeMs for five different basis
  functions (with 160 parameters). The maximum and width of the
  Thomson visibility function have been marked in all figures.}
\label{multibasis} 
\end{figure*}
\subsubsection{Eigenmodes with varying standard parameters}
\label{sec:PCA_std_VAR}
Above we assumed that the background cosmology was fixed and the
parameter space only included perturbation parameters, i.e., $ q_i$'s.
However, as we will see in \S~\ref{mcmc_std}, the eigenmodes
constructed this way do not necessarily stay uncorrelated with one
another when standard cosmological parameters are also being varied.
Moreover, due to the correlation of the XeMs with one another as well
as with the standard parameters, the constraints on the eigenmodes
will possibly increase compared to the forecasted errors (reported in
Table~\ref{eval}).

To avoid this issue, the eigenmodes of perturbations should be
constructed in the presence of varying standard parameters. In this
case the Fisher information matrix has the following general form
\begin{align} 
\label{Fmarg}
\bf{F}&=
\left( {\begin{array}{cc}
F_{ss} & F_{sp}\\
F_{ps} & F_{pp} \\
\end{array}} \right),
\end{align}
where we have ${\bf F}_{ss}=-\left<\frac{\partial^2 \ln {\cal
    L}}{\partial s \partial s}\right> $, ${\bf F}_{sp}={\bf
  F}_{ps}-\left<\frac{\partial^2 \ln {\cal L}}{\partial s \partial
  p}\right> $ and ${\bf F}_{pp}=-\left<\frac{\partial^2 \ln {\cal
    L}}{\partial p \partial p}\right> $, with $s$ and $p$ symbolically
representing the standard and perturbation parameters.  To find the
eigenmodes for perturbations after marginalization over the standard
parameters, hereafter extended eigenmodes, or eXeMs, under the
assumption of ellipsoidal parameter contours, we need to
eigendecompose the perturbation block of inverse of the Fisher matrix,
i.e., to find the eigenvectors of $({\bf F}^{-1})_{pp}$. However, the
inversion of the Fisher matrix is numerically problematic as it is
ill-conditioned due the non-constrainable parameters which exist in
the parameter space under study.

In the case of fixed standard parameters, the Fisher matrix reduces to
the perturbation block only, and since the eigenvectors of a symmetric
matrix and its inverse are the same (with inverse eigenvalues), there
is no need to invert the Fisher matrix before its eigendecomposition.
Similarly to get the eXeM, i.e., the eigenvectors of $({\bf
  F}^{-1})_{pp}$ with ${\bf F}$ as in Eq.~\eqref{Fmarg}, we avoid the
direct full inversion of ${\bf F}$ by noting that
\begin{equation}
({\bf F}^{-1})_{pp} = ( {\bf F}_{pp} - {\bf F}_{ps} {\bf F}_{ss}^{-1}{\bf F}_{sp} )^{-1}.
\end{equation}
The eigendecomposition of ${\bf F}_{pp} - {\bf F}_{ps} {\bf
  F}_{ss}^{-1}{\bf F}_{sp} $ then only requires the inversion of the
well-behaved standard parameter block.  It is straightforward to
directly check that $({\bf F}^{-1})_{pp}$ properly describes the
marginal likelihood of the perturbation parameters:
\begin{eqnarray*}
{\cal L}({\bf p}| d)\propto e^{-{\bf p}^T {\bf F}_{pp} {\bf p} /2} \int e^{-{\bf s}^T {\bf F}_{ss} {\bf s} /2} e^{-{\bf p}^T {\bf F}_{ps} {\bf s}} {\rm d}{\bf s}\\
\propto  e^{-{\bf p}^T ({\bf F}_{pp}-{\bf F}_{ps} {\bf F}_{ss}^{-1}{\bf F}_{sp} ) {\bf p} /2}.
\end{eqnarray*}
Here ${\bf p}$ and ${\bf s}$ are the arrays of the perturbation and
standard parameters.

The eXeMs are then uncorrelated with each other (but not necessarily
to the standard parameters) even in the presence of varying standard
cosmic parameters.  We will use this method later in \S~\ref{eXeM_rec}
to find the eXeMs describing the perturbations to recombination
history.  Hereafter, we use the term eigenmode with the very general
meaning and reserve XeM only for the eigenmodes if they are
constructed with fixed standard parameters.

It is also possible to treat the standard parameters depending on the
way they affect the power spectra. Among the standard parameters,
$Y_{\rm p}$ has this unique property of influencing $C_\ell$'s {\it
  only} through its impact on $\Xe$.  In other words, if we find the
$\Xe$ template in $\Xe$ parameter space corresponding to $Y_{\rm p}$,
i.e., $\intd\Xe/\intd Y_{\rm p}$, small changes in $Y_{\rm p}$ can be
mimicked by properly changing the amplitude of this template while
$Y_{\rm p}$ is left unchanged.  Note that changes in other standard
parameters either directly lead to changes in the CMB with no
influence on $\Xe$ (such as $n_{\rm s}$ and $A_{\rm s}$), or have both
direct and indirect (i.e., through $\Xe$) impacts on the CMB (such as
$\Omega_{\rm b}$).  Therefore, when including a $Y_{\rm p}$-like
parameter in the analysis, we can restrict our search for the
perturbation eigenmodes to the part of $\Xe$ space which is
uncorrelated to the $\Xe$ template corresponding to this parameter as
described above. In this way if perturbations to $\Xe$ are initially
described by $N$ parameters, we will in the end have the $Y_{\rm
  p}$-like parameter or its associated template as one parameter and
$N-1$ eigenmodes which are uncorrelated to the $Y_{\rm p}$-like
parameter. These $N-1$ eigenmodes together with the $Y_{\rm p}$-like
parameter fully parametrize the original $N$-dimensional space of the
perturbations.  However, for the purpose of this work we did not
further explore this possibility.

\section{Perturbation eigenmodes for Recombination}\label{sec:rec_EMs}
In this section we follow the procedure explained in
\S~\ref{sec:em_const} to find the eigenmodes for perturbations in the
ionization fraction at high redshifts.  For the most part we leave the
standard parameters fixed, but return to the question of how these
affect the mode functions later.  We choose the redshift range of
$[200,3000]$ which covers hydrogen and singly ionized helium
recombination ($z\sim 1100$ and $z\sim 1800$ respectively) as well as
part of the dark ages while being well above the standard reionization
scenarios ($z\lesssim 30$). We assume the fiducial recombination
history is given by the SRS, as explained in \S~\ref{sec:SRS}, unless
otherwise stated.

 In the following we compare the eigenmodes generated by using various
 bases and study some of the aspects associated with them, such as
 their convergence and fiducial model dependence. Some consistency
 checks will also be presented.  Special attention will be given to
 perturbation to helium recombination. We will also study how
 including the standard cosmic parameters in the analysis would change
 the eigenmodes of perturbations to the ionization history. At the end
 we propose an information-based criterion for cutting off the
 eigenmode hierarchy to be used in the data analysis.

\begin{deluxetable*}{ccccccc}
\tablecaption{The forecasted standard deviations of the first six 
XeMs from the Fisher analysis for different observational cases.}
\tablehead{\colhead{XeM} & \colhead{1} & \colhead{2}
  & \colhead{3} & \colhead{4} &
  \colhead{5} & \colhead{6} }
\startdata
CVL($\ell_{\rm max}=3500$)     &0.003  & 0.009 & 0.013 & 0.016 & 0.022 & 0.047 \\[1mm]
CVL($\ell_{\rm max}=2000$)     &0.011  & 0.019 & 0.024 & 0.041 & 0.094 & 0.190 \\[1mm]
CVL($\ell_{\rm max}=3500$, T only)         &0.004  & 0.021 & 0.064 & 0.103 & 0.208 & 0.275 \\[1mm]
Planck-ACTPol($\ell_{\rm max}=3500$)  & 0.015  & 0.047 & 0.068 & 0.13 & 0.22 & 0.31
\enddata
\label{eval}
\end{deluxetable*}

\subsection{XeM construction using different bases}
\label{sec:basis}
 In this section we take $\delta u (z)=\delta \ln \Xe$ and try the
 five different sets of basis functions described in
 \S~\ref{sec:basis_functions}: Chebyshev polynomials and Fourier
 series as orthogonal non-local functions of redshift, and $M_4$
 splines, triangular and Gaussian bumps as localized functions. For
 the latter three, the width of the bumps is chosen to be independent
 of redshift. We choose $\sigma_i=\delta z/2$ for Gaussian and
 triangular bumps (Eqs.~\eqref{eq:Gauss_bump} and~\eqref{eq:tri_bump})
 and $h=1.5 \delta z$ for $M_4$ splines (Eq.~\eqref{eq:M4_spline}). In
 all cases, $\delta z=\Delta z /(N+1)$ is the spacing between the
 centers of adjacent bumps, where $\Delta z$ is the redshift range of
 interest and $N$ is the number of basis functions used.  For each set
 of basis functions we calculate the $N\times N$ Fisher information
 matrix as explained in \S~\ref{sec:em_const} where the $N$ parameters
 are the amplitudes of the perturbations in the form of the basis
 functions, i.e., $ q_i$'s in Eq.~\eqref{paramexpand}. The standard
 cosmic parameters are fixed to their fiducial values.  For the {\it
   data} we simulate the $T$, $E$ and $TE$ spectra up to $l=3500$ for
 a full-sky, cosmic variance-limited (hereafter CVL) CMB experiment,
 unless otherwise stated. We then construct the Fisher matrix
 (Eq.~\eqref{fisher}) and from it the $N$ XeMs (Eq.~\eqref{Eiz}).  The
 first six XeMs for $N=160$ are shown in Fig.~\ref{multibasis}.  The
 first row in Table~\ref{eval} shows the forecasted errors of these
 XeMs, obtained from the eigenvalues of the Fisher matrix. Note that
 including standard parameters in the analysis, e.g. MCMC simulations,
 can increase the error bars, as we will see later in
 \S~\ref{sec:measure}.

We can see that the first six XeMs -- which are the most constrained
modes -- all have the strongest variations close to the maximum of the
Thomson visibility function. The freeze-out tail is not perturbed
significantly, and the oscillations around helium recombination
($z\sim 1800$) have much smaller amplitude than those at $z\sim 1100$.
This is expected since the CMB anisotropies are most sensitive to
perturbations during maximum visibility and features at low and high
redshift are not weighing as much in the CMB power spectra, once
uncertainties close to $z\sim 1100$ are allowed.  This in turn implies
that only once the ionization history during hydrogen recombination is
known well can small modifications in the freeze-out tail or at
$z\gtrsim 1400$ be constrained.

We can observe another aspect of the XeMs. The larger the expected
error bar the more small scale structure or high frequency
oscillations the modes have and the further away from the visibility
peak they probe. This is again understandable, since neighbouring ups
and downs in the mode functions lead to partial cancelation of the
effect on the $C_\ell$'s.  Once several oscillations are occurring
close to $z\sim 1100$, signals produced farther away from maximum
visibility can start competing with those from $z\sim 1100$, and hence
become constrainable by the data.

We also see that the first few XeMs are practically the same
independent of the chosen expansion basis (Fig.~\ref{multibasis}),
although individual perturbations in different bases lead to totally
different $C_\ell$ responses (Figs.~\ref{m4_dlnxdlncl} and
\ref{c_dlnxdlncl}). Moreover, these modes are converged and do not
change by including higher modes, as we will see in
\S~\ref{sec:converg}.  However, as we go to modes with higher
uncertainty (not shown here), the XeMs from different bases start to
slightly differ from each other.  A larger number of basis functions
are required to make these higher XeMs agree as well.  Moreover, we
found that the higher (poorly constrained) XeMs, in particular from
the extended basis expansions such as Fourier, become dominated by
numerical noise. The reason is that for weakly constrained modes where
the higher frequencies start to play a more important role, the impact
of adjacent ups and downs from the high frequency perturbations
(e.g. sine functions) may not be well resolvable in the $C_\ell$'s,
resulting in their net effect being dominated by numerical noise. For
the localized basis functions, as long as the individual bumps are
numerically resolvable, we do not find this issue, because each
perturbation has just one bump with no destructive neighbour.

For more precise computations of the higher XeMs improvements of the
numerical treatment in {\sc Camb} would become necessary. We tried
several obvious modifications, as well as different settings for the
accuracy level, but were unable to stabilize the results for very high
frequency modes.  However, since in the analysis we are hardly using
more than a few XeMs, for the purpose of this work this was
sufficient.
 
\begin{figure*} 
\begin{center}
\includegraphics[scale=0.87]{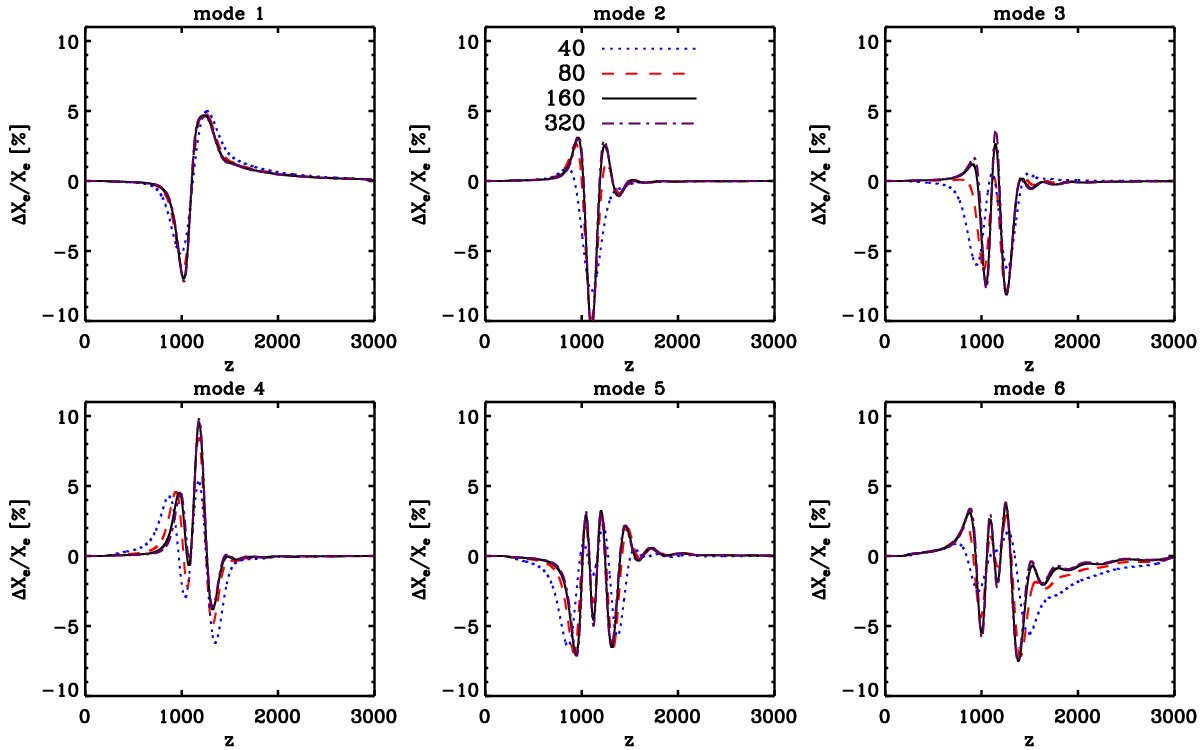}
\end{center}
\caption{Testing the convergence of the eigenmodes. Here, the six most
  constrained XeMs are shown for cases with different number of
  parameters (40, 80, 160 and 320) and with $M_4$ splines as the basis
  functions. We see that the modes for 160 and 320 parameters are
  basically the same, indicating that these modes have already
  converged with 160 parameters.}
\label{conv}
\end{figure*}
\begin{figure*} 
\begin{center}
\includegraphics[scale=0.88]{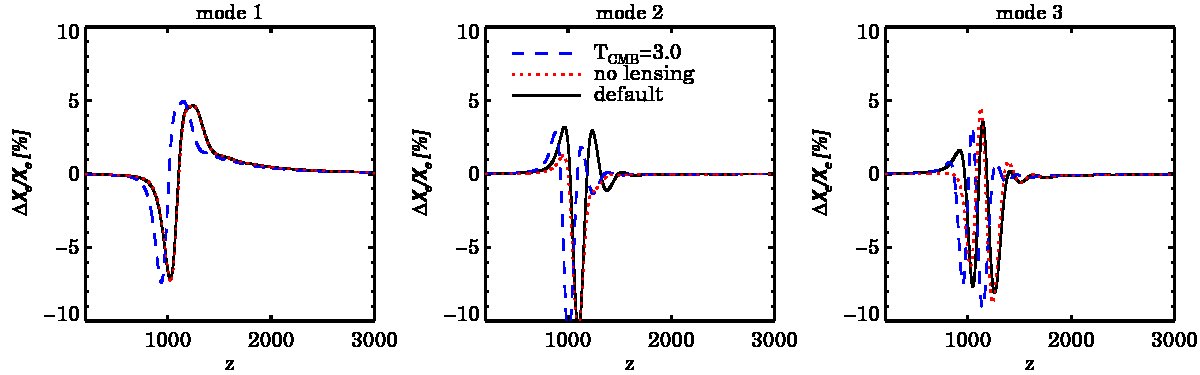}
\end{center}
\caption{The three most constrained XeMs for three different fiducial
  models. The default model corresponds to the SRS and the effect of
  gravitational lensing on the CMB anisotropies has been included. One
  model corresponds to a recombination history with a different CMB
  temperature and in the other model lensing is not included.  For the
  case of the two different CMB temperatures, the major difference is
  the shift in the eigenmodes associated to the shift in the fiducial
  $\Xe$ and visibility functions (see Fig.~\ref{xevis_fid}). }
\label{fid_dependence}
\end{figure*} 
\subsection{Convergence test}\label{sec:converg}
 In all cases considered above the number of modes was chosen to be
 $N=160$. We tested the convergence of the XeMs by trying $N=40$,
 $80$, $160$ and $320$ in different bases and found that by $N=160$
 the first few modes are converged. Fig.~\ref{conv} shows a
 representative example of this convergence with $M_4$ splines as the
 basis. Thus the few constrainable XeMs are independent of the very
 high frequency (or highly localized) perturbations and including
 basis functions of higher order will not affect these eigenmodes.

For the case of $M_4$ spline functions the robustness of the results
should also be checked against increasing the width of the kernel. By
comparing the (first six) XeMs with $h=1.5 \delta z$ (as explained in
\S~\ref{sec:basis}) to those with $h=3 \delta z$, we conclude that the
modes have already converged for $h=1.5 \delta z$ and thus we will
adopt $h=1.5 \delta z$ in the rest of this paper.

\subsection{Fiducial model and dataset dependence}\label{sec:fid_data}
It is important to note that the XeMs are by construction fiducial
model dependent. In principle, the observables (such as $C_\ell$'s)
for different fiducial models respond differently to the same
perturbations depending on the strength of the signals, at different
redshifts, from the unperturbed fiducial model.

As an example, in Fig.~\ref{fid_dependence} we compare the eigenmodes
for three fiducial $\Xe$ histories. Two of the models have different
CMB temperatures and in the third one lensing has not been
included. In the first two, the different $ T_{\rm CMB}$'s lead to
different fiducial $\Xe$'s. Here, the main difference in the
eigenmodes is their shift towards lower $z$'s for the case with higher
CMB temperature. This is consistent with the delayed recombination
shown in Fig.~\ref{xevis_fid}, remembering that XeMs are primarily
localized around the maximum of visibility where the $C_\ell$'s are
most sensitive to.  For the latter case with no lensing, although
$\Xe$ and the physics around recombination have not changed, there are
still slight changes in some of the XeMs as seen in
Fig.~\ref{fid_dependence}.

 We also checked the robustness of the eigenmodes against changes in
 the fiducial value of the parameters and the assumed reionization
 scenario. We tried a different value for $\Omega_{\rm b}$, as the
 parameter most strongly affecting the ionization fraction, $1\sigma$
 away from its fiducial value. For the late reionization we tried an
 extended reionization scenario (i.e., $\Xe=1$ for $z \leq 6$;
 $\Xe=0.5$ for $6 < z\leq 30$ and $\Xe =0$ elsewhere) radically
 different from our sharp fiducial reionization model (the default in
 {\sc Camb}).  For both of these tests the first six eigenmodes were
 found to be the same as our main eigenmodes (Fig~\ref{multibasis})
 with tiny differences in the fifth and sixth modes for the latter
 case.

This implies that, although the eigenmodes are fiducial model
dependent, the constrainable ones are not practically sensitive to
changes in the fiducial model or its parameters in the limits
currently allowed by the data for the standard model of cosmology.
That is because small changes in the fiducial parameters and the
corresponding small changes in the ionization history only affect the
XeMs at second order. Here by small we mean changes that lead to
(smaller than or) the same order of magnitude signal in the simulated
data as the (few best) XeMs. The higher XeMs with larger uncertainties
are more affected by the same changes in the fiducial parameters, as
these changes are no longer considered small relative to these poorly
constrained XeMs. This non-sensitivity of the best modes to the
fiducial values of parameters does not contradict their significant
correlation once the standard parameters are also allowed to vary, as
we will see later in \S~\ref{mcmc_std}.

\begin{figure*} 
\begin{center}
\includegraphics[scale=0.88]{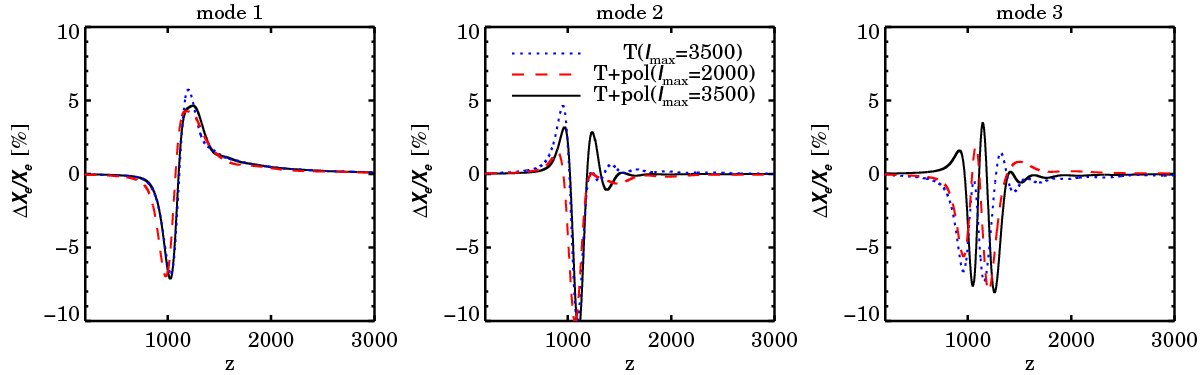}
\end{center}
\caption{The three most constrained XeMs with and without
  polarization and with $\ell_{\rm max}=2000$ and $3500$.}
\label{T_vs_T_pol_lmax}
\end{figure*}
\begin{figure*} 
\begin{center}
\includegraphics[scale=0.88]{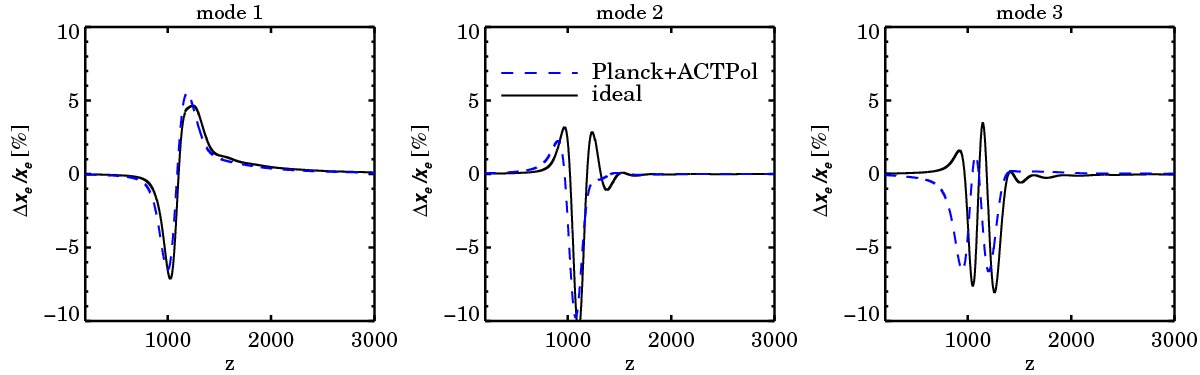}
\end{center}
\caption{The three most constrained XeMs for a Planck-ACTPol-like
  experiment compared to
  a CVL experiment.}
\label{pl_vs_ideal}
\end{figure*}
\begin{figure*} 
\begin{center}
\includegraphics[scale=0.88]{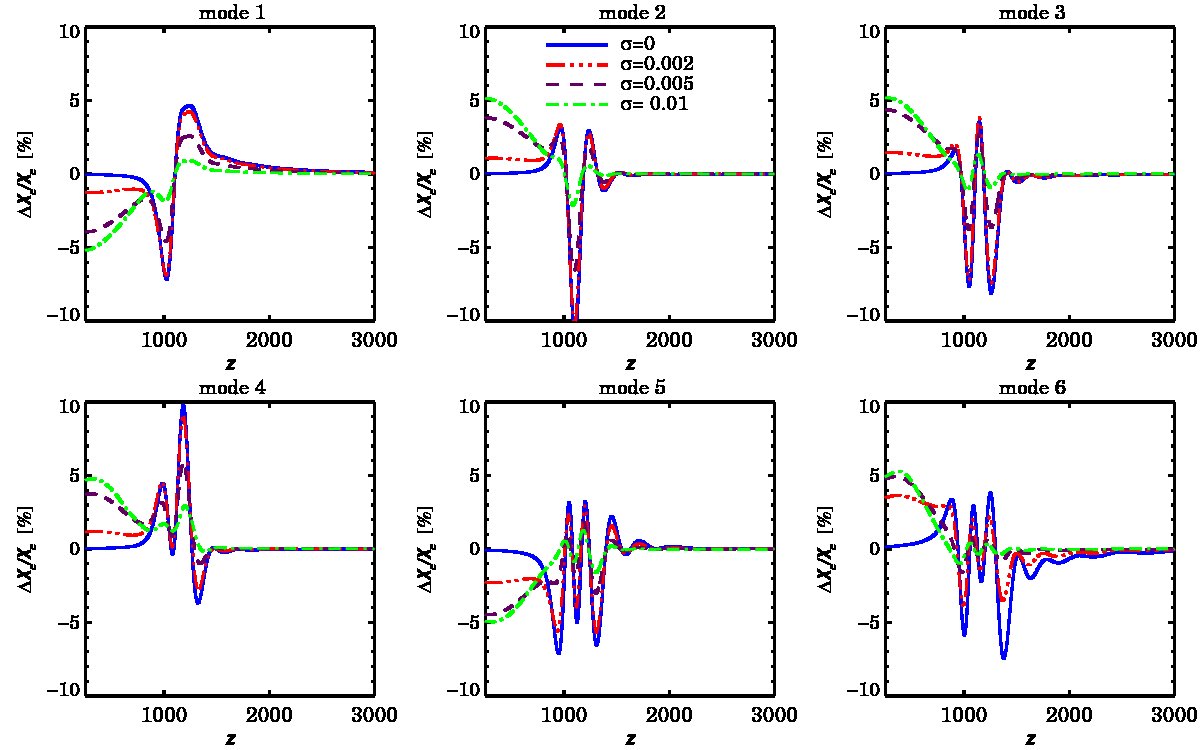}
\end{center}
\caption{The six most constrained XeMs with $\delta u (z)=\delta \ln(x+\sigma)$ as the parameter, having three different values for 
  $\sigma$, and with Gaussian bumps
  as the basis functions.}
\label{multisig}
\end{figure*}
\begin{figure*} 
\begin{center}
\includegraphics[scale=0.49]{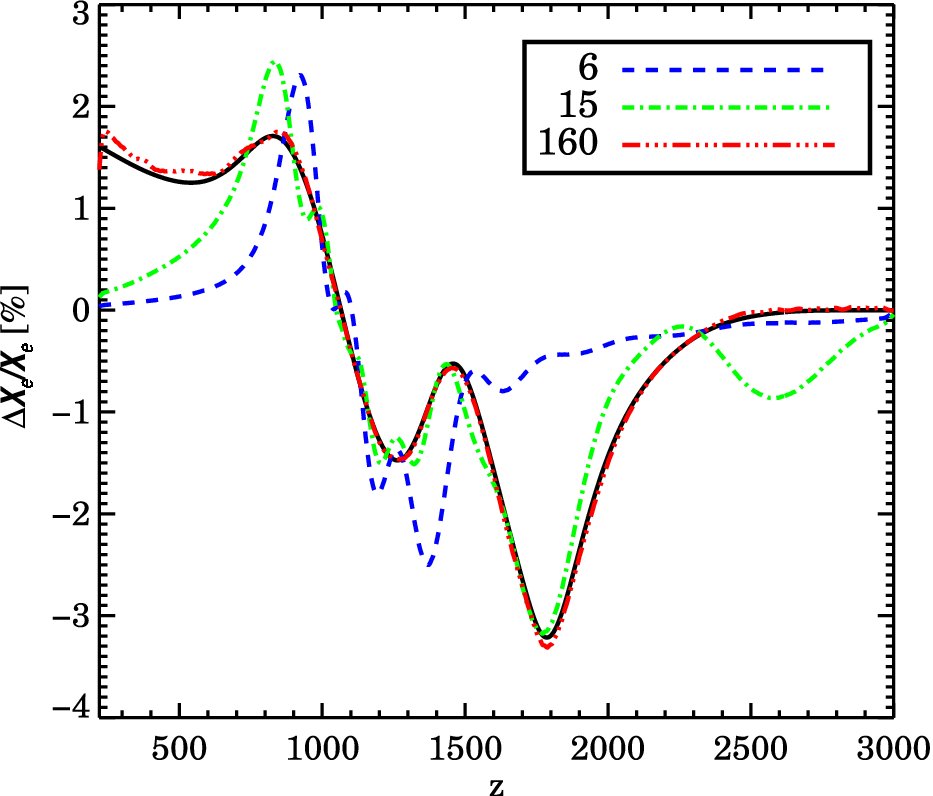}
\includegraphics[scale=0.46]{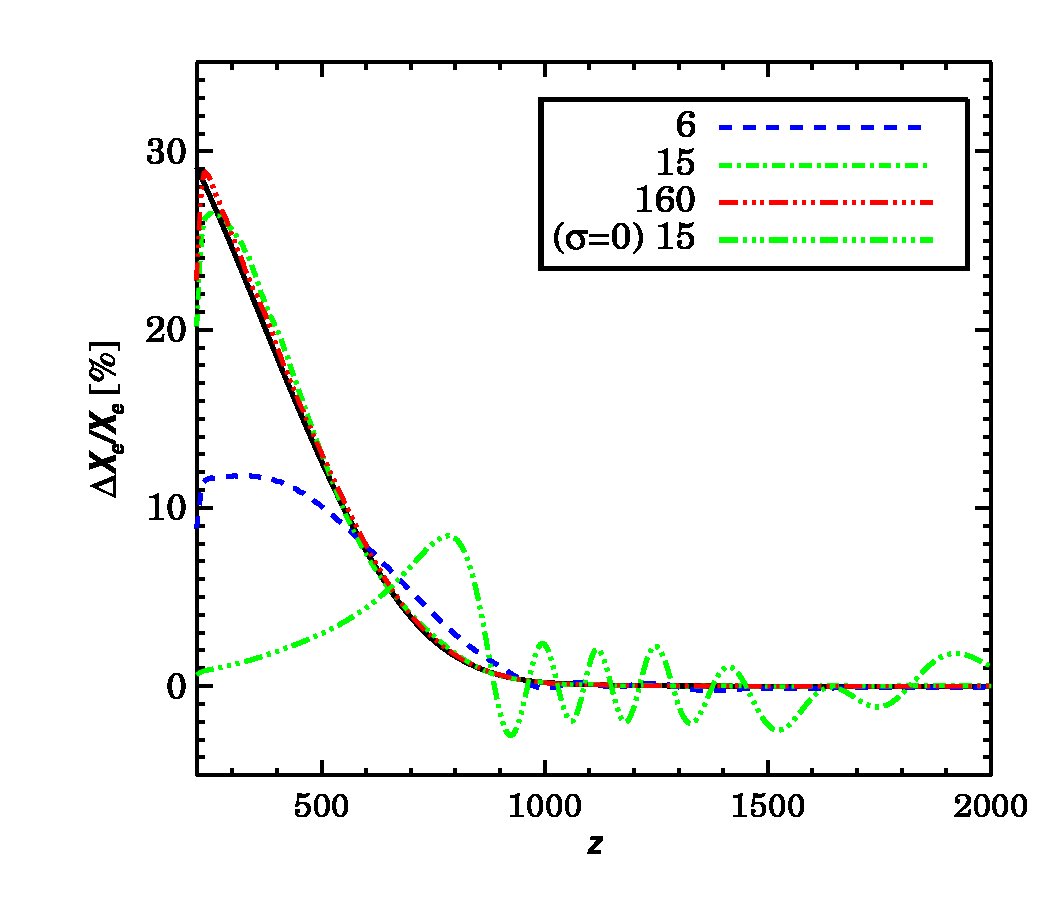}
\includegraphics[scale=0.49]{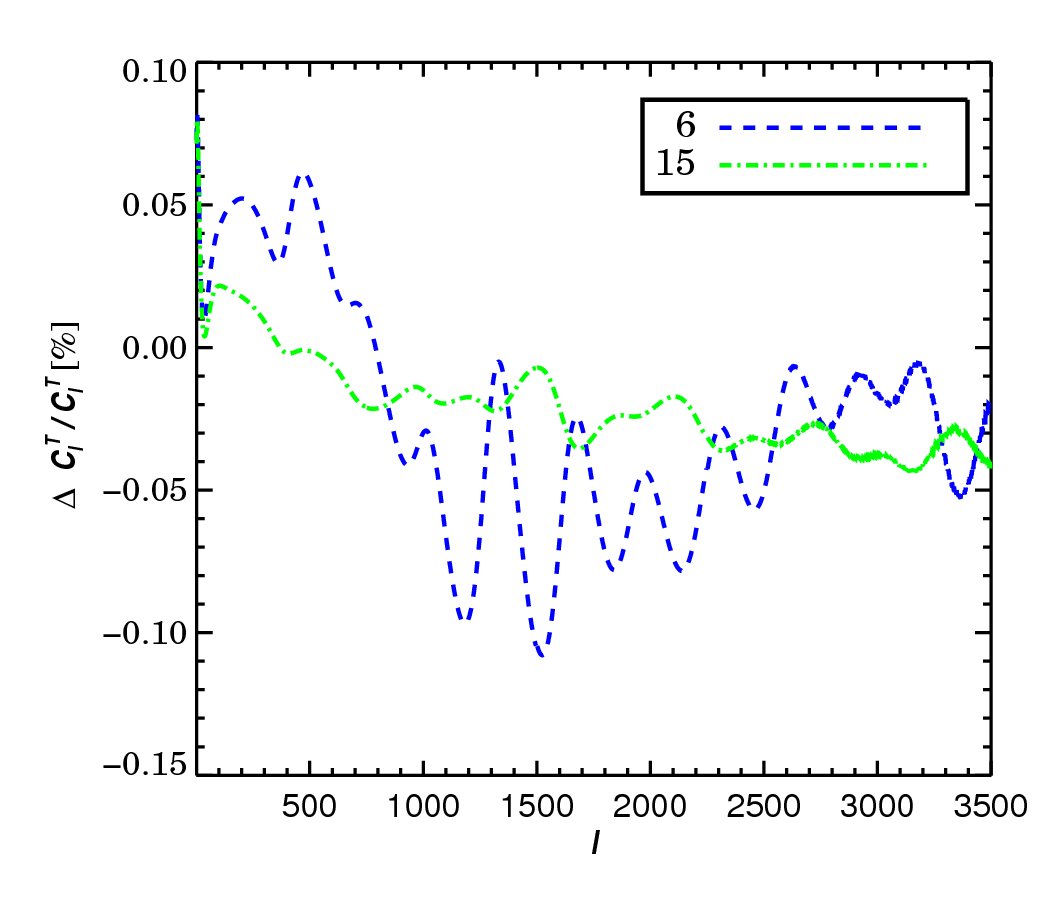}
\includegraphics[scale=0.49]{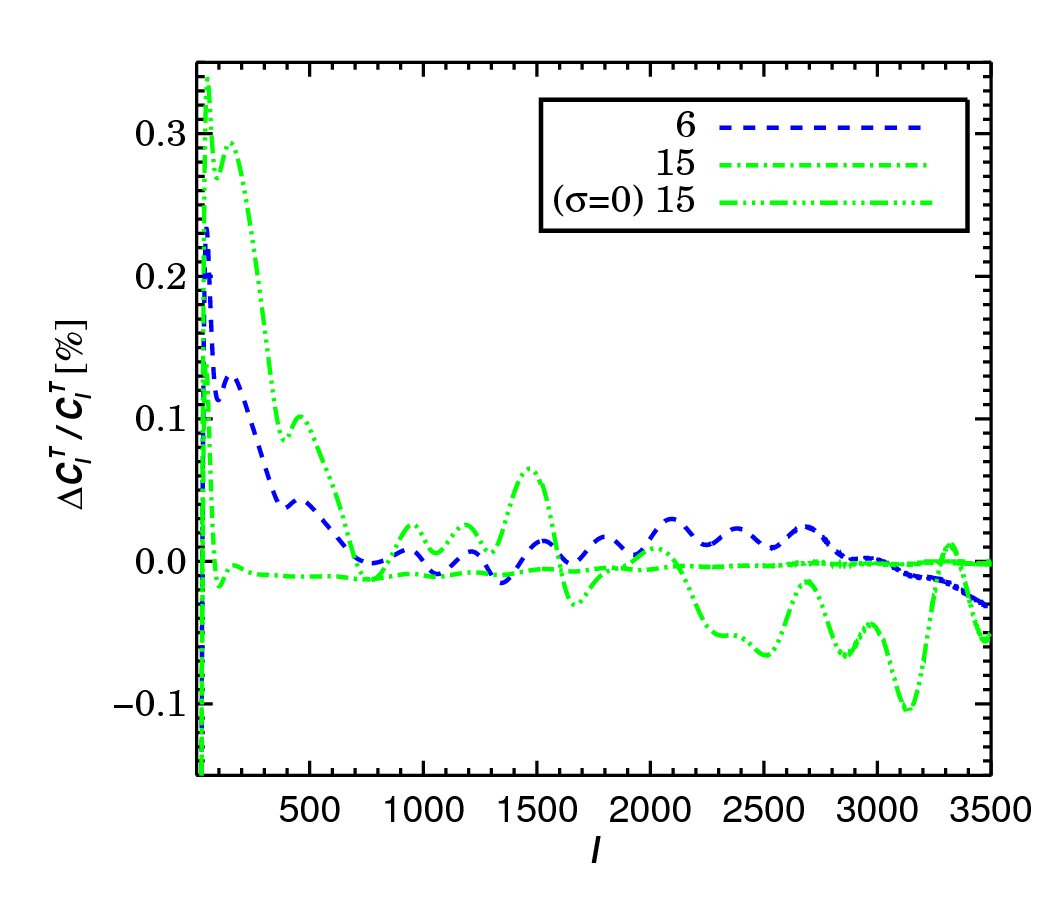}
\end{center}
\caption{The reconstruction of two physically motivated $\Xe$
  perturbation scenarios on (different number of) XeMs generated with
  Gaussian bumps (top) and the relative difference in the temperature
  power spectrum between the reconstructed perturbations and the full
  corrections (bottom).  Right: The perturbations come from deviation
  from physical corrections to the recombination process
  (CT2010). Here the perturbation parameter is $\delta
  \ln(\Xe)$. Left: The perturbations are due to a model of dark matter
  annihilation. As the perturbation parameter we used $\delta
  \ln(\Xe+0.01)$ to better accommodate for the freeze-out
  perturbation. A case with $\delta u (z)=\delta \ln(\Xe)$, i.e.,
  $\sigma=0$, is shown for comparison. }
\label{recon}
\end{figure*}

We also studied the dependence of the XeMs on some properties of the
simulated CMB datasets used for their construction, such as different
$\ell_{\rm max}$ corresponding to the smallest scale information
present in the data, and different experimental noise levels. The
results for a CVL experiment up to $\ell =2000$ in temperature and
polarization and also a CVL experiment only sensitive to temperature
(up to $\ell =3500$) are shown in Fig.~\ref{T_vs_T_pol_lmax}.  As a
more experimentally motivated case, we calculated the XeMs for
simulated Planck-like
data\footnote{http://www.rssd.esa.int/SA/PLANCK/docs/Bluebook-ESA-SCI282005291\_V2.pdf}
(using $100$, $143$ and $217$ GHz channels, with effective galaxy-cut
sky coverage of $75\%$) and ACTPol-like data, including both wide and
deep surveys \citep{nie10}. As shown in Fig.~\ref{pl_vs_ideal}, there
is a tiny shift in the first mode relative to the mode for an ideal
experiment and the changes grow as we proceed to higher modes.

More significant than the small changes in the XeMs constructed with
different assumptions about data, are the forecasted error bars in
different cases (see Table~\ref{eval}). By removing the temperature at
high $\ell$'s or the polarization spectrum, the constraints on the
amplitudes of the modes, determined from the eigenvalues of the Fisher
matrix, become considerably larger. All these errors are calculated
with the standard parameters fixed.  However, the considered cases
illustrate the general behaviour of the method.  Taking into account
the correlation between the perturbations and the standard cosmic
parameters leads to relatively higher error bars, depending on the
dataset used.  We will return to this point later
(table~\ref{evaleXeM}).

\subsection{Alternative parametrizations for perturbations}\label{sec:param}
We now turn to the parametrization defined by Eq.~\eqref{eq:sig},
$\delta u (z)=\delta \ln(\Xe+\sigma)$ with $\sigma >0$.  As mentioned
earlier, this parametrization allows us to focus the perturbations on
the freeze-out tail which are more difficult to recover with $\delta u
(z)=\delta \ln(\Xe)$. This is because choosing $\sigma >0 $ results in
over-weighting the signal from the perturbations at low $z$ (with low
$\Xe$) compared to a case with $\sigma=0$ for the same value of
$\delta u (z)$.

Fig.~\ref{multisig} shows the six most constrained XeMs with $\delta u
(z)=\delta \ln(\Xe+\sigma)$ for different values of $\sigma$, all
constructed by using 160 Gaussian bumps as the basis functions and
assuming CVL experiment up to $\ell_{\rm max}=3500$.  We see that the
amplitude of the XeMs increases in the freeze-out tail as $\sigma$
increases. It will be illustrated in \S~\ref{sec:recon} that a
relatively high value for $\sigma$ is the preferred choice for
studying perturbations that most significantly alter the freeze-out
tail.\\

\subsection{Consistency check: eigenmodes as a complete basis}
\label{sec:recon}
Any function $\Xe(z)$ (in the redshift range under consideration) can
be expanded in terms of these XeMs unless it has highly localized
features compared to the highest frequency present in the basis
functions or to the width of the bumps in the case of localized modes.
That is because the XeMs are just linear combinations of the original
basis functions, and thus cannot have frequencies higher than the
maximum frequency present in the basis functions.  On the other hand,
as is clear from Fig.~\ref{multibasis}, strongly localized features in
possible perturbations to recombination history are not constrained
with CMB datasets. Therefore the lost features of an ionization model
via expansion by these eigenmodes will not be measurable even if modes
with higher frequencies are included in the analysis. In other words,
the XeMs serve as a complete basis for the expansion of {\it
  constrainable} features in the possible perturbations in the
recombination history.

To demonstrate the reconstruction of perturbations using the XeMs we
choose two physically motivated ionization perturbations, one
associated with physical corrections to the recombination process
(\cite{Chluba2010b}, hereafter CT2010) and the other due to energy injection
coming from a model of dark matter annihilation \citep[using the
  description of][]{Chluba2010a}.

\subsubsection{Standard recombination corrections}
\label{sec:recon_CT2010}
The modification to $\Xe$ corresponding to CT2010 is shown in the top
left panel of Fig.~\ref{recon} (black solid line).  This correction
should be added to the $\Xe$ from the original version of {\sc
  Recfast} (or the $\Xe$ from {\sc Recfast} v1.4.2 setting ${\rm
  He_{Switch}}=0$).  At high redshifts one can see the effect of
accelerated helium recombination caused by absorption of photons in
the Lyman continuum of hydrogen. During hydrogen recombination the
corrections are caused by detailed radiative transfer effects as well
as two-photon and Raman scattering events.  The freeze-out tail is
slightly higher than obtained with {\sc Recfast} because of deviations
from statistical equilibrium in the angular momentum sub-states.  We
note that with {\sc Recfast v1.5} a large part of all these
corrections can be accounted for, however, these corrections are not
explicitly modelled using a physical description but have been fudged
to reproduce the results obtained with detailed recombination codes.

\begin{deluxetable}{ccccccc}
\tablecaption{The projection of the modifications to recombination
  history on the first six XeMs.}
\tablehead{\colhead{XeM} & \colhead{1} & \colhead{2}
  & \colhead{3} & \colhead{4} &
  \colhead{5} & \colhead{6} }
\startdata
CT2010  &-0.32  & 0.08 & 0.16 & 0.02 & -0.09 & 0.25 
\\[1mm]
DM annihilation  &-0.31  & -0.30 & 0.46 & -0.14 & 0.33 & 0.88 
\enddata
\label{CT_proj}
\end{deluxetable}
We project this $\delta \ln( \Xe)$ on the 160 XeMs constructed from
perturbations in the recombination history in the form of Gaussian
bumps and with the perturbation parameter being $\delta u (z)=\delta
\ln(\Xe)$ for the CVL case with $\ell_{\rm max}=3500$, as described in
\S~\ref{sec:basis}.  The figure compares the reconstructed
perturbation for three cases with different number of XeMs
included. First note that by including all 160 XeMs the original
perturbation is practically fully recovered. If only the 15 most
constrained modes are included, the helium correction ($z\sim 1800$)
and also hydrogen correction around $z\sim 1100$ are well restored
while for lower $z$ regions higher modes are required. The
reconstruction by six XeMs, however, is most sensitive to variations
around $z\sim 1100$ and cannot tell much about the helium correction.
The projection coefficients for the first six XeMs are shown in
Table~\ref{CT_proj}. For this particular model of corrections to the
perturbation scenario we see that the XeM $1$, $3$ and $6$ are
strongly dominant among the first six modes.

The lower left panel in Fig.~\ref{recon} illustrates the relative
difference between the temperature power spectrum for the
reconstructed perturbations and the original full corrections. We see
that with only six modes the error in the recovered $C_\ell$'s is less
than $0.1\%$. Remembering that the changes in the $C_\ell$'s due to
the full corrections are about a few percent, this shows that the main
corrections to the CMB power spectra can be captured by just
introducing a small number of modes. The CMB data indeed are not
sensitive to all the details in the freeze-out tail of recombination
and during helium recombination, unless prior knowledge renders
uncertainties at $z\sim1100$ very small. As we will see below, part of
the corrections from higher modes are compensated for by biasing the
XeMs included in the analysis.

\begin{figure*} 
\begin{center}
\includegraphics[scale=0.78]{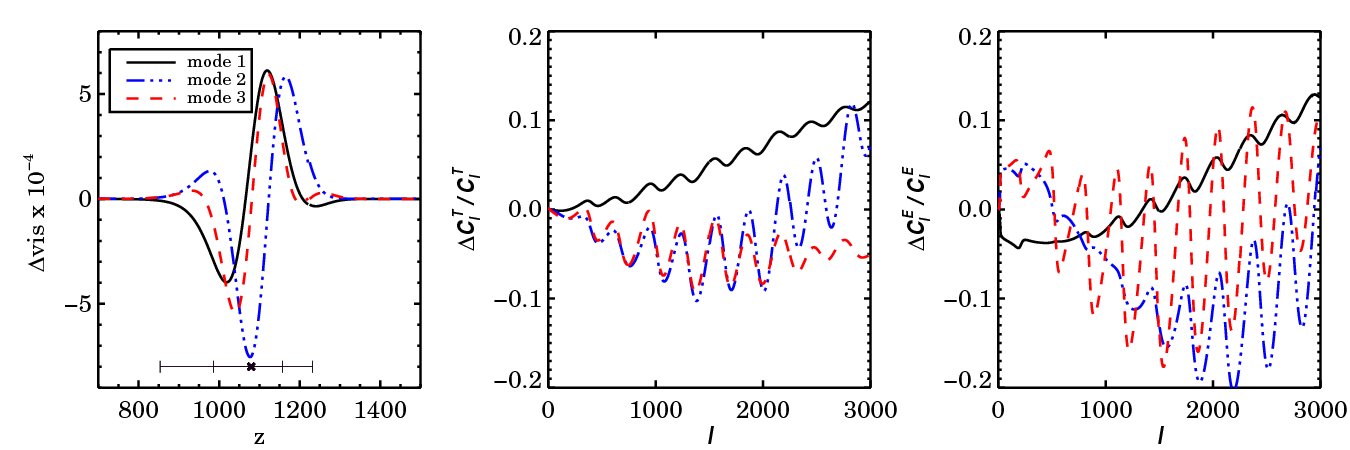}
\includegraphics[scale=0.78]{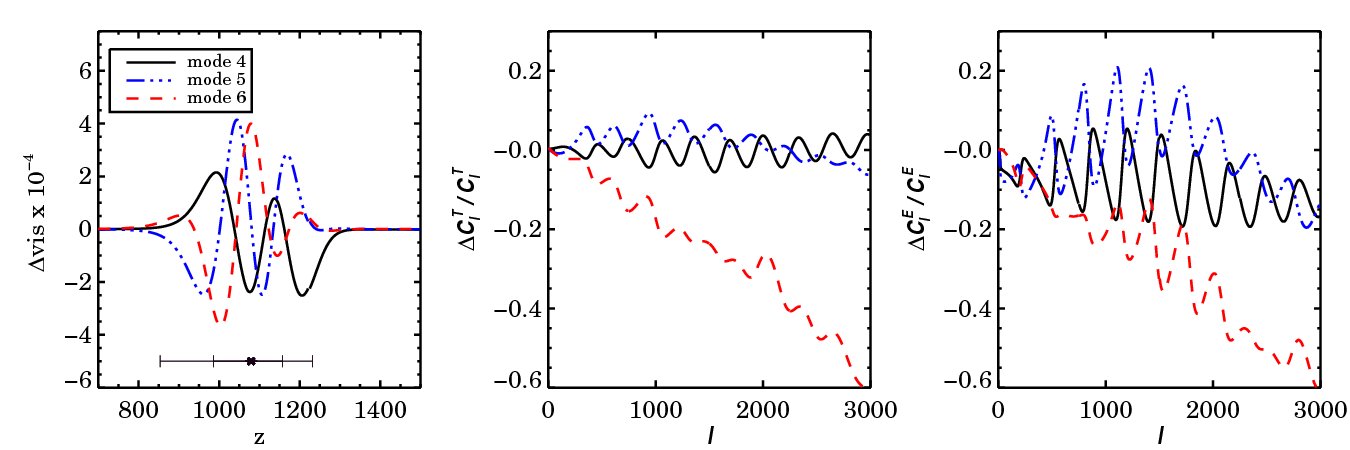}
\end{center}
\caption{The $\delta {\rm vis}= ({\rm vis} -{\rm vis}_{{\rm fid}})$
  (left) and the relative changes in the $TT$ and $EE$ power spectra
  (middle and right) for the six most constrainable XeMs.}
\label{mode_vis_cl} 
\end{figure*}
\subsubsection{Dark matter annihilation scenario}
\label{sec:recon_DM}
As the second example we chose the perturbations arising from a model
of dark matter annihilation. It was computed using the description of
\citet{Chluba2010a} with an annihilation efficiency $f_{\rm DM}\sim
2\times 10^{-24}\,\rm eV/s$. The difference with respect to {\sc
  Recfast} is shown in the right panel of Fig.~\ref{recon}.  In
contrast to the previous case, the perturbations here are not
concentrated around the maximum of differential visibility but are
most significant at lower redshifts.  Therefore, for the decomposition
of the DM perturbations we choose $\delta u (z)=\delta \ln(\Xe+0.01)$
(see \S~\ref{sec:param} and Fig.~\ref{multisig}) to allow a better
recovery of the relatively large perturbations in the freeze-out tail
without the need to include too many modes.  This procedure can be
interpreted as placing a strong prior on (physically) expected changes
in the freeze-out tail.

The top right panel of Fig~\ref{recon} shows the reconstructed
perturbation including three different number of XeMs.  Here the
recovered curve becomes very close to the original perturbation by
including the first 15 XeMs, while six XeMs have a poor recovery of
the low-$z$ part.  Note that the plots are illustrating $\delta
\ln(\Xe(z))$ although the XeMs and thus the decomposition of the
perturbation are all performed with $\delta \ln(\Xe+0.01)$.

For comparison the reconstruction of the perturbation with $\delta u
(z)=\delta\ln(\Xe)$, i.e., with $\sigma=0$, and with 15 XeMs taken
into account is also shown. As expected, this reconstruction is much
poorer compared to the previous case with $\delta u (z)=\delta
\ln(\Xe+0.01)$ due to its lack of coverage of corrections in the
freeze-out tail. This demonstrates that when there is prior knowledge
in favour of the freeze-out tail of recombination being affected, a
parametrization with $\sigma > 0$ should be used in the analysis.
However, it is still correct that the main signal is produced by the
modifications close to $z\sim 1100$, even if the freeze-out tail
apparently has the largest deviation from the SRS.  This is why the
first few mode functions for $\sigma=0$ do not have any strong low
redshift tails. The eigenvectors naturally order the perturbations in
the strength of the associated change in the CMB power spectra, as
explained in \S~\ref{sec:basis}.  This point is visible from the lower
right panel where the $C_\ell$ difference is plotted for reconstructed
perturbations with different number of modes included compared to the
full perturbations.  Similar to the previous case, these differences
are several times smaller than the changes in the $C_\ell$'s caused by
this model of DM annihilation, again meaning that these few modes can
well capture the constrainable features of the perturbations.

Also if we look at the decomposition of the recombination correction
into the first six XeMs (see Table~\ref{CT_proj}) we see that they all
have comparable contributions. This seems reasonable if we remember
that the mode functions, despite being weighted toward the low
redshift part, still have a significant component at high redshift
which need to be cancelled out to recover this pattern of perturbation
with its low redshift modification. Therefore the neighbouring modes
have the same order of magnitude amplitude to properly cancel out the
high redshift perturbations.  This difference in the amplitude of the
modes in principle allows us to distinguish this type of perturbation
from those of CT2010.

\subsection{Impact of the eigenmodes on differential visibility and CMB power spectra}
It is worthwhile to see how the XeMs affect the visibility function
and the CMB power spectra.  The left panel in Fig.~\ref{mode_vis_cl}
shows the change in the visibility function (normalized to the maximum
of the fiducial visibility) for the first six XeMs.  It is remarkable
that relative changes in $\Xe$ of a few percent close to the maximum
of visibility, which lead to measurable effects in the CMB power
spectra, only cause relative changes in the visibility of the order of
$10^{-4}$. This confirms the high sensitivity of the $C_\ell$'s to
tiny changes in the visibility.

From Fig.~\ref{mode_vis_cl} we also see that the most constrained
mode, XeM $1$, has an effect on the CMB power spectra consistent with
changes the width of the visibility function and a slight shift of its
peak position. The second mode corresponds primarily to a shift in the
visibility peak.  The higher XeMs, with their several peaks and
valleys, lead to less trivial changes in the width and position of the
visibility function.

\begin{figure*} 
\begin{center}
\includegraphics[scale=0.88]{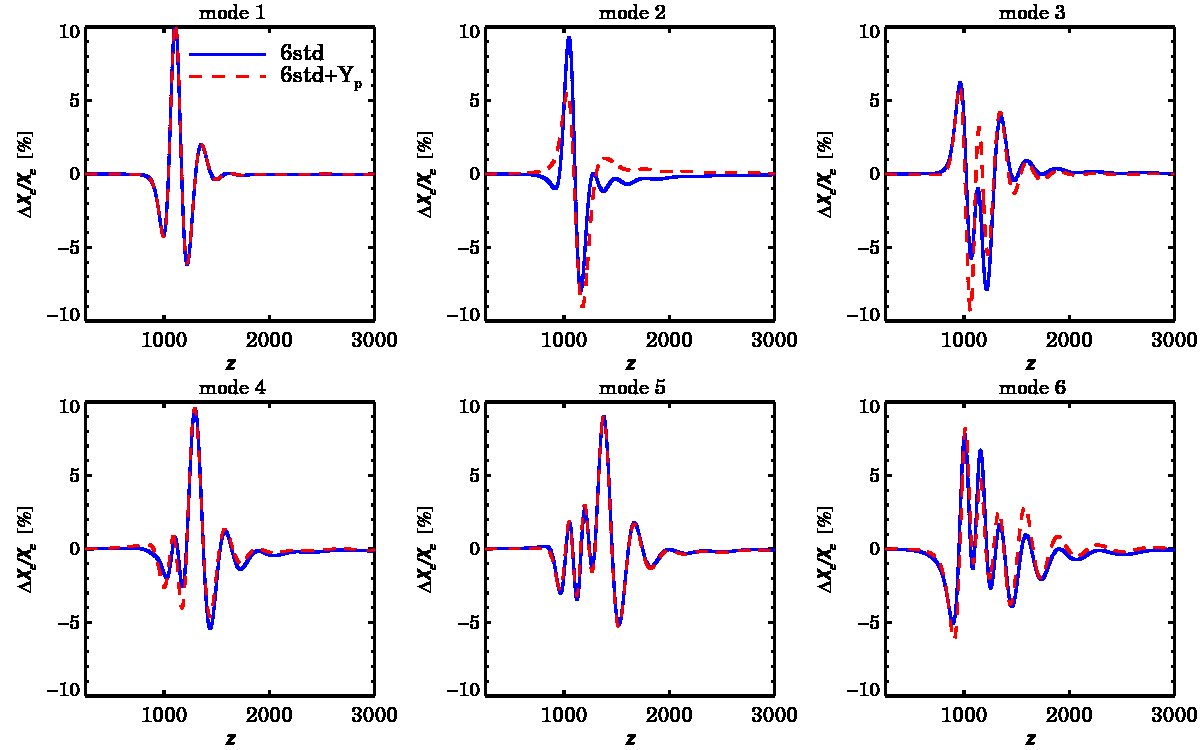}
\end{center}
\caption{The six most constrained eXeMs. The solid blue lines
  correspond to modes constructed after marginalization over six
  standard parameters while for the dashed red curves $Y_{\rm p}$ is
  marginalized over in addition. }
\label{eXeM_6_7} 
\end{figure*}
The relative changes in the CMB temperature and $E$-mode polarization
power spectra due to the first six XeMs are illustrated in the middle
and right panels of Fig.~\ref{mode_vis_cl}. To aid visual comparison,
the amplitudes of the XeMs have been chosen equal to their associated
$1\sigma$'s so that they would lead to comparable changes in the
$C_\ell$'s.  It is interesting to note that for XeM $1$, due to its
narrower visibility width compared to the fiducial model, the high
$\ell$ damping in the temperature and polarization anisotropies is
smaller. At the same time, because of fewer scattering opportunities
for the photons, this mode leads to less polarization (negative
$\delta C_\ell^E$ at low $\ell$'s).  The $C_\ell$ signal associated
with the second mode has the strongest oscillatory behaviour,
consistent with a shift in the position of the visibility function.
The mainly oscillatory impact of XeM $4$ and $5$ on the $C_\ell$'s
also suggests that an effective shift in the position of the
visibility function is present, while for mode XeM $3$ and $6$ the
tail of the power spectrum is strongly damped (thus the negative
$\delta C_\ell^{T,E}$ at small scales), again corresponding to a
change in the effective width of the visibility.

\subsection{Perturbations to helium recombination}
\label{he_pert}
As mentioned early in this section, the redshift range chosen in our
analysis of perturbations to ionization fraction includes the
recombination of singly ionized helium. Some of the most constrained
XeMs we found also extend up to $z\sim 1600$. These, therefore imply
some impact from the helium recombination epoch on the XeMs.

One way to confirm this statement is to limit the redshift range of
perturbations to mainly include singly ionized helium recombination,
e.g., $[1500,3000]$, while the total $\Xe$ (from both hydrogen and
helium) is perturbed.  We observe that the XeMs constructed this way
have comparably large values at the lower redshift boundary ($z=1500$)
and would steeply go to zero if enforced by the imposed boundary
conditions. This indicates that despite being restricted to the helium
recombination epoch, the XeMs are still most sensitive to changes in
the signal from the hydrogen recombination and changes in $\Xe$ due to
the helium recombination are hardly constrainable, unless a properly
chosen non-uniform prior on $\delta \Xe$ is imposed.
 
 As already emphasized, the parameters $\delta u (z)$ only
 characterize relative changes in $\Xe$ and the full description of
 the ionization fraction depends also on the standard cosmic
 parameters as well as the relevant theoretical assumptions about the
 physics of recombination. Among the standard parameters, $Y_{\rm p}$
 has a distinct role in describing an aspect of the ionization
 fraction complementary to $\delta u (z)$ by determining the maximum
 total number of electrons available at each redshift: $N_{\rm e,
   max}=N_{\rm e,max}^{\rm H}+N_{\rm e, max}^{\rm He} \approx
 (1-Y_{\rm p}/2)N_{\rm b}$, where $N_{\rm b}$ is the baryon number
 density.  Therefore, although $Y_{\rm p}$, in the first instance,
 requires to be marginalized over when constructing the eXeMs due to
 its intimate relation with the ionization fraction, it is also
 legitimate to treat recombination perturbations and the maximum
 number of electrons available at each redshift separately.

In \S~\ref{mcmc_std} we will use MCMC to measure constraints on
$Y_{\rm p}$ alongside the six standard parameters and the first few
XeMs using (simulated) CMB data. Also in the next section, we explore
how the eigenmodes change if they are marginalized over $Y_{\rm p}$.

\begin{figure*} 
\begin{center}
\includegraphics[scale=0.88]{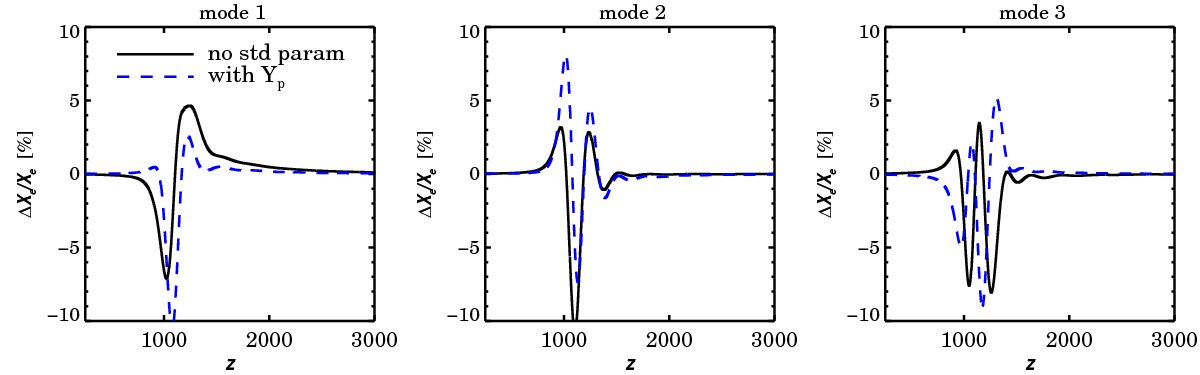}
\end{center}
\caption{The first three eXeMs with only $Y_{\rm p}$ being
  marginalized over (dashed blue curves). For comparison, the first
  three XeMs are also plotted (solid black curves). }
\label{EM_eXeMyhe}
\end{figure*}
\begin{deluxetable*}{ccccccc}
\tablecaption{The forecasted standard deviations of the first six eXeMs
  from the Fisher analysis constructed by marginalization over different number of
  standard cosmic parameters and for different observational cases. In
all cases, $\ell_{\rm max}=3500$.}
\tablehead{\colhead{eXeM} & \colhead{1} & \colhead{2}
  & \colhead{3} & \colhead{4} &
  \colhead{5} & \colhead{6} }
\startdata
CVL,  marg: six std   &0.011  & 0.012 & 0.029 & 0.052 & 0.059 & 0.064 \\[1mm]
CVL, marg: six std + $Y_{\rm p}$   &0.011  & 0.027 & 0.029 & 0.052 & 0.059 & 0.071 \\[1mm] 
Planck-ACTPol, marg: six std   &0.058  & 0.074
& 0.189 & 0.308 & 0.439 & 0.532  \\[1mm]
CVL, marg: $n_{\rm s}$, $A_{\rm s}$   &0.009  & 0.011 & 0.016 & 0.018 & 0.033 & 0.059 
\enddata
\label{evaleXeM}
\end{deluxetable*}

 \subsection{eXeM construction}\label{eXeM_rec}
The XeMs we discussed so far were constructed with non-varying
standard parameters and therefore can be considered as the limiting
case of zero errors on the standard parameters.  However, as mentioned
earlier in \S~\ref{sec:PCA_std_VAR}, the eigenmodes become correlated
when they are simultaneously being measured with the standard
parameters, due to their degeneracy with the standard parameters.  The
strength of the impact of these correlations on the XeM estimation
depends on the (prior) constraints on the standard parameters.  It is
therefore worthwhile to see how the modes and their rank ordering
change if the standard parameters are allowed to vary as well.

Figure~\ref{eXeM_6_7} illustrates the first six eXeMs constructed
after marginalization over the main six and seven (including $Y_{\rm
  p}$) standard parameters, constructed as described in
\S~\ref{sec:PCA_std_VAR}. The eXeMs have stronger high redshift
features compared to the XeMs.  This implies that the degeneracy
between the standard parameters and some features in the perturbations
of the ionization fraction has pushed back some patterns of high
significance to lower levels opening up the room for some high
redshift or higher frequency patterns which only had the chance to
show up at lower significance XeMs.

The modes in the shown two cases (i.e., marginalized over six and
seven standard parameters) differ only slightly. That is because
$Y_{\rm p}$ is rather weakly constrained using CMB data alone and in
the presence of other standard parameters its role in shaping the
eigenmodes is only secondary.  If, on the other hand, we hold the six
standard parameters fixed and only let $Y_{\rm p}$ vary, the
eigenmodes will be more significantly affected (see
Fig~\ref{EM_eXeMyhe}). The reason is that $Y_{\rm p}$ is comparable in
significance to small changes in the ionization fraction and
marginalizing over it, without the dominance of the standard
parameters, would lead to marked changes in the eigenmodes.  The
forecasted errors on the first six eXeMs with and without $Y_{\rm p}$
included is compared in Table~\ref{evaleXeM}. We see that the errors
are mostly the same, again implying the subdominant role of $Y_{\rm
  p}$.  In terms of the errors the most affected modes are eXeM 2 and
6.

\begin{deluxetable*}{ccccccccc}
\tablecaption{The coefficients of projection of the six most constrained  eXeMs on the first eight XeMs.}
\tablehead{\colhead{} & \colhead{XeM 1} & \colhead{XeM 2}
  & \colhead{XeM 3} & \colhead{XeM 4} &
  \colhead{XeM 5} & \colhead{XeM 6} & \colhead{XeM 7} & \colhead{XeM 8}  }
\startdata
eXeM 1     & -0.00  & -0.90 & 0.21 & -0.29 & -0.12 & 0.02 & 0.02 & 0.01       \\[1mm]
eXeM 2     & -0.76  & -0.05 & -0.48 & -0.35 & 0.16 & -0.02  & -0.02 & -0.01   \\[1mm]
eXeM 3     & -0.31  & 0.34 & 0.42 & -0.38 & -0.54 & -0.41  & 0.02 & -0.11      \\[1mm]
eXeM 4     & -0.35  & 0.13 & -0.48 & -0.31 & -0.50 & 0.27  & -0.28  & 0.37     \\[1mm]
eXeM 5     & 0.24   & -0.01 & -0.15 & -0.20 & 0.06 & -0.48 & 0.63  & 0.49      \\[1mm]
eXeM 6     & -0.19   & -0.22 & -0.07 & 0.41 & 0.00 & -0.41 & -0.55  & 0.49      
\enddata
\label{eXeMonEM} 
\end{deluxetable*}
\begin{figure*} 
\begin{center}
\includegraphics[scale=0.88]{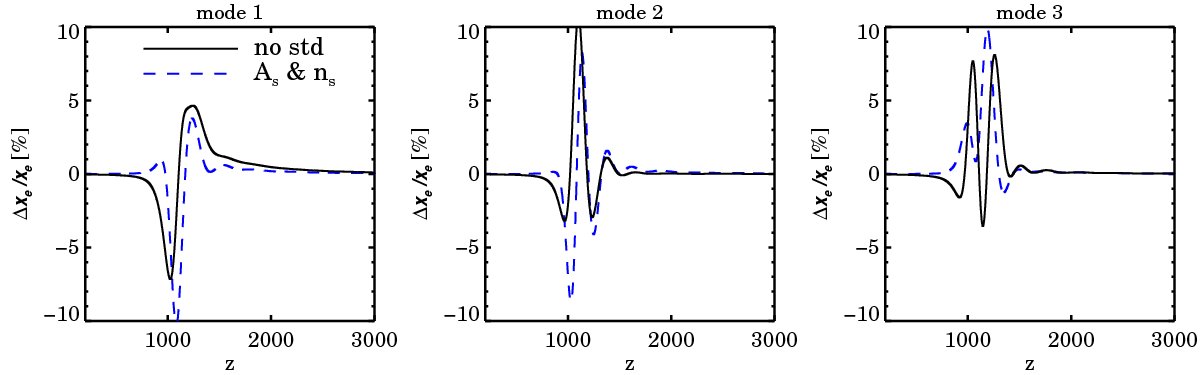}
\end{center}
\caption{The best three eXeMs with the two
  inflationary parameters $A_{\rm s}$ and $n_{\rm s}$ being marginalized over, compared to the
  XeMs, where all standard cosmic parameters are held fixed in the
  construction process of perturbation eigenmodes.}
\label{inf_eXeM}
\end{figure*}
\subsubsection{Projecting eXeMs onto XeMs}
\label{eXeM_project}
It is instructive to see how the eXeMs can be constructed from the
XeMs. Table~\ref{eXeMonEM} shows the coefficients of the projection of
the first six eXeMs (Fig.~\ref{eXeM_6_7}) on the best eight XeMs
(Fig.~\ref{multibasis}).  Note that the most constrained eXeMs have
their strongest projections along these first few XeMs and the
contribution from all other modes is at most about a percent for these
first six eXeMs. This means that allowing the standard parameters to
vary mixes and rearranges the first few XeMs with negligible leakage
from higher neglected XeMs.

More specifically, the first eXeM is very similar to XeM 2, with some
contribution from XeM 3 and XeM 4. The second eXeM is mainly a
combination of the first, third and fourth XeMs, and eXeM 3 and eXeM 4
have comparable contributions from the first six and eight XeMs,
respectively. The eXeM 5 and 6 are dominated by higher modes (XeM
6-8).
 As a result the space covered by the first few XeMs is not
 practically hugely different from the space covered by the first few
 eXeMs if proper number of modes (here eight XeMs for six eXeMs) are
 included in the analysis.  This implies that using similar number of
 XeMs and eXeMs in an analysis of possible recombination perturbations
 should give similar results for the reconstructed modification in the
 ionization history, at least for the CVL case where relatively large
 number of eigenmodes are included.  The main advantage of the eXeMs
 is that one can obtain more realistic estimates for the error bars
 directly after the construction of these modes. Also the eigenmode
 measurement is robust against including new eigenmodes in the
 analysis if the eXeMs are used as the modes.  However, it also turns
 out that the eXeMs perform better than the XeMs in the simulated
 analysis of {\sc Planck} data, where only 1-3 modes seem to be
 constrainable.

\subsection{From eXeMs to XeMs}
The two sets of ionization perturbation eigenmodes introduced and
constructed so far, i.e., the XeMs and the eXeMs, allow us to best
describe and measure the uncertainties in the ionization fraction in
the two extreme ends of our knowledge of the standard cosmic
parameters. The eXeMs present a case where the tightest constraints on
the standard (six) parameters are from CMB data alone. Therefore a
simultaneous measurement of the standard parameters and the
uncertainties in the ionization fraction, using the CMB dataset at
hand, is required. The construction of the XeMs, on the other hand,
assumes the standard cosmic parameters are measured with high accuracy
from other cosmological probes and the CMB data are only used for the
direct measurement of the ionization history. In other words, the
XeMs, by ignoring the uncertainties in the standard parameters,
extract the maximal amount of information that the CMB data would ever
have to offer about the ionization fraction.

Between these two limiting cases, there is a {\it gray} region where,
depending on the dataset at hand, tight priors from non-CMB surveys
can be imposed on some of the standard parameters while the rest are
marginalized over when constructing the eigenmodes. For example, if
all standard parameters, but the inflationary ones $A_{\rm s}$ and
$n_{\rm s}$, are measured to very high precision by other probes, such
as large scale structure, baryonic oscillation, lensing and supernova
surveys (e.g. LSST\footnote{http://www.lsst.org/lsst/},
Pan-STARRS\footnote{http://pan-starrs.ifa.hawaii.edu/public/home.html},
BigBOSS, WFIRST\footnote{http://wfirst.gsfc.nasa.gov/},
EUCLID\footnote{http://sci.esa.int/science-e/www/object/index.cfm?\\ fobjectid=42266}),
the corresponding eigenmodes would be constructed after
marginalization only over these inflationary parameters.

Fig~\ref{inf_eXeM} compares the first three XeMs with eigenmodes
marginalized over $A_{\rm s}$ and $n_{\rm s}$. The forecasted errors
on these eigenmodes (from the Fisher analysis) are presented in the
last row of Table~\ref{evaleXeM}. Not surprisingly, these modes have
smaller errors compared to the eXeMs which have been made after
marginalization over six standard parameters, and have larger errors
compared to XeMs (with no standard parameter varying).  These modes
and similar ones after marginalization over different sets of standard
parameters smoothly bridge the gap between the XeMs and the
eXeMs. Depending on the datasets available at the time of real data
analysis, the proper eigenmodes marginalized over the appropriate
standard parameters must be constructed. With the current (and very
near future) surveys, the most realistic choice are the eXeMs,
constructed according to the experiment under consideration, which
should be quite similar to the Planck-ACTPol-like case studied here.

\subsection{Criteria for truncating the eigenmode hierarchy}\label{num_EM}

For the full reconstruction of perturbations to the ionization
fraction, all eigenmodes are needed in principle since they form a
complete basis set. In practice, sequentially adding modes rank
ordered in the (possibly renormalized) eigenvalues $f_k=
\sigma_k^{-2}$ of ${\bf F}$ from high to low gives a rapidly
diminishing return once one goes beyond a dozen or so. And often we
can learn much from using just the first few. As more modes are added,
the width covered by the allowed $\Xe$ trajectories increases, as
Figs.~\ref{JR_CVL_traj_8EM} to \ref {JR_pl_traj_3eXeM} in
\S~\ref{sec:traj} show. The errors in those standard cosmic parameters
which are correlated with the $\Xe$ eigen-parameters also increase. On
both counts, it behooves us to develop criteria for selecting which
modes to keep, bearing in mind Occam's Razor for minimizing the number
of new parameters to be added.  Thus we show in \S~\ref{sec:measure}
what happens when one mode, a few modes and a handful of modes are
added. To be more quantitative, we explore a criterion based on not
allowing the Shannon entropy to increase too much as the next
eigenmodes in the hierarchy are added.
  
The information action is defined in terms of the {\it a posteriori}
probability of the variables $p_{\rm f}$ and the evidence ${\cal E}$
as ${\cal S}_{\rm I,f}({\bf q})\equiv \ln p_{\rm f}^{-1} -\ln {\cal
  E}$. Recalling from \S~\ref{sec:PCA_std_const}, the {\it a
  posteriori} probability $p_{\rm f} \equiv p ({\bf q} \vert d, {\cal
  T})$ of variables ${\bf q}=(q_1,...,q_N)$ given the theory space
${\cal T}$ and the datasets $d$ is related to the {\it a priori}
probability $p_{\rm i} \equiv p({\bf q} \vert {\cal T})$, the
likelihood ${\cal L}({\bf q}\vert d,{\cal T}) \equiv p(d\vert {\bf q},
{\cal T}) $ and the evidence ${\cal E}\equiv p(d|{\cal T})$ through
Bayes theorem: $p_{\rm f}= {\cal L}({\bf q}|d,{\cal T})p_{\rm i}/{\cal
  E}$.  The information action can then be written in terms of $p_{\rm
  i}$ and ${\cal L}$:
 \begin{equation}\label{selfinfo}
{\cal S}_{\rm I,f}({\bf q}) =  \ln p_{\rm i}^{-1} + \ln {\cal L}^{-1} \, . 
\end{equation}
For basic information theoretic and Bayesian notions and notations
see, e.g., the \cite{MacKay} textbook. The framework given here was
used in a CMB context by \cite{far11}.) For us the $q_k$'s are the
amplitudes of the ordered eigenmodes for XeMs or eXeMs. Generally the
fluctuations in the standard cosmic parameters from their maximum
likelihood values are included along with these eigen-parameters. We
shall assume the prior distribution of the parameters to be uniform in
the $q_k$. The expansion of $S_{\rm I,f}$ to quadratic order is the
basic perturbative approach used throughout this paper, leading to a
Gaussian $p_{\rm f}$: ${\cal S}_{\rm I, f}({\bf q}) \approx {\cal
  S}_{\rm I,m} +{\bf q}^\dagger {\bf F} {\bf q} /2$ in terms of the
Fisher matrix and the information action minimum ${\cal S}_{\rm
  I, m}=-\ln (p_{\rm i} {\cal L}_{\rm max} )$.

The posterior Shannon entropy is related to the final-state
ensemble-average of the information action and the evidence:
 \begin{equation*}
 S_{\rm f}\equiv \avrg{\ln p_{\rm f}^{-1}}_{\rm f} = \avrg{{\cal
     S}_{\rm I, f}({\bf q})}_{\rm f} + \ln {\cal E} \, .
 \end{equation*}  
For the quadratic order expansion it is 
\begin{eqnarray*}
 S_{\rm f} && \approx \frac{1}{2} \rm{Tr} \ln{\bf F}^{-1} +
 \frac{N}{2} \ln (2\pi)+ \frac{1}{2} \rm{Tr}\left( \avrg{{\bf q}{\bf
     q}^\dagger} {\bf F} \right) 
  \\[1mm] 
  && =\frac{1}{2} \rm{Tr} \ln{\bf
   F}^{-1} + \frac{N}{2} (\ln (2\pi)+1) .
\end{eqnarray*}  
The second line follows from the first since the correlation matrix of
the ${\bf q}$ is $\avrg{{\bf q}{\bf q}^\dagger}={\bf F}^{-1}$.  The
associated evidence involves the information action minimum, $\ln
{\cal E} \approx S_{\rm f} - {\cal S}_{\rm I, m} - \frac{N}{2} $.

\begin{figure} 
\begin{center}
\includegraphics[scale=.4]{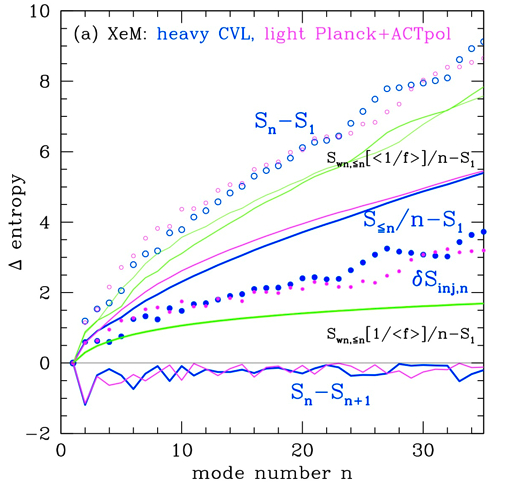}
\includegraphics[scale=.4]{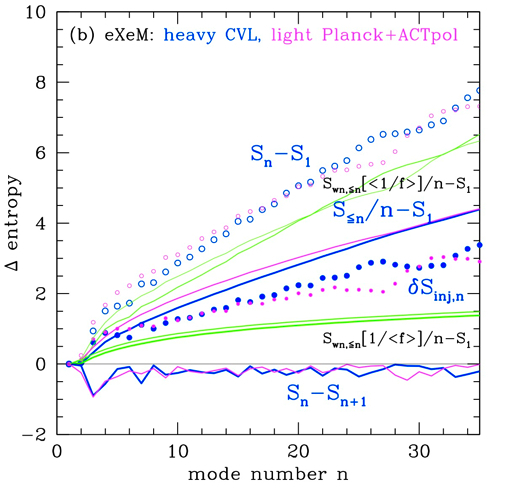}
\end{center}
\caption{Various measures of entropy differences defined in the text
  are plotted against increasing eigenmode number, for (a) the XeM
  case and (b) the eXeM case. The Cosmic Variance Limited mode results
  have heavy lines or points, the Planck+ACTpol forecast has lighter
  lines and points, as indicated. They look quite similar. For this
  figure, the modes are determined by the densely-packed Gaussian bump
  expansion, but the triangular and spline expansions look similar,
  differences becoming notable only at higher $n$. The basic
  information on growing entropy is given by $S_n-S_1$, and the mean
  difference $S_{\le n}/n-S_1$, with the latter bracketed by the two
  white-noise entropies. The criteria for threshold selection
  discussed in the text involve the injection entropy, $S_n-S_{\le
    n}/n$ and $S_{n+1}-S_n$ (plotted with opposite sign for
  clarity). Thresholding at $\Delta S_{\rm t} \sim 1/2$ selects the
  first mode or two, but mode-groups are also evident, suggesting
  modes should be added in blocks rather than singly as we eliminate
  bias, check convergence and demonstrate robustness.  }
\label{fig:Dentropy}
\end{figure}

The entropy associated with mode $n$ is
\begin{equation*}
 S_{n} \equiv -\frac{1}{2}  \ln f_n +(1+\ln 2\pi)/2 = S(\le n)-S(\le n-1)  \, .   
\end{equation*}
It is a finite difference of the total entropy of the first $n$ modes  
in the eigen-hierarchy,   
\begin{equation*}
S(\le n)=\frac{n}{2}(1+\ln 2\pi) - \frac{1}{2} \sum_{k=1}^{n} \ln f_k \, 
\end{equation*}
and 
\begin{equation*}
\avrg{s}_n \equiv  S(\le n) /n \, 
\end{equation*}
is the associated mean entropy-per-mode. Fig.~\ref{fig:Dentropy} shows
how the relative entropy $S_n-S_1$ and the mean entropy $\avrg{s}_n -
S_1= S(\le n)/n-S_1 $ grow with $n$ for the modes derived from the
localized Gaussian expansion. We also plot two versions of
"white-noise" entropy.
\begin{eqnarray*}
&& S({\rm wn},\le n) (\sigma^2) \equiv n (\ln \sigma + (1+\ln 2\pi) /2) ,\\[1mm]
&& {\rm mean-variance} \ \sigma^2 = \sum_{k\le n}  f_k^{-2} /n , \\[1mm]
&& {\rm mean-weight} \   \sigma^2 = [\sum_{k\le n}  f_k^{2} /n]^{-1}.
\end{eqnarray*}
These are entropies maximized subject to the constraint that we only
have knowledge of the integrated $\sigma^2$, whereas $S(\le n)$ is the
maximized entropy given knowledge of the full spectrum $\{ f_k^{-1}
\}$. The mean-variance white noise lies above $S(\le n)$ and the
mean-weight white noise lies below.  The mean-weight behaviour is
dominated by a $\ln (n)$ rise, since the total weight of modes below
$n$, $ \ln \sum_{k\le n} f_k^{2} $, quickly approaches a constant,
reflecting the dominance of the high-weight eigenmodes in the sum.

 We first discuss why we do not use the traditional evidence ratio
 often used in Bayesian theory to decide if a new parameter $q_{n}$
 should be added. The log-evidence difference for the addition of
 $q_{n}$ is
\begin{eqnarray*}
 \Delta \ln {\cal E}_{n} && \equiv \ln {\cal E}(\le n)-\ln {\cal E}(\le n-1) 
 \\[1mm]
&& = S_n -1/2  - \Delta {\cal S}_{\rm I,m} .  
\end{eqnarray*}
This requires evaluation of the change in the information minimum. It
also has the usual disadvantage of depending upon the $f_k$
measure. Although using eigen-parameters ensures the same
dimensionality for the different $f_k$, it does not fully remove this
re-parameterization ambiguity since there can be a $k$-dependent
scaling. (In fact, we have usually renormalized our $f_k$ so that the
associated eigenmodes $E_k(z)$ have unit norm upon $z$-integration.)

Our preferred approach for hierarchy truncation is to use
suitably-defined entropy differences. In particular we wish to set a
threshold control on the injection entropy,
\begin{equation*}
\delta S_{{\rm inj}, n} = S_n -\avrg{s}_n\, , 
\end{equation*}
the entropy from adding mode $n$ relative to the mean-entropy from all
$\le n$ modes.  It is related to the relative increase in phase space
volume $V (\le n) = \exp(S (\le n) -n/2)= \exp (n(\avrg{s}_n-1/2))$
associated with mode additions:
\begin{eqnarray*}
\ln \left[V(\le n+1)/V (\le n)^{(n+1)/n}\right] && = S_{n+1}-\avrg{s}_{n}  
\end{eqnarray*}
We chose $S_{{\rm inj}, n} $ instead because it is zeroed out for mode
one, but $S_{{\rm inj}, n} $ quickly approaches
$S_{n+1}-\avrg{s}_{n}$. For example, if we impose a $\Delta S_{\rm
  t}\sim 1/2$ threshold in Fig.~\ref{fig:Dentropy} on the CVL XeM
case, we would use only one mode, whereas $\Delta S_{\rm t}\sim 1$
picks up about 5, $\sim 3/2$ harvests about 10, and 2 gives about a
dozen.  Similar tales can be told for the eXeM CVL case and for both
Planck+ACTpol forecasts.

Another more erratic measure is relative injection jumps, which is
nearly $S_{n+1}-S_n$. In Fig.~\ref{fig:Dentropy}, the negative of this
is plotted for clarity of presentation.  Either reading off from the
figure, or the using the lists of errors in Tables~\ref{eval} and
\ref{evaleXeM}, the sample threshold $\Delta S_{\rm t}=1/2$ again
yields only a mode or two.

The fluctuating nature of $S_{n+1}-S_n$ implies we can use it to split
the modes into groups of similar information content which arise by
thresholding it. If a mode is selected to be included in the analysis
by, say, sharp-thresholding the injection entropy, it is logical that
all of its co-modes be included, which is akin to softening the
threshold. The groupings found with $ \Delta S_{\rm t}=1/2$ imposed
upon $S_{n+1}-S_n$ creates boundaries at one mode, five modes, and so
on. These are, not surprisingly, similar to mode numbers obtained as
we move the threshold on injection up, hence that criterion can be
used instead to define mode groups.

Although these entropy difference criteria imply that relatively
little additional information is gained by including more than a
handful of higher modes, in real data analysis the situation is
subtler, with other criteria important to consider. For example,
depending on how close the assumed model is to the true underlying
history, our measurements of standard cosmic parameters might be
biased. In that case one would like to add enough modes to remove the
bias, sequentially checking if the recovered values of the standard
parameters are robust against the introduction of the next
eigenmode. A reasonable strategy is to add one {\it mode-group} at a
time to the analysis until the biases are removed.  In the next
section, we show how varying the mode number cutoff affects our
results, roughly following this grouping procedure.

\begin{figure} 
\begin{center}
\includegraphics[scale=1]{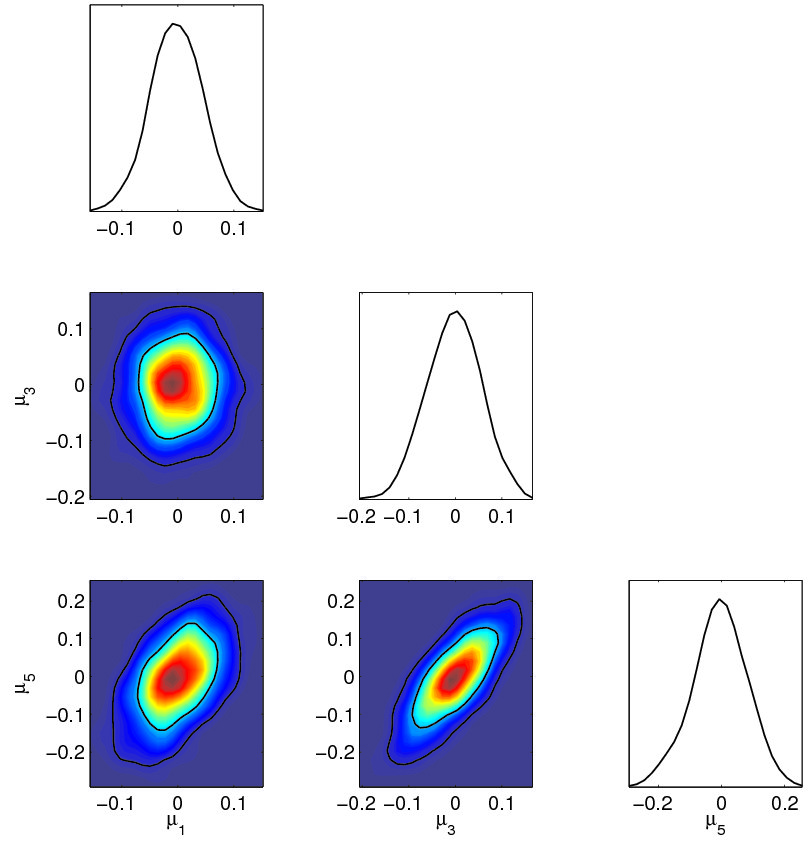}
\end{center}
\caption{ $2$D contours for the amplitudes of some of the best five
  XeMs as measured by a CVL experiment and with the standard
  recombination history. Here the six standard cosmic parameters were
  also allowed to vary. That is why the measured eigenmodes are
  correlated.}
\label{EM5_std}
\end{figure} 

\begin{figure} 
\begin{center}
\includegraphics[scale=1]{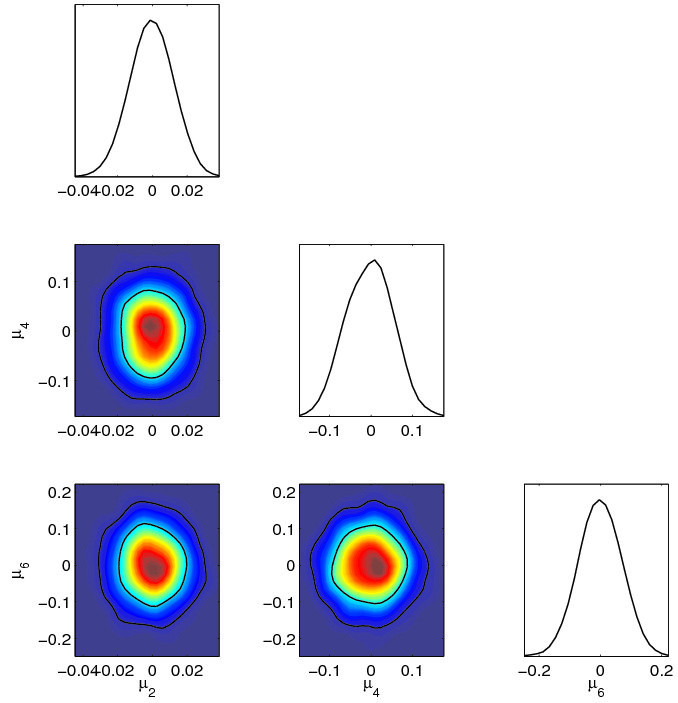}
\end{center}
\caption{$2$D contours for the amplitudes of some of the best six
  eXeMs as measured by a CVL experiment and with the standard
  recombination history.  The six standard cosmic parameters were also
  allowed to vary in the simulations.}
\label{eXeM6_std}
\end{figure} 


\section{Measuring the amplitudes of perturbation eigenmodes for simulated data}
\label{sec:measure}
Having constructed the eigenmodes, their amplitudes can now be
considered as additional parameters to be plugged into {\sc
  CosmoMC}\footnote{http://cosmologist.info/cosmomc/}.  In this
section we investigate how well the most constrained XeMs and eXeMs
can be measured by simulated data.  To study the impact of these new
variables on the standard parameter estimation, we first consider the
case in which the data are both simulated and analyzed using the SRS
(\S~\ref{sec:SRS}).  We then study the case for which the effects of
physical corrections to the recombination history (CT2010, see
\S~\ref{sec:recon}) are included in the constructions of the mock
data, but are neglected in the fiducial recombination model used in
the analysis.  Here the question is how well the eigenmodes compensate
for the deviations from the fiducial model and how much the data are
telling us about the amplitudes of the modes.  We also briefly discuss
how the eigenmodes should be used in a more general case where little
prior knowledge about the recombination perturbations is available.

\subsection{Case 1: The standard recombination scenario}
\label{mcmc_std}
As the first example we choose the fiducial recombination model (here
the SRS) to be identical to the ionization history used in the
simulation of the data.  We ran {\sc CosmoMC} to estimate the best-fit
values and errors of the six standard parameters together with those
for the perturbation eigenmodes.  We tried the two sets of eigenmodes
described before: the first five XeMs and the first six eXeMs.  The
number of modes in each case was chosen in rough agreement with the
mode cutoffs described in \S~\ref{num_EM}.  The simulations were
carried out for a CVL experiment.

\begin{figure} 
\begin{center}
\includegraphics[scale=1.]{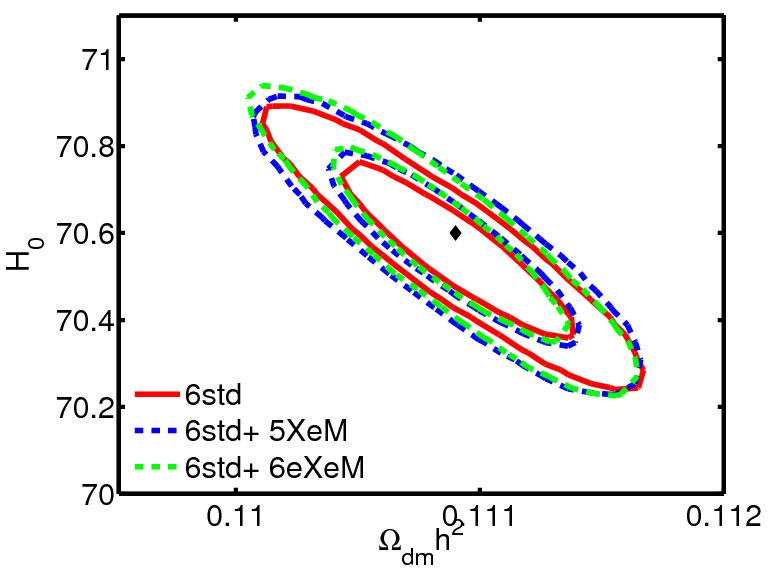}
\\[1mm]
\includegraphics[scale=1.]{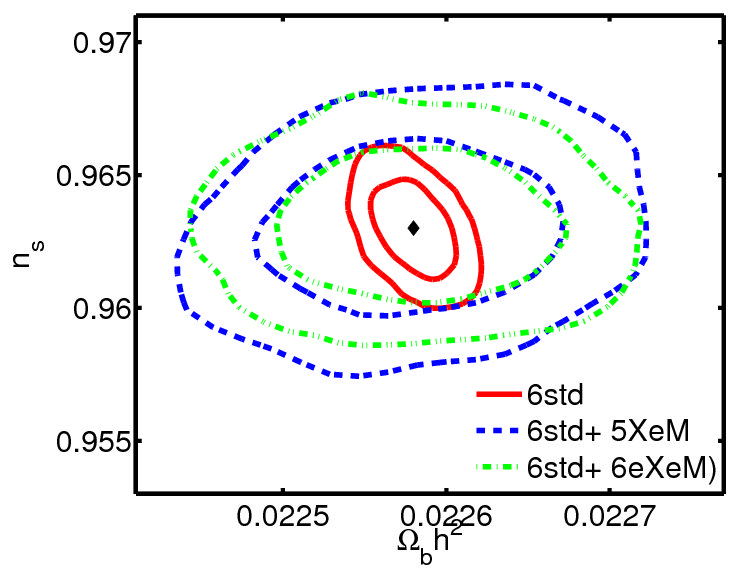}
\\[1mm]
\includegraphics[scale=1.]{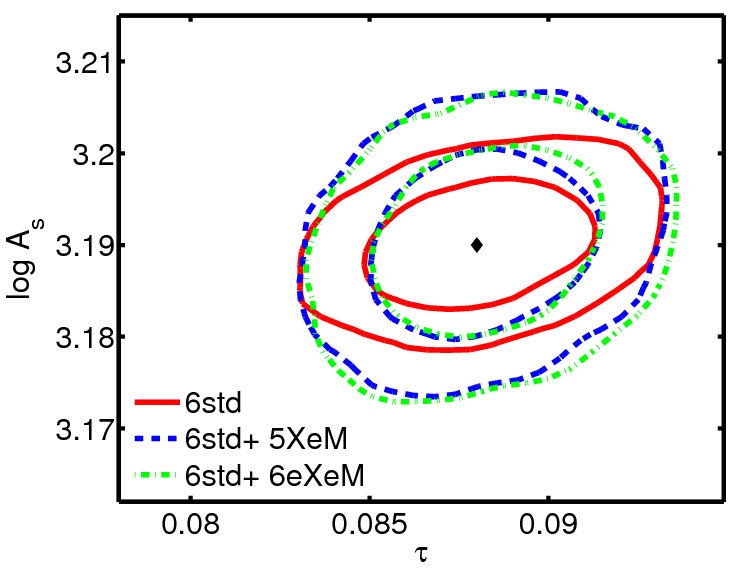}
\end{center}
\caption{The contours for the standard parameters as measured by an
  ideal experiment in the presence of five (six) XeMs (eXeMs) compared
  to the case with no eigenmodes included. The input value of the
  parameters is shown by the black diamond. }
\label{std_std_cont}
\end{figure} 
\begin{figure*} 
\begin{center}
\includegraphics[scale=0.9]{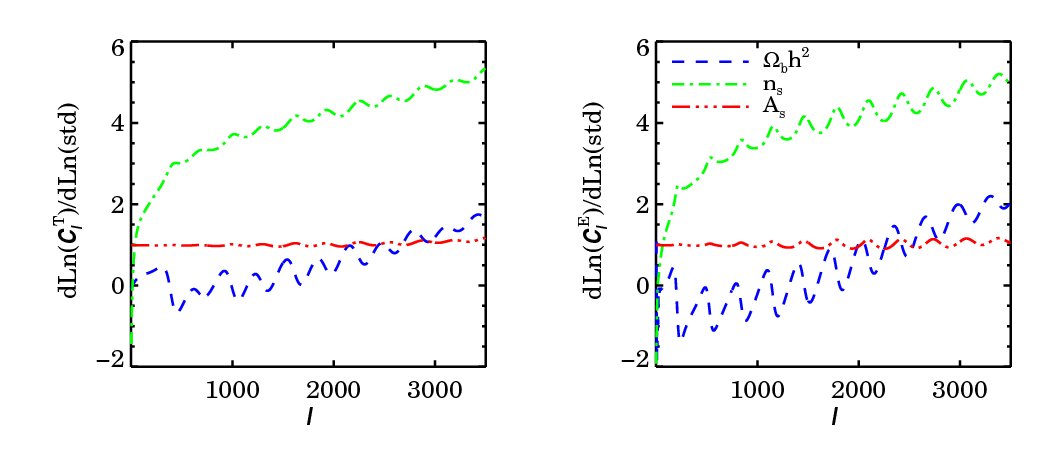}
\end{center}
\caption{The derivatives of the $C_\ell^{T,E}$'s with respect to some
  of the standard parameters. }
\label{dcl_std}
\end{figure*} 

\begin{deluxetable}{cccccc}
\tablecaption{The standard deviations of the first five XeMs
  from chains produced by CosmoMC, marginalized over the main six
  standard parameters.}
\tablehead{\colhead{XeM} & \colhead{1} & \colhead{2}
  & \colhead{3} & \colhead{4} &
  \colhead{5} }
\startdata
$1\sigma$   & 0.046  & 0.030 & 0.057 & 0.088 & 0.086 
\enddata
\label{error_cosmomc}
\end{deluxetable}

One expects no detection of eigenmodes since the fiducial model for
$\Xe$ and the underlying model used to simulate data are the same, as
verified by Figs.~\ref{EM5_std} and \ref{eXeM6_std}.  Also, by
construction there is almost no visible correlation between the
measured parameters for the eXeMs, at least sufficiently close to the
best-fit model where the assumptions of the Gaussianity for the
likelihood surface approximately holds.  However, Fig.~\ref{EM5_std}
indicates that the XeMs become partially correlated with each other,
although by construction these were initially
uncorrelated \footnote{We confirmed this statement by running MCMC
  with non-varying standard parameters.}.  The reason is that the
standard parameters were held fixed during the process of XeM
construction, but now that they are allowed to vary, their degeneracy
with the XeMs induces correlations. These new correlations lead to
larger errors than those deduced from the simple Fisher analysis
(Table~\ref{eval} cf. Table~\ref{error_cosmomc}) and can also change
the rank ordering of the modes, e.g., the error on XeM 2 is smaller
than XeM 1 (Table~\ref{error_cosmomc}).

\begin{figure*} 
\begin{center}
\includegraphics[scale=1.1]{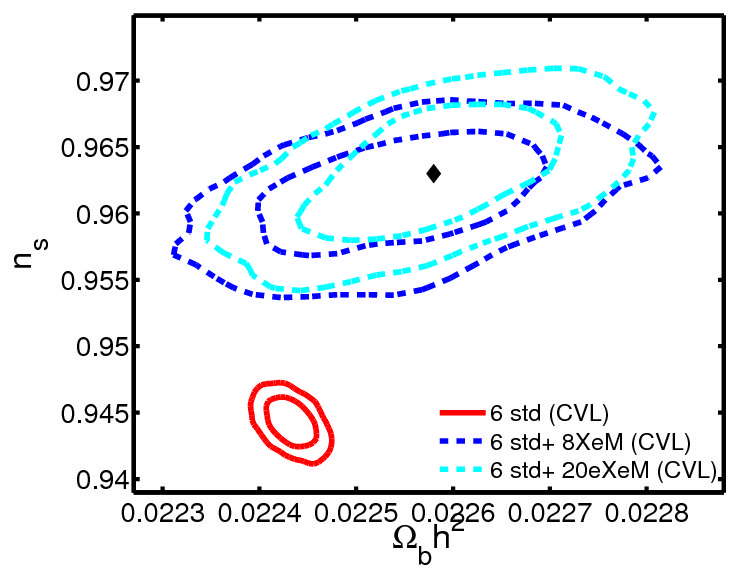}
\hspace{5mm}
\includegraphics[scale=1.1]{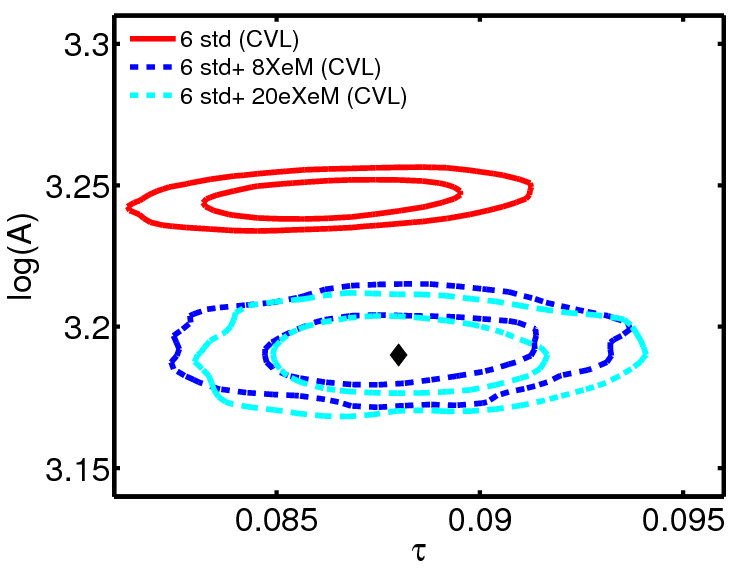}
\end{center}
\caption{Contours of some of standard parameters for CT2010 case, with
  eight XeMs in one case and 20 eXeMs in the other case included in
  the analysis, compared to a case where no perturbation eigenmodes
  (of any kind) has been included (the solid red curves). The
  simulations are performed for a CVL experiment. The input value of
  the parameters is shown by the black diamond.}
\label{JR_CVL}
\end{figure*}

The standard parameters remain unbiased, as the model used for
simulating data and the theoretical model used in the analysis were
the same. This is no longer true once recombination corrections to the
SRS are added (see Fig.~\ref{JR_CVL}).  However, the correlations of
the eigenmodes with some of the standard parameters increase the
errors of the standard parameters.\footnote{It should also be noted
  that the correlations between the standard parameters themselves
  also change when the eigenmodes are introduced.} From
Fig.~\ref{std_std_cont} we see that among the standard parameters,
$\Omega_{\rm b}h^2$, $n_{\rm s}$ and $A_{\rm s}$ are the ones most
affected by the introduction of the eigenmodes into the analysis. This
can be understood by noting the relatively high degeneracy between
these parameters and some of the eigenmodes. The most evident one is
the correlation of $n_{\rm s}$ with the first XeM which by changing
the width of the visibility function leads to a tilt in the power
spectra (compare Figs~\ref{dcl_std} and \ref{mode_vis_cl}). For the
case of $\Omega_{\rm b}h^2$ and $A_{\rm s}$ it is harder to give a
visual interpretation. $\Omega_{\rm b} h^2$, leading to both tilt
changes and oscillations in the $C_\ell$'s, correlates with most of
the first five XeMs (the highest being with XeM 1), while $A_{\rm s}$,
being an amplitude multiplier, mainly correlates with XeM~1.  These
correlations between the standard parameters and the eigenmodes
emphasize the fact that uncertainties in the recombination scenario in
particular undermine our ability to measure the precise values of
$\ns$ and $\Omega_{\rm b} h^2$ \citep[see e.g.,][]{sha11}.  Also note
that the changes in the error bars of the standard parameters are
actually practically independent of which set of eigenmodes are used
(Fig.~\ref{std_std_cont}).  This suggests that in terms of the
standard parameter estimation, the use of XeMs or eXeMs should not
lead to vastly different results in the parameter estimation. However,
the perturbations are measured to higher accuracy with the eXeMs
(Table~\ref{evaleXeM}) than XeMs (Table~\ref{error_cosmomc})
especially if only a few modes are included in the
analysis. Therefore, as long as only CMB data are used, the eXeMs are
the more appropriate choice of eigenmodes.

\begin{deluxetable*}{cccccc}
\tablecaption{$Y_{\rm p}$ and its measured error from simulations for
  a CVL and a Planck-ACTPol-like experiment, with XeMs and eXeMs taken
  into account as perturbation eigenmodes, compared to the case with
  no eigenmodes. }
\tablehead{\colhead{}  & \colhead{CVL(std)}  & \colhead{Planck-ACTPol(std)}  &  \colhead{CVL(std + 5 XeMs)} 
  & \colhead{CVL(std + 6 eXeMs)} & \colhead{Planck-ACTPol(std + 3 XeMs)} }
\startdata
$Y_{\rm p}$  & $0.240 \pm 0.0016$   & $0.240 \pm 0.006$ &  $0.239\pm 0.005$  & $0.240\pm 0.004 $ & $0.238 \pm 0.017$ 
\enddata
\label{table:yp}
\end{deluxetable*}
\begin{figure*} 
\begin{center}
\includegraphics[scale=1.1]{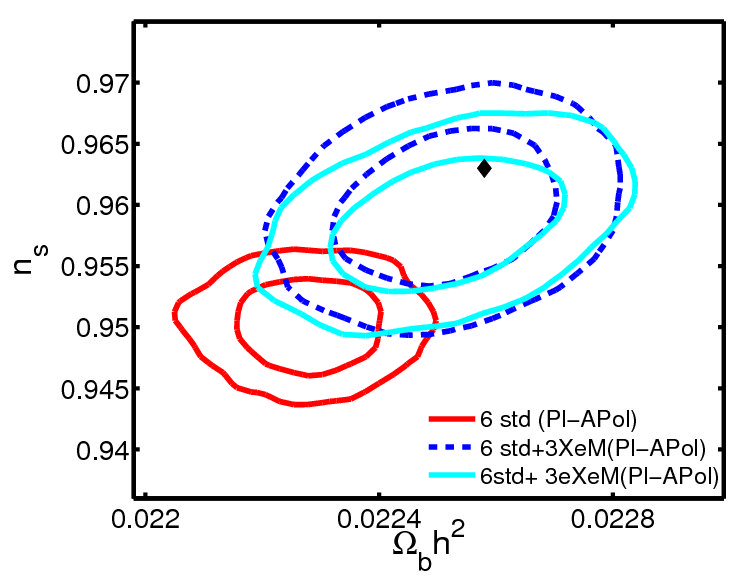}
\hspace{5mm}
\includegraphics[scale=1.1]{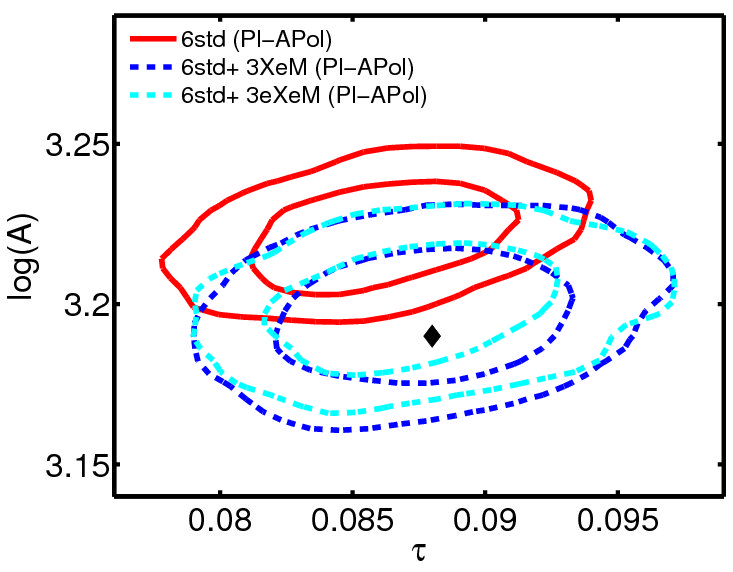}
\end{center}
\caption{Similar to Fig.~\ref{JR_CVL} but for a Planck-ACTPol-like
  experiment. Here three eigenmodes were added for both the XeM and
  eXeM case.}
\label{JR_PlApol}
\end{figure*}

Finally, we studied how much the presence of perturbations to
recombination could affect our ability to determine the precise value
of $Y_{\rm p}$.  The abundance of helium affects the CMB anisotropies
mainly because more helium implies fewer free electrons during
hydrogen recombination.  Consequently, $Y_{\rm p}$ should also couple
significantly to the perturbation eigenmodes.  We therefore performed
simulations in which $Y_{\rm p}$ was also allowed to vary.  The
analysis was performed with three and five XeMs in the
Planck-ACTPol-like and CVL case, and with six eXeMs for the simulated
CVL data.  Table~\ref{table:yp} compares the $1\sigma$ error bars on
$Y_{\rm p}$ in these cases.  We see that for the CVL case similar
number of XeMs and eXeMs used as the eigenmodes lead to similar
constraints on helium abundance.  However, a Planck-ACTPol-like
observation gives a few times larger error due to lack of very high
sensitivity to very small scales, although fewer XeMs compared to the
CVL case have been used.
\subsection{Case 2: A perturbed recombination scenario}\label{sec:JR_MCMC}
As the second example of parameter estimation and perturbation
reconstruction, we simulate {\it data} assuming the recombination
calculation of CT2010 (Fig.~\ref{recon}), while we take the fiducial
model to be as of {\sc Recfast} v1.4.1 or older (equivalent to SRS
with ${\rm He_{switch}}=0$, to remove the helium correction which has
been assumed as part of the {\it perturbations} in the data). The
purpose here is to find out how well the biases in the standard
parameters due to this {\it lack} of knowledge about the physical
corrections can be removed by including the perturbation eigenmodes,
and whether or not, {\it data} can reconstruct part of the true
recombination history.

Constructed from {\sc CosmoMC} chains for a CVL experiment,
Fig.~\ref{JR_CVL} illustrates the $2$D-contours of some of the
standard parameters.  The large biases in the estimated values of the
parameters when only the six standard parameters are measured is due
to the mismatch between the ionization histiry in the theoretical
model and the {\it data}. Here only contours for parameters with the
largest biases are shown. See also \cite{sha11}.  To compensate for
this mismatch we separately add to the parameters the two different
sets of the eigenmodes, the XeMs and eXeMs, as the new parameters.

As Fig.~\ref{JR_CVL} demonstrates, this eliminates the biases in the
standard parameters, however, at the cost of increased error bars.  In
particular for $\Omega_{\rm b} h^2$ the difference is large.  The
error (corresponding to $\sim 5\sigma$ without eigenmodes) increases
by a factor of $\sim 5$ when the (e)XeMs are included.  For $n_{\rm
  s}$, similarly, the error (of $\sim 7\sigma$) decreases by a factor
of 2. Also for $A_{\rm s}$ we find a similar degradation, while for
$\tau$ the difference is rather small.

Our computations indicate that with the XeMs as the eigenmodes and for
a CVL observation, the minimum number of modes required to remove the
bias from these standard parameters is six. However, we included eight
modes in the analysis to take into account the mode-selection
criterion of \S~\ref{num_EM} (determined by the relative injection
jumps).

We also observe that the recovered values for the amplitude of the
XeMs are biased (compared to the theoretically expected values from
direct projection on the XeMs, Table~\ref{CT_proj}), and change by
varying the number of modes included in the analysis. That is due to
the correlation of the XeMs in the presence of the standard
parameters, and the fact that not all XeMs are included into the
parameter estimation.  As a results, parts of the perturbation that
project on the neglected higher XeMs leak into the lower XeMs. The
bias in the measured XeM amplitudes is similar to the bias in the
standard parameters when there are no eigenmodes in the analysis, but
with a much lower significance.

For the same reason, the errors on the XeMs also change when the
number of modes included in the analysis changes.  However, as
mentioned before, due to the low significance of the perturbation
detection for most of the XeMs this is not as important as for the
main cosmic parameters.  For the CVL simulations with six and eight
XeMs included, we see that the most significant contribution comes
from the first mode (respectively $\mu_1= -0.23\pm 0.05 $, $\mu_1=
-0.18\pm 0.04$) while the other modes are consistent with zero. This
is also true for a Planck-ACTPol-like case, which we will come to
shortly, where the first XeM is measured to be $\mu_1= -0.22\pm 0.06 $
$\mu_1= -0.24\pm 0.12 $ for one and three XeM measurements.

\begin{figure*} 
\begin{center}
\includegraphics[scale=0.7]{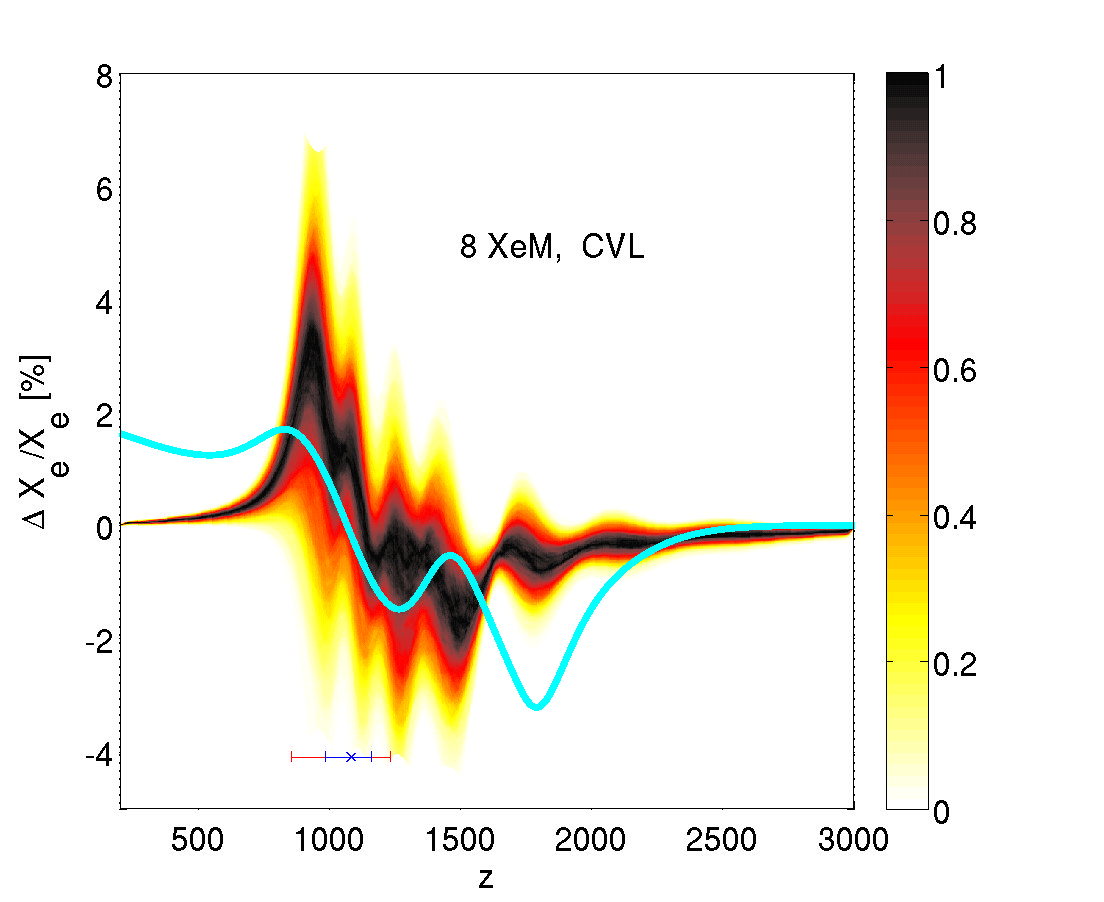}
\includegraphics[scale=0.7]{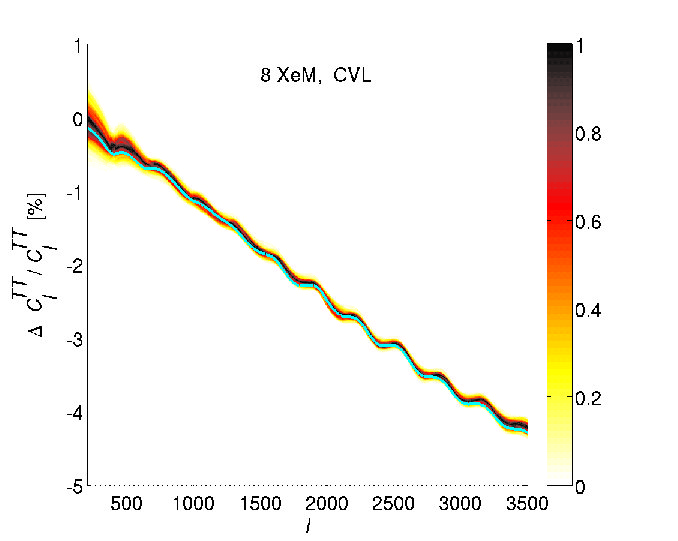}
\includegraphics[scale=0.7]{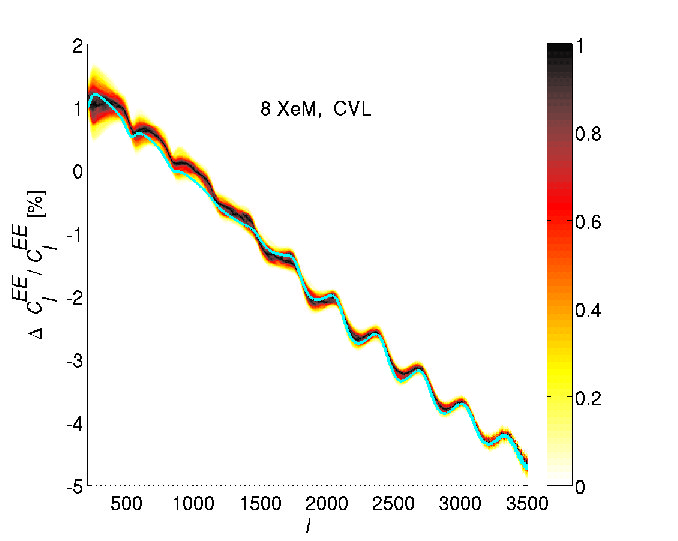}
\end{center}
\caption{Left: The $\delta \Xe /\Xe$ as measured by a CVL experiment
  by including eight XeMs (and six standard parameters) in the
  analysis. The colors show the density of trajectories going through
  each point in the $z$-$\delta \Xe/\Xe$ space, normalized to one at
  each $z$. The maximum and $1$ and $2\sigma$ widths of the Thomson visibility function have been marked at the bottom of the plot.  As this plot and the next ones indicate, the main recovery of $\Xe$ is the slope of the cure around this visibility peak. Middle and right: similar to the left figure, but for
  $\delta C_\ell^{TT,EE}/C_\ell^{TT,EE}$ trajectories.}
\label{JR_CVL_traj_8EM}
\end{figure*}
\begin{figure*} 
\begin{center}
\includegraphics[scale=0.7]{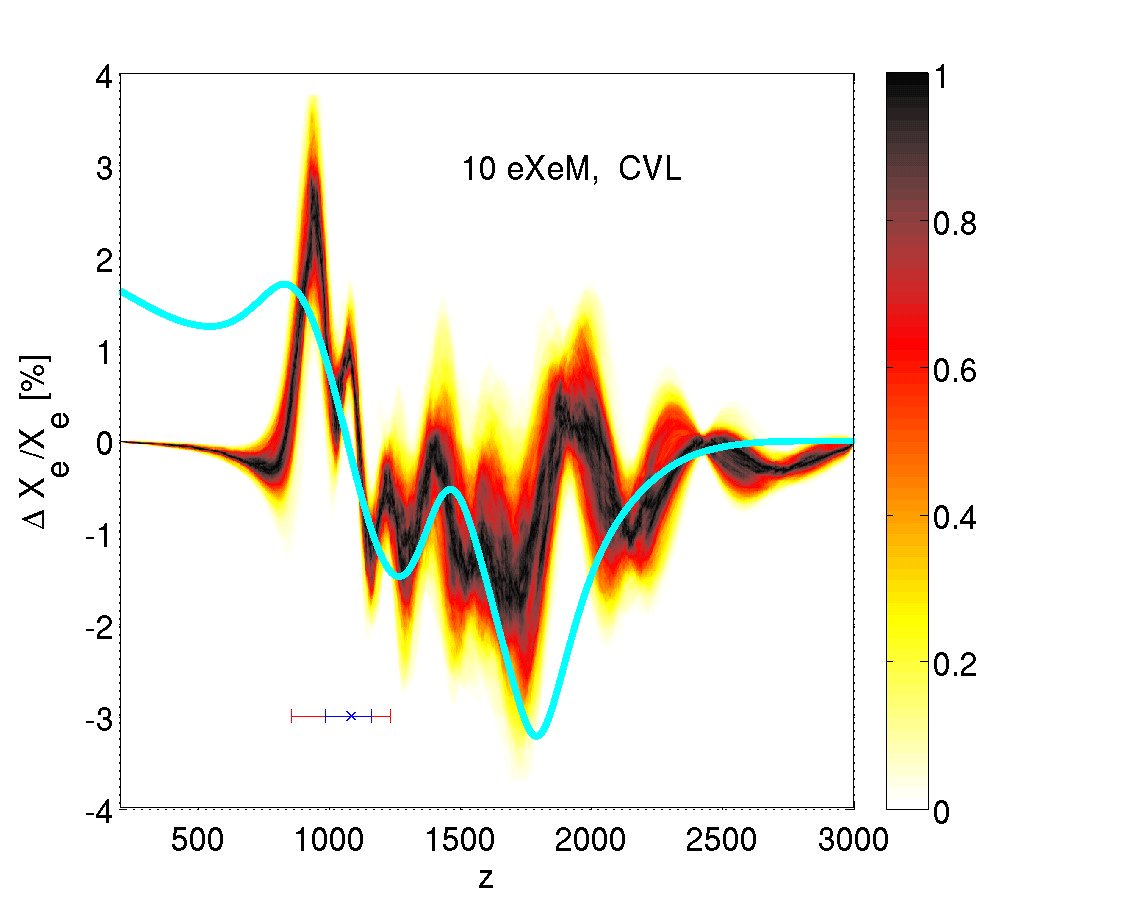}
\includegraphics[scale=0.7]{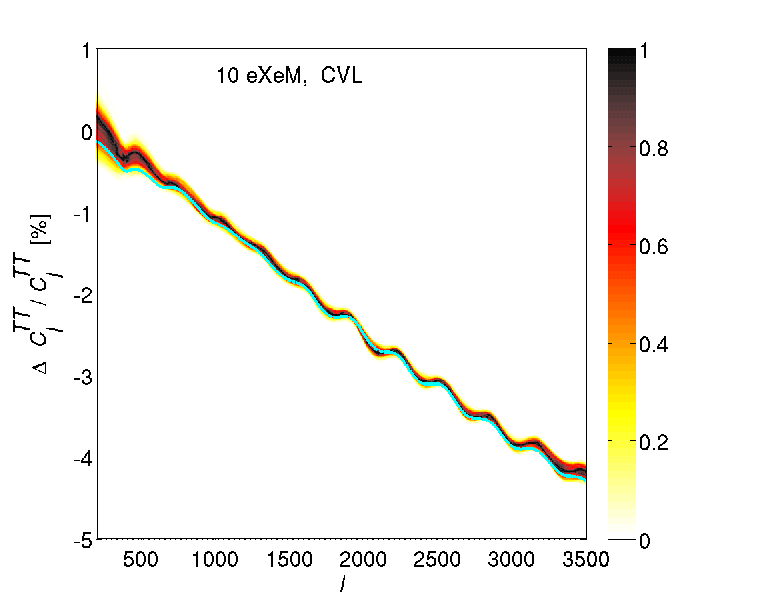}
\includegraphics[scale=0.7]{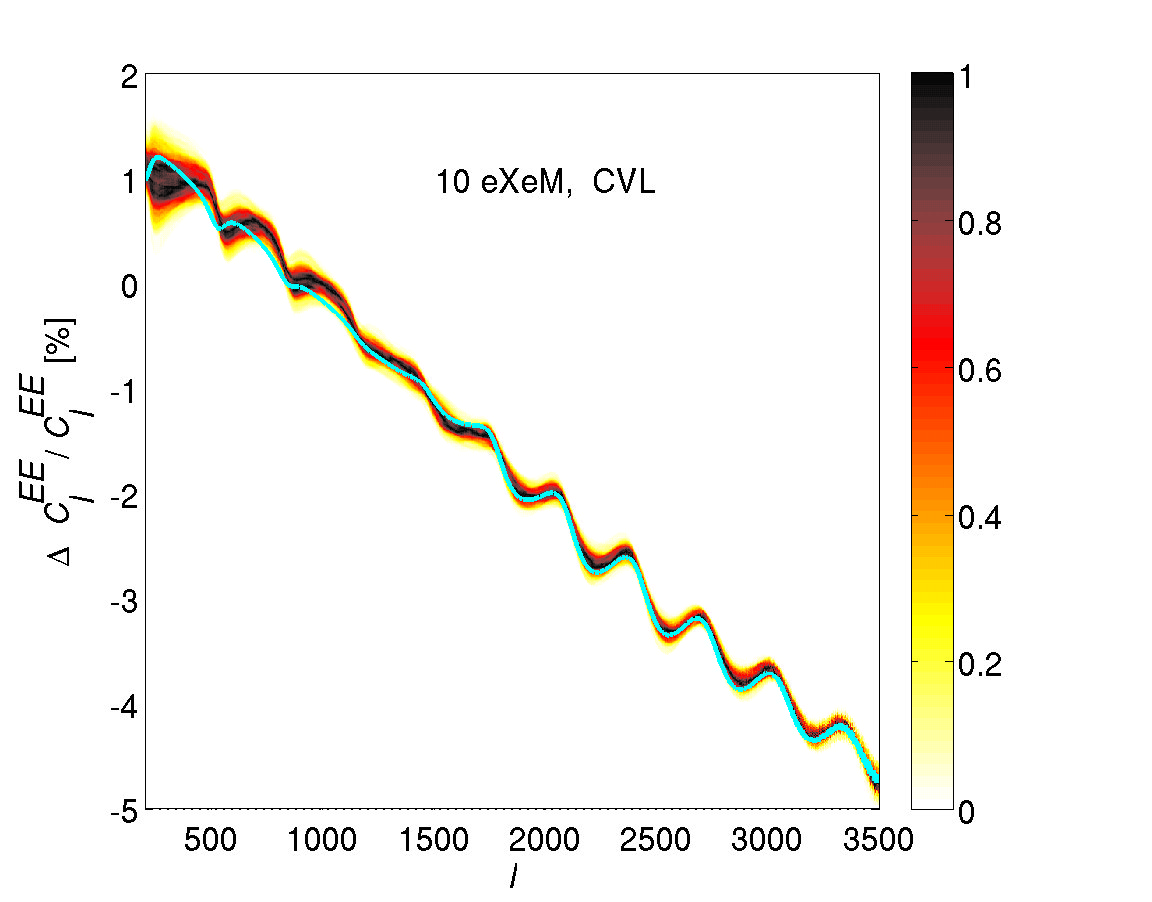}
\end{center}
\caption{Similar to Fig.~\ref{JR_CVL_traj_8EM} but with the first ten eXeMs.}
\label{JR_CVL_traj_10eXeM}
\end{figure*}
\begin{figure*} 
\begin{center}
\includegraphics[scale=0.7]{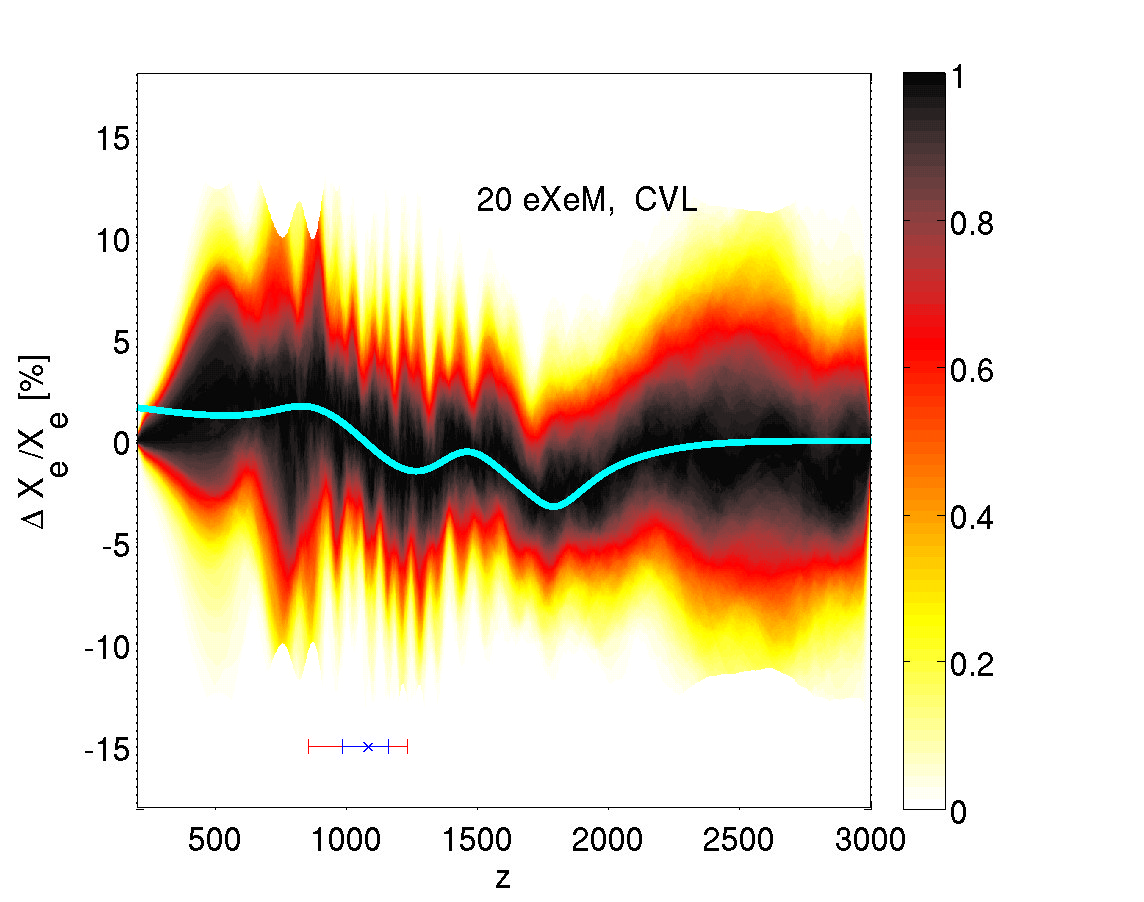}
\includegraphics[scale=0.7]{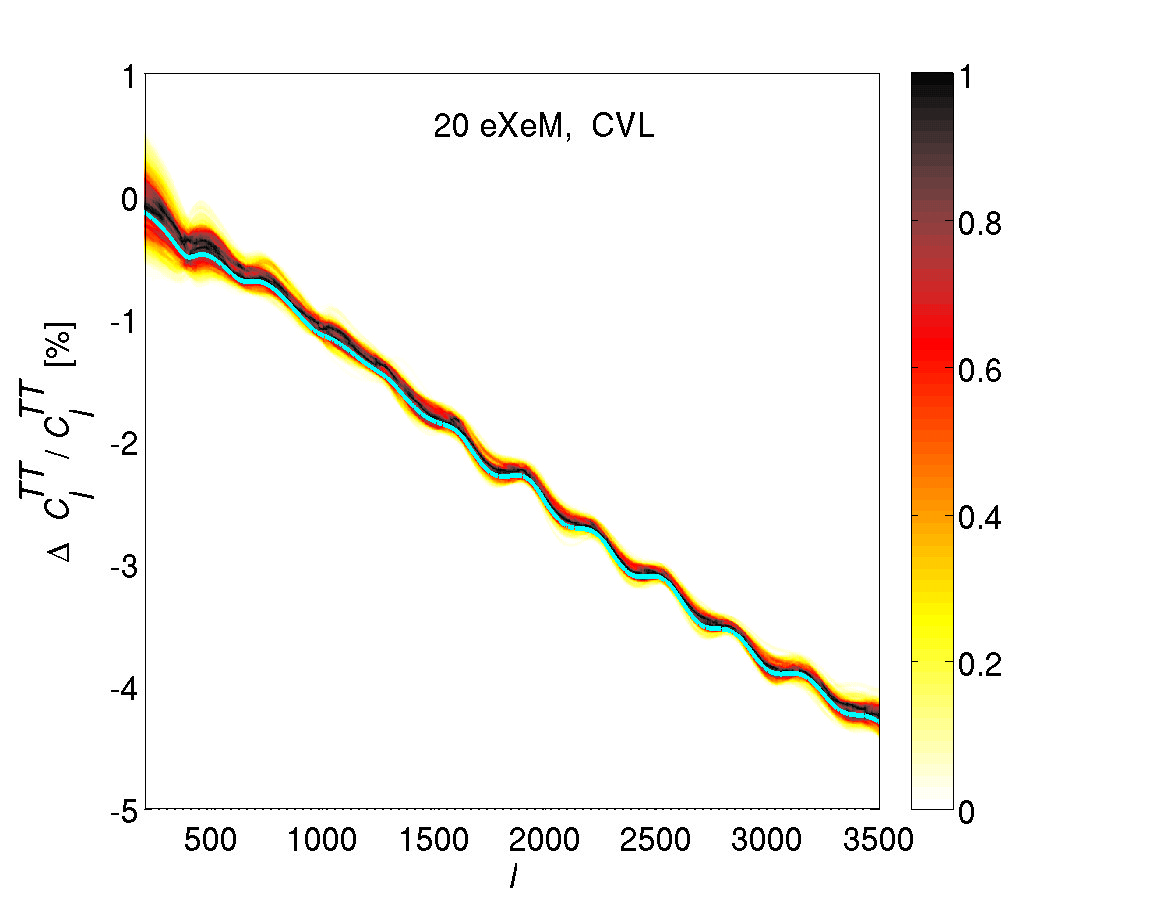}
\includegraphics[scale=0.7]{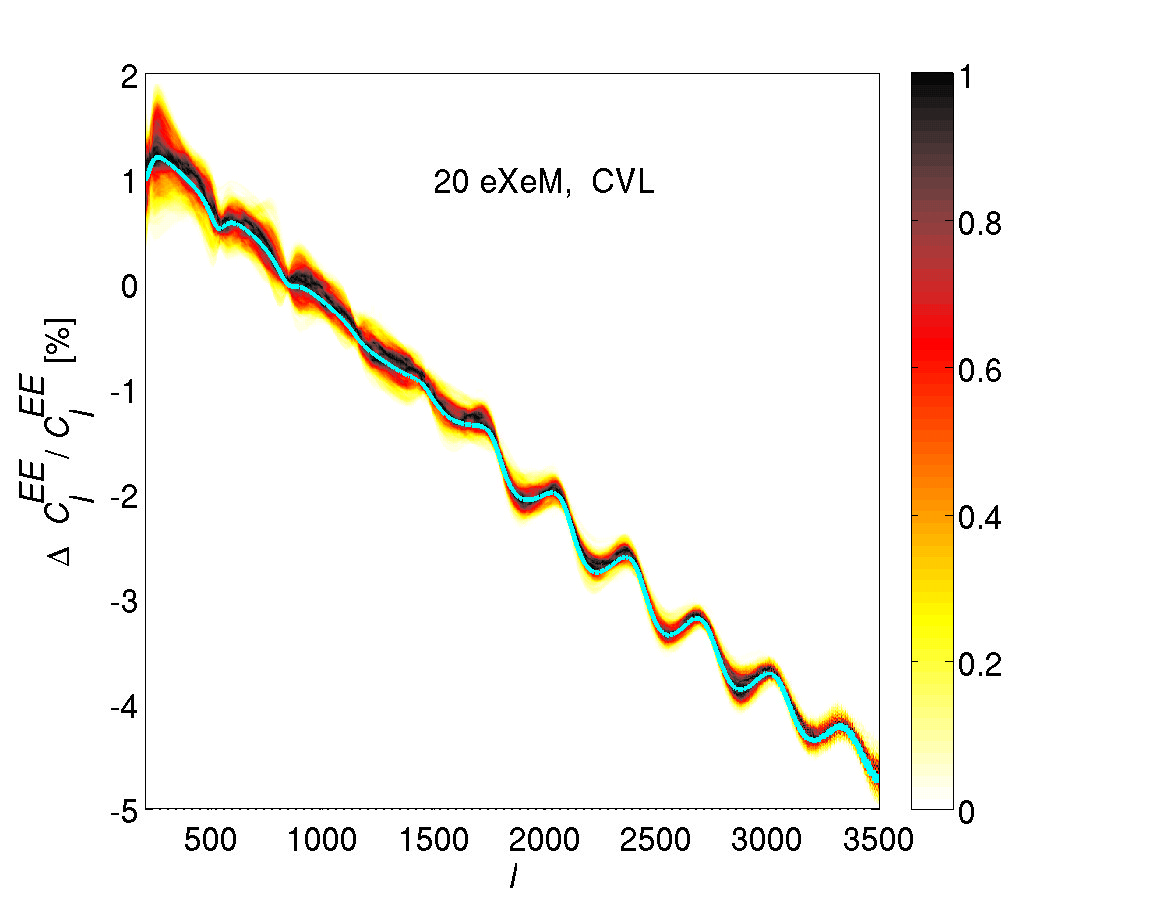}
\end{center}
\caption{Similar to Fig.~\ref{JR_CVL_traj_8EM} but with the first
  twenty eXeMs. As this figure demonstrates, including a higher number
  of modes does not necessarily lead to better $\Xe$ recovery. Here
  the recovered $\Xe$ becomes noisier compared to the case with only
  ten modes included, while the $C_{\ell}$ trajectories do not change
  significantly except for the diminished oscillations around the
  input model, as discussed in the text. }
\label{JR_CVL_traj_20eXeM}
\end{figure*}

If instead eXeMs are used as the perturbation eigenmodes, our
computations show that at least ten modes should be added to get rid
of the bias for a CVL case.  However, as a test case, we tried
including the best 20 eXeMs (see Fig.~\ref{JR_CVL}).

We also found that although the errors on the standard parameters keep
increasing by adding more eXeMs to the analysis up to around the tenth
mode (which is required to remove the bias) it stays more or less the
same afterwards. This suggests that in terms of the constraints on the
standard parameters, we do not lose much by increasing the number of
eXeMs.  Besides, including more eXeMs does not affect the measurement
of the previously included eigenmodes, as they are by construction
uncorrelated (in the presence of standard parameters).  Including more
eXeMs, on the other hand, makes the reconstructed perturbations closer
to the input model of perturbations (as in Fig.~\ref{recon}). However,
as the errors of modes increase by going to higher orders, the error
on the reconstructed curve increases.  We will address this point in
the next section.

Among the first 20 eXeMs for a CVL experiment, the modes with the most
significant contributions (i.e., with at least $1\sigma$ detection)
are $\mu_2=0.11\pm 0.02$, $\mu_3=0.10\pm 0.03$, $\mu_9=-0.31\pm 0.16$
and $\mu_{11}=-0.36\pm 0.24$ (compare to their theoretical prediction
from direct projection of the perturbations on the eXeMs:
$\mu_2=0.14$, $\mu_3=0.10$, $\mu_9=-0.33$ and $\mu_{11}=-0.39$).  The
reason that the recovered value, though close, is not exactly the same
as the forecast is that the assumption of the Gaussianity of the
distributions of the eXeMs and the standard parameters is only
approximate. Also the eigenmodes have been slightly smoothed in the
construction process which may cause numerical inaccuracy and induce
slight correlation between the smoothed modes. By comparing the
theoretical values of projection of the perturbation on the eXeMs and
their forecast errors (from Fisher analysis) we do not expect any
perturbation detection after eXeM 11.

Fig.~\ref{JR_PlApol} shows similar contours but for a simulated
Planck-ACTPol-like observation.  For the analysis we used the
eigenmodes (both eXeMs and XeMs) constructed with the Planck-ACTPol
simulated noise.  The results from the two sets of eigenmodes are very
similar. For both XeMs and eXeMs, one mode was sufficient to remove
the bias ($\mu_1=-0.22\pm 0.06$ and $\mu_1=-0.20\pm 0.06$
respectively).  This happens to be in agreement with the cutoff mode
for the XeMs while with the eXeMs the second mode should also be
included.  The lower number of modes required for the
Planck-ACTPol-like case compared to the ideal experiment is expected
due to higher sensitivity of the data in the latter to deviations from
the underlying $\Xe$ history.  We also tried three modes, with no
significant detection of the new modes, while the error on the XeM 1
increases by a factor of 2.
\subsection{Trajectories}\label{sec:traj}
In this section we investigate the reconstruction of the
$\Xe$-perturbations using the simulated data to illustrate the
corresponding uncertainty at different redshifts.  The left plot in
Fig.~\ref{JR_CVL_traj_8EM} shows the redshift interval covered by
$500$ $\delta \ln \Xe$-trajectories corresponding to an ideal
observational case with eight XeM included, for the CT2010 model.  The
color indicates the number of trajectories passing through each
$(z,\delta \Xe /\Xe)$ bin, normalized to one at each redshift
snapshot.  The trajectories clearly show deviations from the SRS,
slowly morphing into the correction obtained by CT2010 (the cyan
curve). However, the recovery is not perfect, as the model of CT2010
has non-zero (and relatively significant) projection on higher XeMs
which are not well constrained by data, and therefore were not
included into the analysis. Most obviously, corrections to helium
recombination are not captured well when using only the first few
XeMs.  These trajectories do not recover the analytical projection of
the CT2010 corrections on the first eight XeMs very well either. The
reason, as discussed before, is that the correlation of the XeMs
induced by the standard parameters draws some contribution from the
higher absent modes which biases the measurement of the first few XeMs
included in the measurement.

To test this impact of higher, excluded modes on the recovered (low
XeM) trajectories, we ran simulations with the data that only
accounted for the contributions from these low modes.  As expected, in
the absence of higher modes in the data, the measured XeMs were
non-biased and thus the highest probability region of the trajectories
covered the $\delta\ln \Xe$ curve of the input model.

Although our basic target is $\Xe$-recovery, the relevant space for
determining how well we have done is that of the CMB data, reduced to
the power spectra, $C_\ell^{TT}$ and $C_\ell^{EE}$.  The central and
right panels of Fig.~\ref{JR_CVL_traj_8EM} show the $\delta
C_\ell^{TT}/C_\ell^{TT}$ and $\delta C_\ell^{EE}/C_\ell^{EE}$
trajectories, where $\delta C_\ell^X/C_\ell^X=(C_\ell^X-C_{\ell}
^{X,{\rm fid}})/C_{\ell}^{X,{\rm fid}}$ and $C_\ell^{\rm fid}$ is the
fiducial power spectrum without any perturbations. The transformation
from $\Xe$ trajectories to $C_\ell$ trajectories shows a much tighter
band around the input signal. This is a visual confirmation of the
point that some features in the $\delta \ln\Xe$ which make the $\Xe$
trajectories thick do not leave a measurable imprint on the
$C_\ell$'s.  Notice there are small residual oscillations. They
coincide with the peaks and troughs of the $C_\ell$ curves for both
$TT$ and $EE$ (which is out of phase with $TT$). One source for the
oscillations seems to be the eigen mode truncation, as we will see
later.  Using only a limited number of the modes in the analysis
causes the non-$\Xe$ cosmic parameters to try to match the injected
$\Xe$ perturbations.  There is also an issue of accuracy of the
$C_\ell$ code for some of the distortions.

\begin{figure*} 
\begin{center}
\includegraphics[scale=0.7]{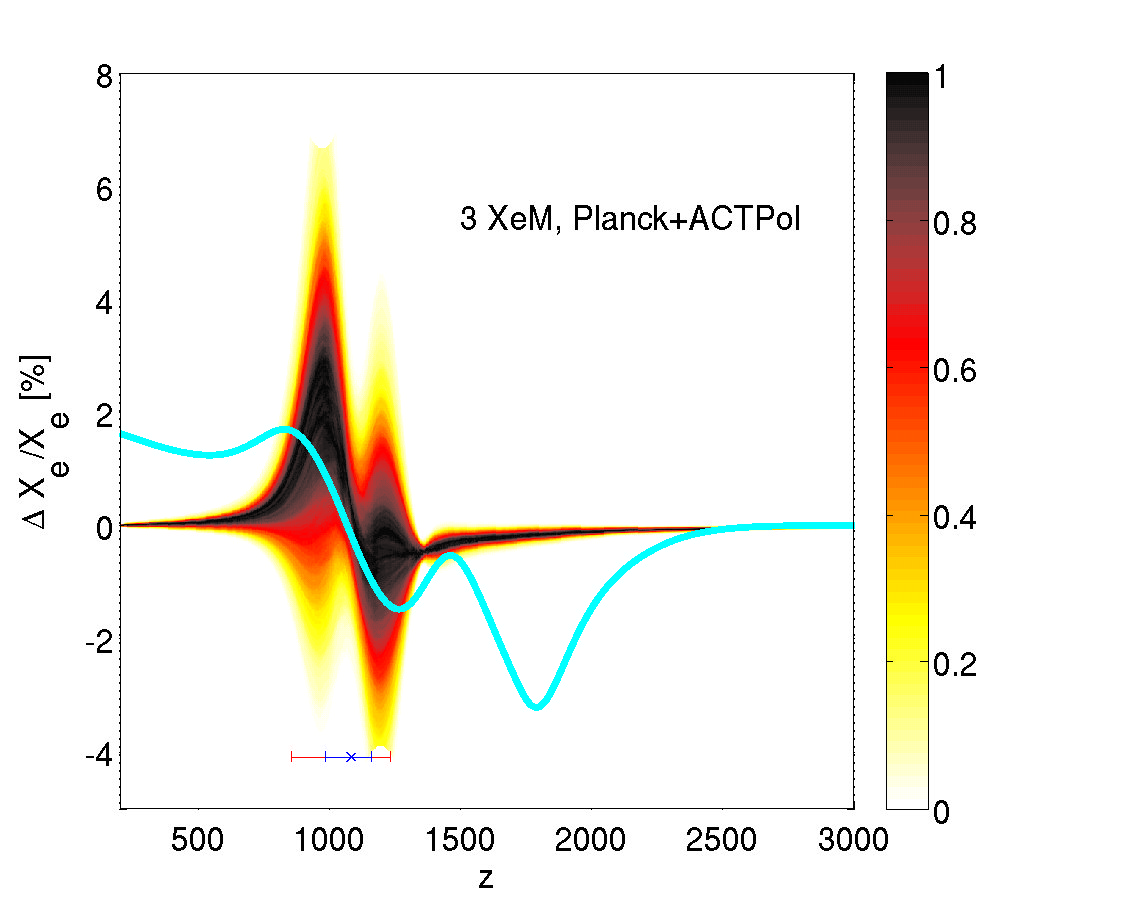}
\includegraphics[scale=0.7]{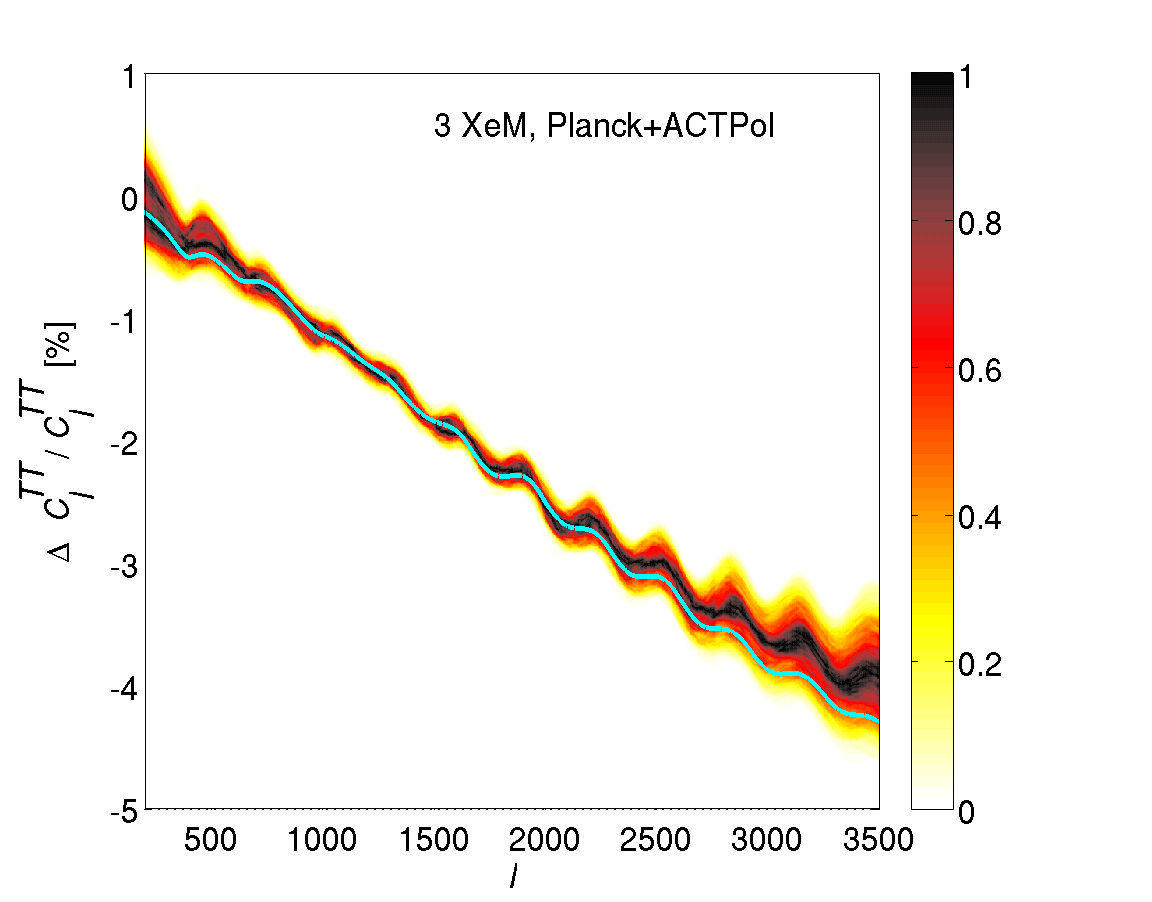}
\includegraphics[scale=0.7]{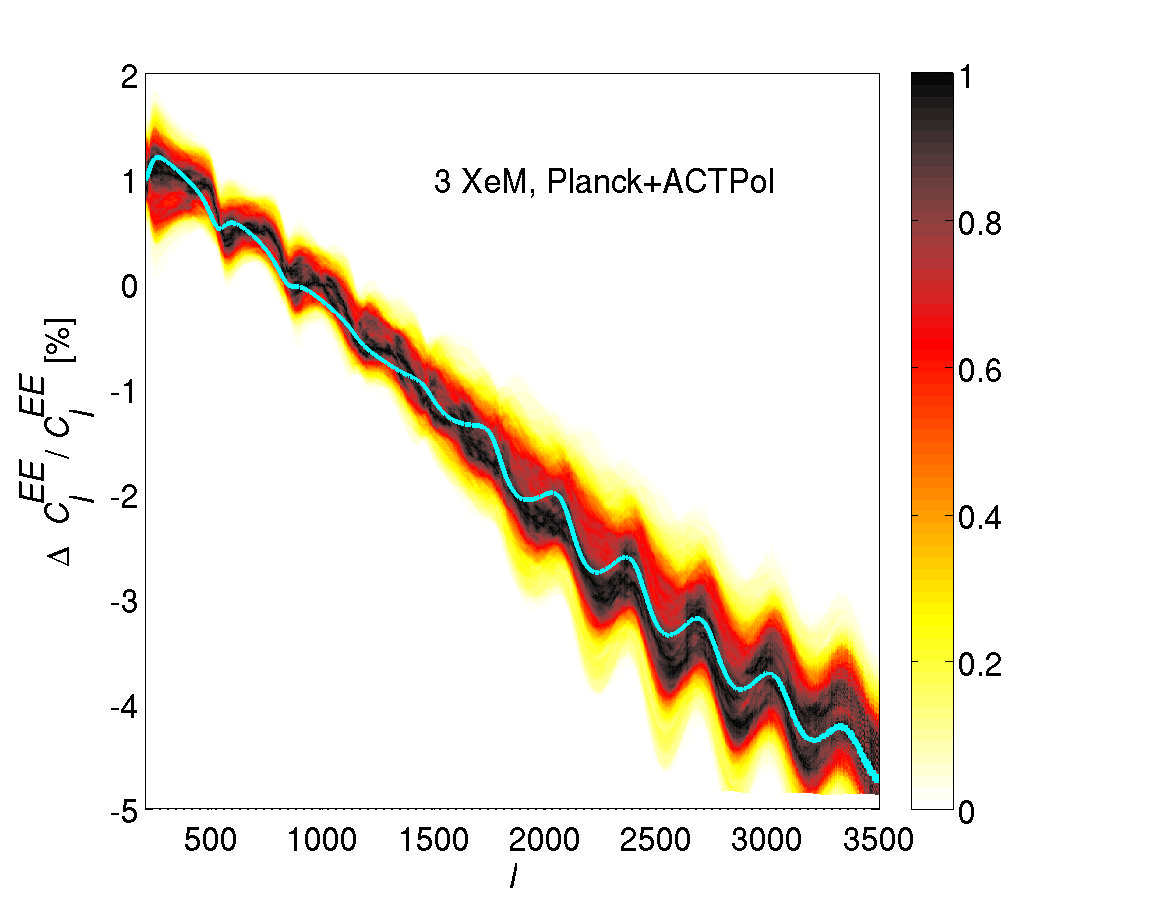}
\end{center}
\caption{Similar to Fig.~\ref{JR_CVL_traj_8EM} but for a Planck-ACTPol-like
 experiment and with only three XeMs taken into account.}
\label{JR_pl_traj_3EM}
\end{figure*}
\begin{figure*} 
\begin{center}
\includegraphics[scale=0.7]{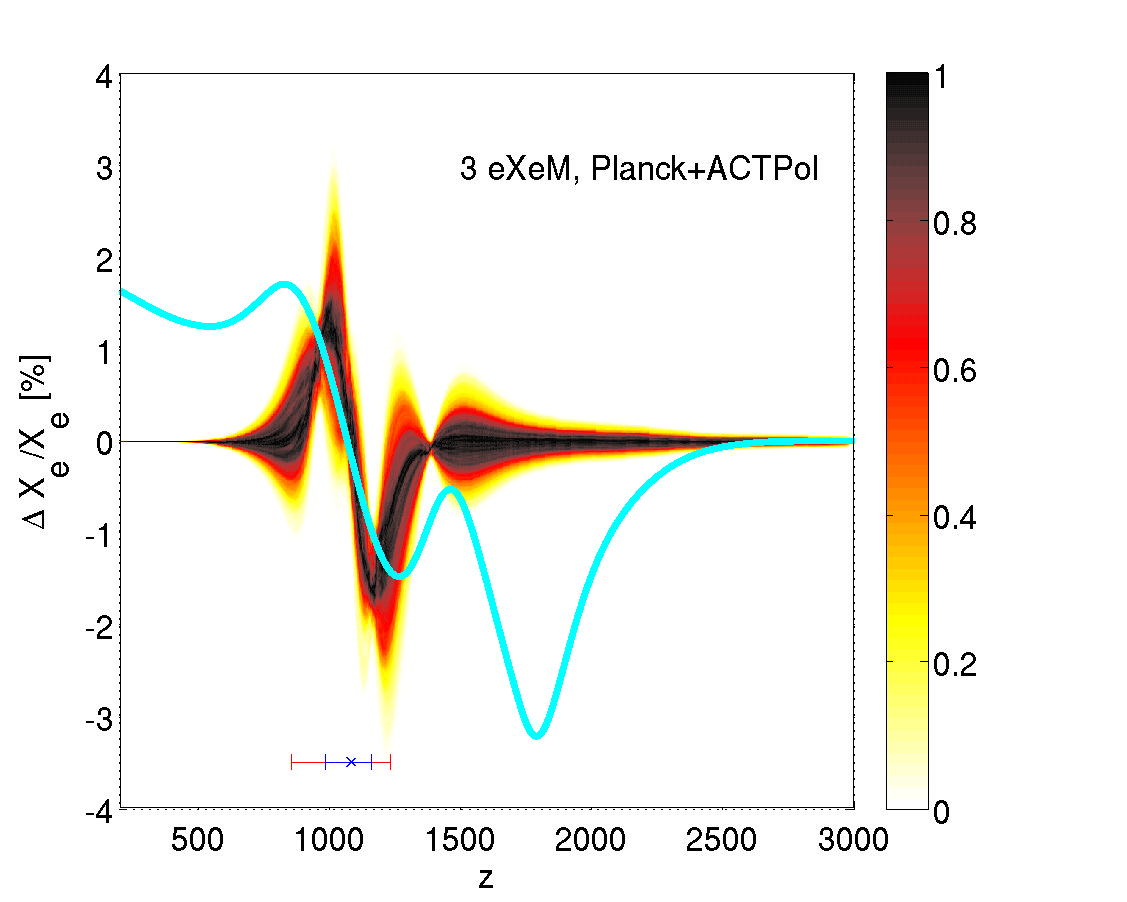}
\includegraphics[scale=0.7]{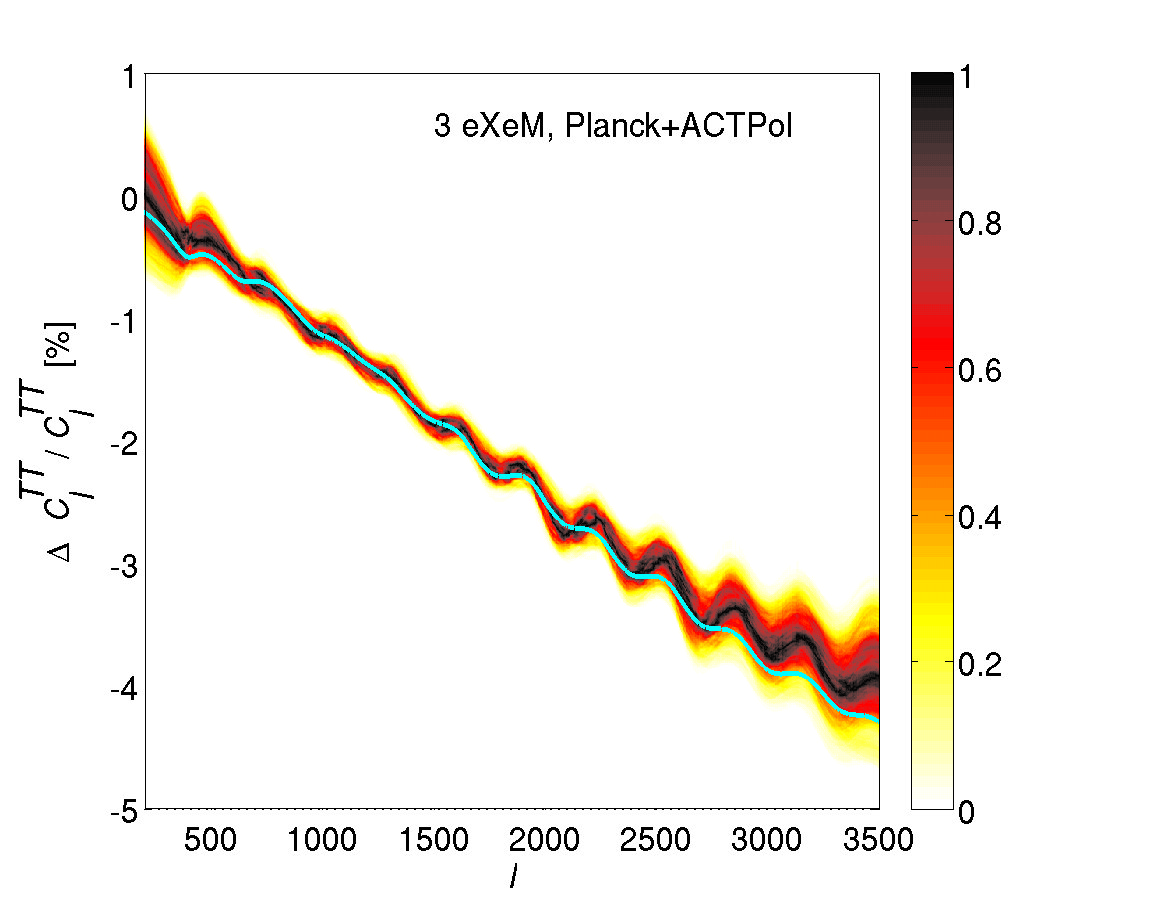}
\includegraphics[scale=0.7]{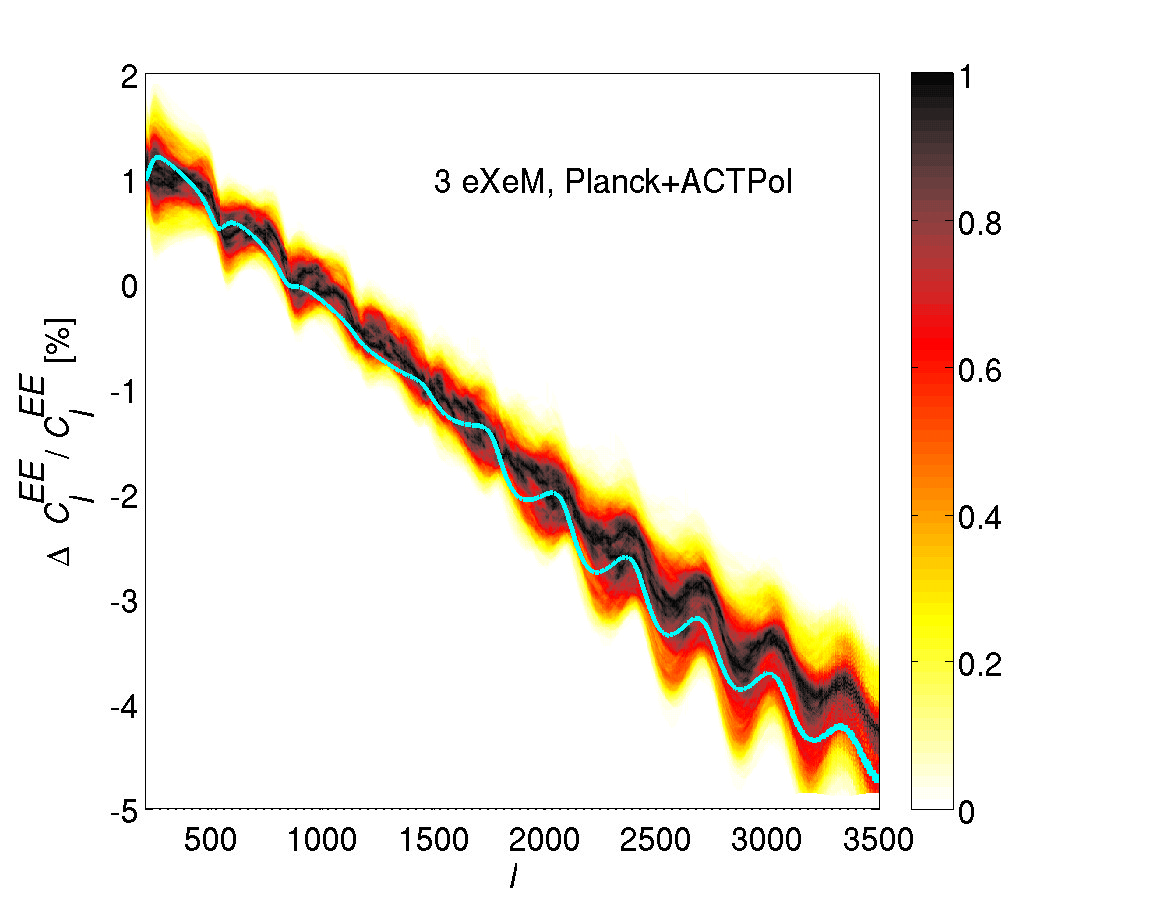}
\end{center}
\caption{Similar to Fig.~\ref{JR_CVL_traj_10eXeM} but for a Planck-ACTPol-like
 experiment and with only three eXeMs taken into account.}
\label{JR_pl_traj_3eXeM}
\end{figure*}

Fig.~\ref{JR_CVL_traj_10eXeM} similarly shows the $2$D histograms of
trajectories for the case with the first ten eXeMs included. Around
the maximum of the Thomson visibility function the $\Xe$
reconstruction is slightly stronger and less fuzzy than in
Fig.~\ref{JR_CVL_traj_8EM}, with part of the helium recombination
correction being recovered.  The improvement in the reconstruction is
because for the computation of these eigenmodes their correlation with
the standard parameters have been optimally taken care of.  In
contrast, the XeMs used in the previous case are non-optimal if no
strong additional priors can be placed on the standard parameters,
leading to confusion in the errors and the rank-ordering of the modes.
Fig.~\ref{JR_CVL_traj_20eXeM}, constructed with 20 eXeMs included,
shows that the oscillation effect mentioned above around the input
$C_{\ell}$ signal is diminished (and also partially swamped by the
slightly higher dispersion around the input curve) for the 20-mode
compared to the ten-mode case.  We also see that including a higher
number of modes does not necessarily lead to better $\Xe$ recovery.

Similar trajectories for a Planck-ACTPol-like experiment are shown in
Figs.~\ref{JR_pl_traj_3EM} and ~\ref{JR_pl_traj_3eXeM}, with three
XeMs and three eXeMs as the eigenmodes respectively.  The trajectories
for the XeMs are more widely spread and blurred due to experimental
noise.  The eXeMs perform slightly better. However, the overall
reconstruction is clearly lacking detailed agreement with the full
recombination correction of CT2010. In particular, most of the
modification during helium recombination is not captured, as the
corresponding signals can only be picked up with higher modes, which
in the considered case are not constrainable at a significant level. 
In the $\delta C_\ell/C_\ell$ plots of these Planck-ACTPol-like cases, there is a small disagreement at high multipoles  between the theoretical curve and the highest probability region of the chains. That is mainly due to mode truncation at a relatively low mode number, i.e., three. We tested this by including eight modes and as expected, observed a wider spread around the input signal with the disagreement diminished.

Although we do not plot the equivalent $\delta C_\ell/C_\ell$ for the
DM case discussed in \S~\ref{sec:recon_DM}, very similar plots result,
namely good recovery of the power spectra with a dispersion around the
input perturbation signal.

\subsection{Beyond small perturbations}
\label{sec:beyond_small}
In this paper it was explicitly assumed that the model best explaining
the ionization fraction (or the true model underlying the ionization
history) is only slightly different from our fiducial model,
justifying our choice of parameter $\delta \ln \Xe$. Therefore, the
eigenmodes constructed for the {\it fiducial} model are also very
close to the eigenmodes for the perturbations to the {\it true} $\Xe$,
the corrections to the eigenmodes arising from the difference between
the fiducial and true $\Xe$ model being only of second order.  Under
this assumption, a one step search for the best-fit parameters
suffices to extract the available relevant information from the data,
provided that the minimum required number of modes are included in the
analysis.  Finding the minimum number of required modes can by itself
involve several parameter estimation steps in parameter spaces with
different dimensions, the criterion being that the best fit values for
the standard parameters stop changing.  That is what was done in the
examples in this work (\S~\ref{sec:JR_MCMC}), to illustrate how the
method works.
 
 However, if the fiducial model is {\it very} far from the true $\Xe$
 history, such that the eigenmodes are expected to be affected at a
 significant level, an iterative approach toward finding the best
 modes with their associated amplitudes and errors is required:
 starting with our best guess for the fiducial model, we estimate its
 deviation from the true ionization history using the dataset
 available and the eigenmodes constructed based on this fiducial
 model.  We then update the model by adding to it the measured
 deviations in the eigenmodes (and the standard parameters, if
 required). This process is repeated until the convergence of the
 model and its eigenmodes.
 
However, current constraints seem to indicate that such an iterative
procedure will not be necessary within the standard picture. For
example, as shown by \citet{sha11}, the recombination corrections of
CT2011 are readily incorporated using one calibrated redshift
dependent correction function relative to the original recombination
model of \citet{sea99}. Even for CVL errors a second
update of the correction functions leads to minor effects.
Nevertheless, if something more surprising occurred during
recombination, an iterative approach might be required.

\section{conclusion and discussion}
CMB data today are becoming so precise that small modifications in
standard ionization history are important. This impressive progress
not only implies that measurements of the main cosmological parameters
are becoming increasingly accurate, but also means that remaining
uncertainties in the recombination dynamics, e.g., caused by neglected
standard or non-standard physical processes, should be quantified.  In
this work we discuss a novel approach to constrain this remaining
ambiguity with future CMB data.  We performed a principal component
analysis to find parameter eigenmodes that can be used to describe
uncertainties in the ionization fraction.  We constructed $\Xe$
eigenmodes over the redshift range of $[200,3000]$, performing several
consistency checks to prove the correctness of our method.  This
approach automatically delivers a hierarchy of mode functions that can
be selected according to their error and then are added to the
standard cosmological parameters when analyzing CMB data.

Due to the strong CMB signal imprinted by hydrogen recombination, the
most constrained modes are mainly localized around $z\sim 1100$, with
some extensions to lower and higher redshift regions (see
Fig.~\ref{multibasis} and \ref{eXeM_6_7}).  This emphasizes that CMB
data are very sensitive to small changes during hydrogen
recombination, while details of helium recombination or small changes
in the freeze-out tail are hard to constrain, unless strong priors on
the reliability of the hydrogen recombination model are imposed.  With
the method described here it is possible to construct mode functions
for different experimental situations, also folding in prior knowledge
on the recombination history using appropriate weight functions and
fiducial $\Xe$ models.  For example, if there are physically motivated
and experimentally supported hints toward (significant) changes in the
freeze-out tail of recombination, e.g., due to energy injection from
dark matter annihilation, we propose a parametrization which weights
the low redshift part more strongly (see Fig.~\ref{multisig}).

After we completed this work, we received a preprint \citep{fin11}
which investigated the use of CMB data to constrain details of energy
injection scenarios related to decaying or annihilating
particles. They also used parameter eigenmodes, but these were
constructed based on an energy release history which is in our
language akin to the imposition of a strong prior on the recombination
dynamics around $z\sim 1100$ and a focus on the freeze-out tail of the
recombination.

We applied the method to different simulated datasets with the aim to
assess how well future CMB experiments will be able to constrain
modifications to the standard recombination scenario. (Current WMAP
plus ACT and SPT data will provide only relatively weak constraints,
but Planck plus ACTPol and SPTPol will considerably improve the
situation.)  As a working example we used the refined recombination
calculations of {\sc CosmoRec}.  For simulated CMB datasets
corresponding to Planck-ACTPol-like experiments we found that the
first 3 eigenmodes can be rather well constrained.  The addition of
these modes allows us to partially morph from the old {\sc Recfast}
$\Xe$ model to the new recombination history computed with {\sc
  CosmoRec} without actually directly using the recombination
corrections in the analysis.  However, because the first few mode
functions are strongly localized around $z\sim 1100$ details during
helium recombination and in the freeze-out tail are not captured
(Fig.~\ref{JR_pl_traj_3eXeM}).  The addition of the first 3 eigenmodes
is sufficient to remove the biases in the standard parameters,
however, at the cost of increased error bars.  We also show that for
CVL limited experiments up to $l\sim 3500$ up to 10 modes might be
constrainable, in this case allowing us to pick up part of the details
during helium recombination (Fig.~\ref{JR_CVL_traj_10eXeM}).

The significance of the detection of any perturbation obviously
depends on the underlying ionization history of the real data. In the
specific {\sc CosmoRec} example for Planck-ACTPol-like experiments,
all three modes but the first one are consistent with zero.  A
significant source for large errors on the eigenmodes is their
correlation with the standard parameters. If tight constraints are
imposed on the standard parameters by non-CMB experiments such as BAO
or SN data, the errors on the eigenmodes will be correspondingly
reduced.  Comparing the first rows of Table~\ref{eval} (where all
standard parameters are held fixed) and Table~\ref{evaleXeM} (where
all standard parameters are being marginalized over) illustrates the
effect of this correlation in the extremes.

This also shows how important one's knowledge in how well elements of
recombination are known, expressed through prior probabilities, will
be.  If the uncertainty in the ionization history during hydrogen
recombination can be reliably reduced by other methods then the
sensitivity to small perturbations at higher or lower redshifts is
enhanced.  We note that measurements of the cosmological recombination
radiation \citep[e.g., see][]{Chluba2006b, Sunyaev2009} could in
principle provide an alternative way of constraining the recombination
dynamics in the future.  In particular, the recombination radiation
could exhibit significant features if something more unexpected
occurred during different cosmological epochs \citep[e.g.,
  see][]{Chluba2008c, Chluba2010a}.

In principle, for a complete study of ionization history, late
reionization should also be included in the analysis. Ambiguities in
the low redshift part of the ionization history may affect the
measurements of high redshift perturbations and vice versa. However,
the main signal from the reionization epoch is measurable from the
very large scale CMB polarization, and the high redshift perturbations
of $\Xe$ affect anisotropies with smaller angular scales. Therefore
the signals from these two regions are rather uncorrelated.  A more
complete analysis for the whole ionization history or where different
parts of it are considered simultaneously is for future work.

An aspect requiring decision when analyzing real data is the choice of
parametrization.  For most of this work we weighted the perturbations
in $\Xe$ by the fiducial history.  If, for example, the recovered
perturbations point towards modifications in the freeze-out tail of
recombination, or if there is strong belief that no sign of
significant deviations around the maximum of visibility are present,
an alternative parametrization which allows better reconstruction of
the tail can be constructed, using appropriate weight functions that
quantify our belief in the underlying fiducial model.

As discussed in \S~\ref{sec:beyond_small}, our semi-blind XeMs are
designed to only probe small perturbations about the fiducial model
$\Xefid$. When it comes to real CMB data analysis, iterations of
$\Xefid$ are required to ensure no leftover bias remains.  We look
forward to the application of iteratively-improved eigenmodes to the
coming high resolution CMB data from Planck, ACTPol and SPTPol.

We thank Doug Finkbeiner, Tongyan Lin and Olivier Dor{\'e} for useful
discussions and Eric Switzer for his comments on the text. Support
from NSERC and the Canadian Institute for Advanced Research is
gratefully acknowledged.

\bibliography{bibtex/xe_pca}
\bibliographystyle{bibtex/apj}

\clearpage

\end{document}